\documentclass{jpp}
\usepackage{graphicx}
\usepackage{subfigure}
\usepackage{physics}
\usepackage{soul}

\usepackage[utf8]{inputenc}
\usepackage[T1]{fontenc}
\usepackage{amsmath}
\usepackage{color}

\usepackage{tikz}
\usetikzlibrary{positioning}
\usetikzlibrary{shapes.arrows}

\newcommand{\vp}{v_{\parallel f}}
\newcommand{\vpu}{v_{\parallel f}+u_f}
\DeclareMathOperator\erfi{erfi}

\shorttitle{Neoclassical transport in strong gradient regions of large aspect ratio tokamaks}
\shortauthor{S. Trinczek, F. I. Parra, P. J. Catto, I. Calvo and M. Landreman}

\title{Neoclassical transport in strong gradient regions of large aspect ratio tokamaks}

\author{Silvia Trinczek \aff{1}
  \corresp{\email{strincze@pppl.gov}},
  Felix I. Parra \aff{1},
  Peter J. Catto \aff{2},
  Iván Calvo \aff{3}
 \and  Matt Landreman \aff{4}}

\affiliation{
\aff{1}Princeton Plasma Physics Laboratory, Princeton, NJ 08543, USA
\aff{2}Plasma Science and Fusion Center, Massachusetts Institute of Technology, Cambridge, MA, USA
\aff{3} Laboratorio Nacional de Fusión, CIEMAT, Madrid, 28040, Spain
\aff{4} University of Maryland, College Park, MD 20742, USA}

\begin{document}

\maketitle
\begin{abstract}
We present a new neoclassical transport model for large aspect ratio tokamaks where the gradient scale lengths are of the size of the ion poloidal gyroradius. Previous work on neoclassical transport across transport barriers assumed large density and potential gradients but a small temperature gradient, or neglected the gradient of the mean parallel flow. Using large aspect ratio and low collisionality expansions, we relax these restrictive assumptions. We define a new set of variables based on conserved quantities, which simplifies the drift kinetic equation whilst keeping strong gradients, and derive equations describing the transport of particles, parallel momentum and energy by ions in the banana regime. The poloidally varying parts of density and electric potential are included. Studying contributions from both passing and trapped particles, we show that the resulting transport is dominated by trapped particles. We find that a non-zero neoclassical particle flux requires parallel momentum input which could be provided through interaction with turbulence or impurities. We derive upper and lower bounds for the energy flux across a transport barrier in both temperature and density and present example profiles and fluxes.
\end{abstract}

\section{Introduction}
The pedestal, and transport barriers in general, play an important role in tokamak performance \citep{Wagner84, Greenfield97} and thus it is useful to find a comprehensive transport model for these regions. In pedestals, for example, strong gradients of temperature, density and radial electric field of the order of the inverse ion poloidal gyroradius are observed \citep{viezzer2013}. Moreover, it has been found that the ion energy transport in pedestals is close to the neoclassical level \citep{Viezzer18}. Measurements of H-mode pedestals in Alcator C-Mod \citep{theiler14, churchill15} and Asdex-Upgrade \citep{cruz22} have shown poloidal variations of density, electric field and ion temperature that cannot be explained using standard neoclassical theory. It is thus desirable to extend neoclassical theory for stronger gradients, and logical to choose the ion poloidal gyroradius as the characteristic scale length. Comparisons of experimental data with standard neoclassical theory \citep{Hinton76} such as the one by \cite{Viezzer18} miss finite poloidal gyroradius effects.\par

Setting the scale length in transport barriers to be the poloidal gyroradius implies that the poloidal component of the $E \times B$-drift in large aspect ratio tokamaks becomes of the order of the poloidal component of the parallel velocity. As a result, a strong radial electric field shifts the trapped-passing boundary \citep{shaing94resonance}, and causes an exponential decrease proportional to the radial electric field in plasma viscosity \citep{shaing94resonance} and radial heat flux \citep{kagan10, shaing12}. The mean parallel flow is also affected by a strong radial electric field and can change direction \citep{kagan10}. A strong shear in radial electric field causes orbit squeezing, which reduces the heat flux and increases the trapped particle fraction for increasing radial electric field shear \citep{Shaing92, shaing94effects}.\par

Combining all these effects, \citet{shaing12} calculated the heat flux and mean parallel velocity but they neglected the strong mean parallel velocity gradient and the poloidal variation of the electric potential. \citet{Kagan08} and \citet{catto2013} have likewise developed extensions to neoclassical theory to allow for stronger density gradients to calculate fluxes. In \citep{kagan10, catto10, catto2013}, the density gradient was taken to be steep but the temperature gradient scale length had to be much larger than the ion orbit width. Furthermore, they assumed a quadratic electric potential profile and also neglected the poloidal variation of the potential. \par

Comparisons between analytical solutions and simulations have been carried out by \citet{landreman14}, which demonstrated the significance of source terms. \par

We will assume that the gradient length scale of potential, density and temperature is of the order of the poloidal gyroradius and we will retain the poloidal variations of density and potential. Assuming a large aspect ratio tokamak with circular flux surfaces in the banana regime and including unspecified sources of particles, parallel momentum and energy, we find equations for the ion distribution function, and a set of transport relations for ions. \par

In section \ref{sec:phase_space}, we justify our choice of orderings physically, and we motivate our choice of sources of particles, momentum and energy by considering the transition from the core into a transport barrier. A more detailed discussion of trapped and passing particles follows in section \ref{sec:banana_variables}, where the shift of the trapped-passing boundary is derived and a new set of variables based on conserved quantities is introduced. In section \ref{sec: banana regime} we calculate the ion distribution function in the trapped-barely-passing and freely passing regions. We also calculate the poloidally varying part of density and potential. The solvability conditions for the equation containing the distribution function of the bulk ions are the density, parallel momentum and energy conservation equations, calculated in section \ref{sec: Moment}. The ion transport equations are discussed further in section \ref{sec: Transport}. We find that a non-zero parallel momentum input is required to sustain a neoclassical particle flux and consider the possibility of interaction with turbulence. For the energy flux, we derive upper and lower bounds and relate the gradient lengths of temperature and density to the growth of neoclassical energy flux as one moves into the transport barrier.  We conclude by presenting some example profiles for the "high flow" case and the "low flow" case. A summary of our results is given in section \ref{sec: Results}.

\section{Orderings and phase space outline}\label{sec:phase_space}
\begin{figure}
    \centering
     \includegraphics[width=0.7\textwidth]{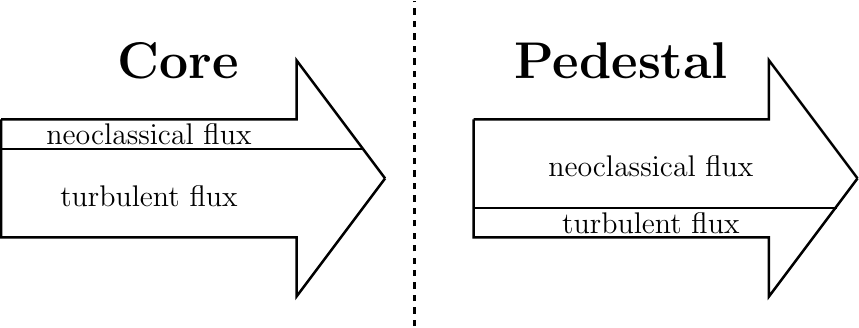}
    \caption{The total flux must be kept constant across the core and pedestal. The neoclassical contribution increases in the pedestal whereas the turbulent fluxes decrease as turbulence quenches. There is the possibility of interaction between turbulent and neoclassical transport in the pedestal.}
    \label{fig: CoreEdge}
\end{figure}

In this paper we consider the transition from regions with large turbulent transport into strong gradient regions. In a region of large turbulent transport, for example the core, neoclassical transport gives a minor contribution because turbulent transport carries most particles, momentum and energy. With the transition into a regime of low turbulence, like a transport barrier, the same total fluxes must be kept but as turbulence decreases, we anticipate that the turbulent transport goes down, too, and instead the fluxes must be picked up by neoclassical transport. Thus, we expect a rise in neoclassical fluxes at the transition from core to, for example, a pedestal (see figure \ref{fig: CoreEdge}). This argument is consistent with the observation that the energy flux in the pedestal is close to its neoclassical value \citep{Viezzer18}. We will see, however, that this simple picture of the top of a transport barrier has limitations. In section \ref{sec: Particle Flux} we find constraints that prevent the neoclassical fluxes from growing with radius.\par
Turbulence and neoclassical transport could interact in the transport barrier and hence we need to include a source $\Sigma$ in the neoclassical picture. This source represents any possible input from turbulence as well as external injection of particles, momentum and energy. The source must balance the neoclassical fluxes $\Sigma/f \sim n^{-1} \abs{\nabla \psi}(\partial \Gamma/\partial \psi)\sim n^{-1}T^{-1}\abs{\nabla \psi}(\partial Q/\partial \psi)$, where $\Gamma$ is the neoclassical particle flux, $Q$ is the neoclassical energy flux, $n$ is the density, $T$ is the ion temperature, and $\psi$ is the poloidal flux divided by $2\pi$, which we use as a flux surface label. To estimate the size of $\Sigma$, we need the size of the neoclassical particle and energy fluxes. We consider trapped and passing particles separately. \par

We can estimate the contributions from trapped and passing particles to particle and energy transport by making random walk estimates. The diffusion coefficient $D$ for a random walk is $ D \sim  (\Delta x)^2/\Delta t$, with $\Delta x$ and $\Delta t$ the random walk size and time, respectively. The neoclassical particle flux is thus
\begin{equation}
   \Gamma \sim \frac{(\Delta x)^2}{\Delta t}\frac{n}{L_n},
\end{equation}
where $L_n=|\nabla\ln n|^{-1}$. In a large aspect ratio tokamak, where $r/R\sim \epsilon\ll 1$, $r$ is the minor radius and $R$ is the major radius, the poloidal gyroradius is much bigger than the gyroradius. For passing particles we will show that the orbit widths are $\Delta x\sim \epsilon \rho_p$, where $\rho_p=qR\rho/r$ is the ion poloidal gyroradius, $q$ is the safety factor and $\rho$ is the ion gyroradius. The time between collisions is $\Delta t\sim 1/\nu$, where $\nu$ is the collision frequency. The gradient of density is assumed to be of the order of the poloidal gyroradius and so the particle flux due to passing particles is
\begin{equation}
    \Gamma_p \sim (\epsilon \rho_p)^2\nu \frac{n}{\rho_p}\sim \epsilon q \nu n \rho.
\end{equation}
The orbit width for trapped particles will turn out to be $\Delta x \sim \sqrt{\epsilon} \rho_p$, the collisional time is $\Delta t \sim \epsilon/\nu$ and again the density gradient length is $L_n\sim\rho_p$. The fraction of trapped particles in phase space is only $\sim\sqrt{\epsilon}$, and with that we arrive at a neoclassical particle flux due to trapped particles of order
\begin{equation}
    \Gamma_t \sim \sqrt{\epsilon}(\sqrt{\epsilon} \rho_p)^2\frac{\nu}{\epsilon} \frac{n}{\rho_p}\sim \frac{q}{\sqrt{\epsilon}}\nu n \rho.
\end{equation}
A comparison of the transport contribution from passing and trapped particles shows that the particle flux due to trapped particles is much larger,
\begin{equation}
    \frac{\Gamma_p}{\Gamma_t}\sim\epsilon^{3/2}\ll1.
\end{equation}
The same estimate can be performed for the neoclassical energy flux when substituting the energy gradient $nT/L_T$ for the particle density gradient $n/L_n$, where $L_T\sim L_n\sim\rho_p$. In section \ref{sec: Moment} we find transport equations that are consistent with this estimate and show that transport is dominated by trapped particles. \par
Using the sizes of particle and energy flux above, we can now give an estimate for the source $\Sigma$ that we have to introduce in the kinetic equation to mimic turbulence, particle, momentum and energy sources. The gradient of the particle flux is
\begin{equation}
   \abs{\bnabla \psi} \pdv{\Gamma}{\psi}\sim\frac{\Gamma_t}{\rho_p}\sim \sqrt{\epsilon} n \nu
\end{equation}
and hence we include a source
\begin{equation}
\Sigma \sim \sqrt{\epsilon} \nu f.
\end{equation}

The random walk estimate of fluxes including source terms is accurate in the region of strong gradients but it should be noted that for weak gradients, random walk arguments overestimate the neoclassical particle fluxes due to constrains imposed by intrinsic ambipolarity. Intrinsic ambipolarity \citep{ sugama98, parra09, calvo12} is a property of neoclassical and turbulent particle fluxes in perfectly axisymmetric tokamaks: these particle fluxes give zero radial current to lowest order in an expansion in $\rho/r$ regardless of the value of the radial electric field. This property is only satisfied when the gradient length scales are much larger than the ion poloidal gyroradius. When the gradient length scales are of the order of the ion poloidal gyroradius and sources are included, the intrinsic ambipolarity constraint is relaxed as is found in this work and before in \citep{landreman14}. We will find the ion neoclassical particle flux to be non-vanishing to lowest order in the presence of a parallel momentum source and discuss these effects in more detail in section \ref{sec: Particle Flux}.


\section{Fixed-$\theta$ variables}\label{sec:banana_variables}
To calculate the particle orbits, we introduce a new set of variables: the fixed-$\theta$ variables, which are based on the conserved quantities energy $\mathcal{E}$, canonical angular momentum $\psi_\ast$, and magnetic moment $\mu$,
\begin{align}
     \mathcal{E} = \frac{1}{2}v^2+\frac{Ze\Phi}{m},&&\psi_\ast=\psi-\frac{I v_\parallel}{\Omega},  && \mu=\frac{v^2_\perp}{2B}.
\end{align}
Here, $v$ is the ion velocity, $m$ is the ion mass, $Ze$ is the charge, $\psi$ is the flux function, $\Omega$ is the Larmor frequency, and $B$ is the magnetic field strength.  The electric potential is $\Phi=\phi+\phi_\theta$. The piece $\phi$ is a flux function, $\phi=\phi(\psi)$, and its size is given by $e\phi/T \sim 1$, whereas $\phi_\theta$ is the small poloidally varying part of the electric potential, so $\phi_\theta=\phi_\theta(\psi,\theta)$ and $e\phi_\theta/T\sim\epsilon$. \color{black} Here, $\theta$ is the poloidal angle. Throughout this work we will use that the electric potential is of the form
\begin{equation}
    \phi_\theta=\phi_c\cos\theta,
\end{equation} 
which we will prove to be true in the banana regime for circular flux surfaces in section \ref{sec: poloidal vari}. Energy, canonical angular momentum and magnetic moment are constant in time, so following the trajectory of a single particle, we find 
\begin{equation}\label{energy}
    \frac{1}{2}v_\parallel^2+\mu B+\frac{Ze}{m}\Phi(\psi, \theta)=\frac{1}{2}v_{\parallel f}^2+\mu B_f+\frac{Ze}{m}\Phi(\psi_f, \theta_f)
\end{equation}
and 
\begin{equation}\label{angular}
\psi-\frac{Iv_\parallel}{\Omega}=\psi_f-\frac{I v_{\parallel f}}{\Omega_f},
\end{equation}
where the subscript $f$ indicates the values of the respective quantities at a fixed poloidal angle $\theta_f$, which represents a reference point in the orbit of the particle. It is important to note that $\psi_f$ and $v_{\parallel f}$ are constants for each particle. For example, following the trajectory of a passing particle, its velocity will deviate from $v_{\parallel f}$, but, having assumed the conservation laws above, the particle returns to its initial position $\psi_f$ with the velocity $v_{\parallel f}$ after one complete poloidal turn. Another particle on a different orbit will have a different $v_{\parallel f}$ and $\psi_f$. Hence, the fixed-$\theta$ quantities can be understood as labels of orbits and will be used as new phase space variables later on. The angle $\theta_f$ is left as a choice at this point, because choosing $\theta_f=0$ only captures particles that are trapped on the low-field side whereas setting $\theta_f=\pi$ captures particles trapped on the high-field side. We show in Appendix \ref{sec: C1} that it is important to take both sides into account when calculating trapped particle effects. \par 
Using the standard large aspect ratio, circular flux surface tokamak, we can write the magnitude of the magnetic field as 
\begin{equation}
    B\simeq B_0\left(1-\frac{r}{R}\cos\theta\right)
\end{equation}
to first order in the inverse aspect ratio $\epsilon$. Here, $B_0$ is the magnetic field on the magnetic axis. For $\theta_f=0$, the magnetic field is
\begin{equation}\label{B}
    B\simeq B_f\left[1+\frac{r}{R}(1-\cos\theta)\right],
\end{equation}
with $B_f=B_0(1-r/R)$, whereas for $\theta_f=\pi$ the magnetic field can be written as
\begin{equation}
    B\simeq B_f\left[1-\frac{r}{R}(1+\cos\theta)\right]
\end{equation}
with $B_f=B_0(1+r/R)$.
Changing $\theta_f$ from $\theta_f=0$ to $\theta_f=\pi$ causes a jump in $B_f$ of \textit{O}$(\epsilon)$. It will be important in Appendix \ref{sec: FQ} that this difference is small.\par
In transport barriers, strong gradients in density, pressure and electric potential are observed. We will assume that $L_n\sim L_T\sim L_\Phi\sim \rho_p$. Ordering the characteristic length of the transport barrier to be of the order of the poloidal gyroradius implies that the poloidal component of the $E\times B$-drift is of the same order as the poloidal component of the parallel velocity. The poloidal component of the $E\times B$-drift is
\begin{equation}
 \frac{c}{B}(\boldsymbol{E}\times\boldsymbol{\hat{b}})\bcdot\bnabla \theta=\frac{cI}{B}\pdv{\Phi}{\psi}\boldsymbol{\hat{b}}\bcdot \bnabla \theta\equiv u \boldsymbol{\hat{b}}\bcdot \bnabla \theta+\frac{cI}{B}\pdv{\phi_\theta}{\psi}\boldsymbol{\hat{b}}\bcdot \bnabla \theta.
\end{equation}
Here, $E=-\bnabla\Phi$ is the electric field, $c$ is the speed of light, and $\boldsymbol{\hat{b}}=\boldsymbol{B}/B$, where the magnetic field is $\boldsymbol{B}=I\bnabla \zeta+\bnabla\zeta\times\bnabla\psi$ and $\zeta$ is the toroidal angle. We have defined the velocity
\begin{equation}\label{def u}
    u=\frac{cI}{B}\pdv{\phi}{\psi}.
\end{equation}
Note that we use $\Delta\psi\sim Iv_t/\Omega$ and thus $u\sim v_{t}$, where $v_t$ is the thermal speed. \color{black} Due to our choice of ordering, $u$ and the parallel velocity $v_\parallel$ are of the same size. The poloidal velocity in this case is 
\begin{equation}
\left(v_\parallel \boldsymbol{\hat{b}}+\frac{c}{B}\boldsymbol{E}\times\boldsymbol{\hat{b}}\right)\bcdot\bnabla\theta\simeq\left(v_{\parallel}+u\right) \hat{\boldsymbol{b}}\bcdot \bnabla \theta.
\end{equation}
Particles are trapped on banana orbits if their poloidal velocity goes to zero at any point on their orbit. In the case of strong radial electric field this requires $v_\parallel+u=0$ instead of the usual trapping condition $v_\parallel=0$, as was first argued by \cite{shaing94resonance}. It follows that particles with a parallel velocity close to $-u$, where $u$ is not necessarily small, are trapped. It has been previously shown that in this case the width of the trapped-barely-passing region in velocity space is $\sim \sqrt{ \epsilon} v_t$ \citep{Shaing92}. We re-derive this result by calculating the deviations in radial position and velocity of particles on trapped orbits in Appendix \ref{sec: orbits}. Passing particles do not get reflected. One can divide the phase space into the freely passing region where $v_\parallel+u\sim v_t$ and the trapped-barely-passing region $v_\parallel+u\sim\sqrt{\epsilon}v_t$. \par
For freely passing particles, we show in Appendix \ref{sec: fixed theta passing} that $v_\parallel-v_{\parallel f}\sim \epsilon v_t$ and $\psi-\psi_f\sim \epsilon \rho_p R B_p$, where $B_p$ is the poloidal magnetic field. Thus, the deviations in parallel velocity and radial location are small in $\epsilon$. The deviations become large and diverge when $v_\parallel+u$ becomes small. This is the trapped-barely-passing region. For trapped-barely-passing particles, the differences are still small but larger by $\sqrt{\epsilon}$, so $v_\parallel-v_{\parallel f}\sim \sqrt{\epsilon} v_t$ and $\psi-\psi_f\sim \sqrt{\epsilon} \rho_p R B_p$ as can be found in Appendix \ref{sec: fixed theta trapped}.

From equation \eqref{vplusu trapped}, which was first derived in this form by \citet{shaing94resonance} (see their equation (22)), we can deduce that particles are trapped for 
\begin{equation}\label{trapping condition}
\frac{(v_{\parallel f}+u_f)^2}{2}\leq\left.\left\lbrace S_f\left[\left(\mu B_f-v_{\parallel f}u_f\right)\left(\frac{B}{B_f}-1\right)+\frac{Ze}{m}(\phi_\theta-\phi_{\theta f})\right]\right\rbrace\right\rvert_\text{max}.
\end{equation}
The quantity $S$ is the squeezing factor as defined by  \cite{hazeltine89}
\begin{equation}\label{def S}
    S=1+\frac{cI^2}{B\Omega}\frac{\partial^2\phi}{\partial \psi^2}.
\end{equation} 
Equation \eqref{trapping condition} implies that $v_{\parallel f}+u_f\sim\sqrt{\abs{S_f}\epsilon}v_t$, which is consistent with \cite{Shaing92}. In our case, $S_f\sim 1$ and $\epsilon\ll 1$ and hence $v_{\parallel f }\simeq-u_f$ holds, to lowest order, in the trapped-barely-passing region. We can rewrite \eqref{trapping condition} setting $\vp\simeq-u_f$
\begin{equation}\label{trapping condition2}
\frac{(v_{\parallel f}+u_f)^2}{2}\leq\left.\left\lbrace S_f\left[\left(\mu B_f+u_f^2\right)\left(\frac{B}{B_f}-1\right)+\frac{Ze}{m}(\phi_\theta-\phi_{\theta f})\right]\right\rbrace\right\rvert_\text{max}.
\end{equation}
Now we see that the term on the right hand side containing $u_f^2$ is the centrifugal force that pushes particles towards the outboard midplane and is small in low flow neoclassical theory. Here, both the magnetic mirror force and the centrifugal force can trap particles on the outboard side. For $\phi_c>0$, the electric potential can oppose the magnetic mirror and the centrifugal force and if the electrostatic force is strong enough, it can cause trapping of particles on the inboard side. This will become relevant in Appendix \ref{sec: FQ}.\par
\begin{figure}
\centering
\includegraphics[trim=0 0 0 0,clip,width=0.5\textwidth]{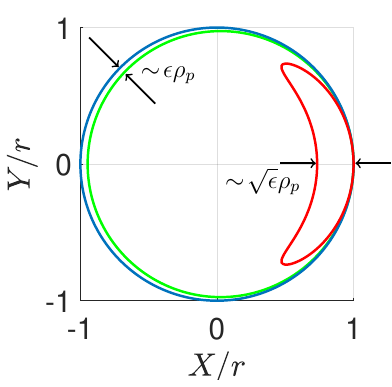}
  \caption{Orbits of passing (green) and trapped (red) particles which follow from \eqref{psibananapassing} and \eqref{vbananatrapped2} are shown for $r/R=0.1$ and circular flux surfaces (blue). We chose $\theta_f=0$, $\phi_\theta=0$, $\mu B_f/v_t^2=1$, $\Omega_f \psi_f/(Iv_t)=1$, $u_f/v_t=1.5$ and $S_f=1.5$. We use $v_{\parallel f}/v_t=-u_f/v_t+5$ for the example passing particle trajectory and $v_{\parallel f}/v_t=-u/v_t+0.2$ for the trapped particle trajectory. The spatial coordinates $X$ and $Y$ determine the position in the poloidal plane with respect to the magnetic axis. To make the orbits visible, we have chosen a flux surface with radius $r=\sqrt{X^2+Y^2}= \Omega \psi_f/(I v_t)$, but note that we assume $r\ll \Omega \psi_f/(I v_t)$ in the rest of the paper. The deviation from the flux surface are much larger for trapped particles than for passing particles.}
   \label{fig:orbits}
\end{figure}

Example orbits for trapped and passing particles for a circular-flux-surface tokamak are shown in figure \ref{fig:orbits}. In the figure, we emphasise the difference between the width of trapped and passing particle orbits.\par

\section{Banana regime}\label{sec: banana regime}
The drift kinetic equation follows from an expansion of the Vlasov equation in $\rho/L$. In our case, this expansion is equivalent to an expansion in $\epsilon$ because $\rho/L\sim\rho/\rho_p\sim\epsilon$, where $\rho/R\ll \epsilon^2$. Keeping only terms of order \textit{O}$(\epsilon^3\Omega f)$, the steady state drift kinetic equation for an ion distribution function $f(\psi,\theta,v_\parallel,\mu)$ is
\begin{multline}\label{drift kinetic equation}
    \left(v_\parallel \boldsymbol{\hat{b}}+\boldsymbol{v}_E\right)\bcdot\bnabla\theta\pdv{f}{\theta}+\left(\boldsymbol{v}_E+\boldsymbol{v}_M\right)\bcdot\bnabla\psi\pdv{f}{\psi} \\
    +\left[\boldsymbol{\hat{b}}+\frac{v_\parallel}{\Omega}\boldsymbol{\hat{b}}\times\left(\boldsymbol{\hat{b}}\bcdot\bnabla\boldsymbol{\hat{b}}\right)\right]\bcdot\left(-\mu \bnabla B+\frac{Ze}{m}\boldsymbol{E}\right)\pdv{f}{v_\parallel}=C[f,f]+\Sigma,
    \end{multline}
where $\boldsymbol{v}_E$ is the $E\times B$-drift, $\boldsymbol{v}_M=\mu \boldsymbol{\hat{b}}\times\bnabla B/\Omega+v_\parallel^2 \boldsymbol{\hat{b}}\times(\boldsymbol{\hat{b}}\bcdot\bnabla\boldsymbol{\hat{b}})/\Omega$ is the magnetic drift, $C[f,f]$ is the Fokker-Planck ion-ion collision operator and we include a source $\Sigma\sim \sqrt{\epsilon}\nu f$, which is consistent with our estimate in section \ref{sec:phase_space}. Note that we neglect terms small in $\epsilon$ and ion-electron collisions that are small in $\sqrt{m_e/m}$, where $m_e$ is the electron mass. \color{black}  
 It is convenient to make a change of variables from $(v_\parallel,\psi)$ to the fixed-$\theta$ variables $(\vp,\psi_f)$. The resulting drift kinetic equation is
\begin{equation} \label{kinetic equation simplified}
   \dot{\theta}\pdv{f}{\theta}\Bigg \vert_{\vp,\psi_f}=C[f,f]+\Sigma,
\end{equation}
where $\dot{\theta}=(v_\parallel \boldsymbol{\hat{b}}+\boldsymbol{v}_E)\bcdot\bnabla\theta$, $f=f(\psi_f,\theta,\vp,\mu)$ and the derivative in $\theta$ is holding $\vp$ and $\psi_f$ fixed. To lowest order in the inverse aspect ratio, one can approximate $\dot{\theta}\simeq (v_\parallel+u)/qR \gtrsim \epsilon^{1/2} v_t/qR$. Assuming that the collisionality is in the banana regime $qR\nu/v_t \ll \epsilon^{3/2}$, the system is, to lowest order in collision frequency, \color{black} described by
\begin{equation}\label{kinetic equation simple simple}
      \frac{v_\parallel+u}{qR}\pdv{f}{\theta}=0
\end{equation}
and, hence, $f$ is to lowest order independent of $\theta$. Thus, any poloidal variations in density, mean flow velocity or temperature must be small.\par
To determine the dependence of $f$ on $\psi_f$, $\vp$ and $\mu$, we define the transit average, which is the average over one orbit of a particle. For passing particles, the transit average is
\begin{align}
    \langle \mathcal{F}\rangle_\tau=\frac{1}{\tau} \int_0^{2\pi}\frac{\mathrm{d}\theta}{\dot{\theta}}\mathcal{F}, 
    \end{align}
    where
    \begin{align}
    \tau=\int_0^{2\pi}\frac{\mathrm{d}\theta}{\dot{\theta}}.
\end{align}
Using the approximate form of $\dot{\theta}$, the transit average for trapped particles is
\begin{align} \label{transit average trapped}
    \langle \mathcal{F}\rangle_\tau=\frac{qR}{\tau} \int_{-\theta_b}^{\theta_b}\frac{\mathrm{d}\theta}{v_\parallel+u}\mathcal{F}(v_\parallel+u>0)+\frac{qR}{\tau}\int_{-\theta_b}^{\theta_b}\frac{\mathrm{d}\theta}{\abs{v_\parallel+u}}\mathcal{F}(v_\parallel+u<0),
\end{align}
where
\begin{equation}
    \tau=2qR \int_{-\theta_b}^{\theta_b}\frac{\mathrm{d}\theta}{\abs{v_\parallel+u}}
\end{equation}
and $\theta_b$ is the bounce angle, determined by $v_\parallel+u=0$. Transit averaging \eqref{kinetic equation simplified} gives
\begin{equation}\label{C Sigma}
    \langle C[f,f]\rangle_\tau=-\langle \Sigma \rangle_\tau.
\end{equation}
To lowest order in $\epsilon$, the source $\langle \Sigma\rangle_\tau\sim\sqrt{\epsilon}\nu f$ is negligible, and the solution is a $\theta$-independent Maxwellian in fixed-$\theta$ variables,
\begin{equation}\label{Maxwellian}
    f_{M_f}=n(\psi_f)\left(\frac{m}{2\pi T(\psi_f)}\right)^{3/2}\exp\left(-\frac{m\left(v_{\parallel f }-V_\parallel(\psi_f)\right)^2}{2 T(\psi_f)}-\frac{m\mu B_f}{T(\psi_f)}\right).
\end{equation}
Note that unlike usual neoclassical theory, we keep the mean parallel velocity $V_\parallel\sim v_\parallel$. To zeroth order in $\epsilon$ \color{black} particles do not leave their flux surface or experience a change in their parallel velocity going through one orbit, that is, $\psi\simeq \psi_f$ and $v_\parallel\simeq \vp$. \par 
The dependence of $T$ on $\psi_f$ might be surprising because strong temperature gradients usually drive deviations away from a Maxwellian equilibrium. If the time scale associated with the ion energy flux $Q$, given by $nT/\abs{\nabla \psi}(\partial Q/\partial \psi)$ is longer than the ion-ion collision time, and the orbit widths are of the same order as the transport barrier, there is no temperature gradient because all particles have reached thermodynamic equilibrium and have been able to sample the entire volume. This is why the temperature gradient was assumed to be small in \citep{kagan10, catto2013}. However, by having introduced the large aspect ratio expansion, the gradient lengths can be of the same size as the poloidal gyroradius whilst still being much larger than the ion orbit width. In this way, we can get a Maxwellian to lowest order and a strong temperature gradient at the same time.
\par
We define the next order solution as
\begin{equation}\label{expansion f}
    f=f_{M_f}+h(\psi_f, v_{\parallel f}, \mu)=f_M+g(\psi,\theta, v_\parallel, \mu),
\end{equation}
where $f_M$ is the Maxwellian in \eqref{Maxwellian} evaluated at the particle variables $\psi, v_\parallel$ and $\mu$,
\begin{equation}
    f_{M}=n(\psi)\left(\frac{m}{2\pi T(\psi)}\right)^{3/2}\exp\left(-\frac{m\left(v_{\parallel  }-V_\parallel(\psi)\right)^2}{2 T(\psi)}-\frac{m\mu B(\psi,\theta)}{T(\psi)}\right)
\end{equation}
and $h\sim g\sim \sqrt{\epsilon}f_M $ are the \textit{O}$(\sqrt{\epsilon})$ corrections to the Maxwellian. 
One \color{black}needs to be careful about the distinction between $h$ and $g$. Whilst $h$ is the distribution function in the fixed-$\theta$ variables and can be interpreted as the distribution of orbits, $g$ is a function of the variables $\psi$, $v_\parallel$ and $\mu$ and it is the distribution function of particles in the classic sense. \par
In the banana regime, the collision frequency satisfies $qR\nu/v_t\ll \epsilon^{3/2}$. The collisionality is small enough that, in both the freely passing and the trapped-barely-passing region, orbits can be completed before particles collide. Consequently, $h$ does not depend on $\theta$ to next order as $\dot{\theta}\partial h/\partial \theta\sim\epsilon^{1/2}v_t h/qR$, while $C[h, f_M]+C[f_M,h]\sim \nu h/\epsilon$. Thus, following \eqref{kinetic equation simple simple}, $h$ does not depend on $\theta$. The large aspect ratio expansion is crucial from here on. We expand $g=h+f_{M_f}-f_M$ in orders of $\sqrt{\epsilon}$,
\begin{equation} \label{expansion g}
    g=g_0+g_1+... \quad \text{where} \quad g_n\sim \epsilon^{\frac{n+1}{2}} f_M.
\end{equation} 

\begin{figure}
  \subfigure[]{\includegraphics[trim=0 0 0 0,clip,width=0.49\textwidth]{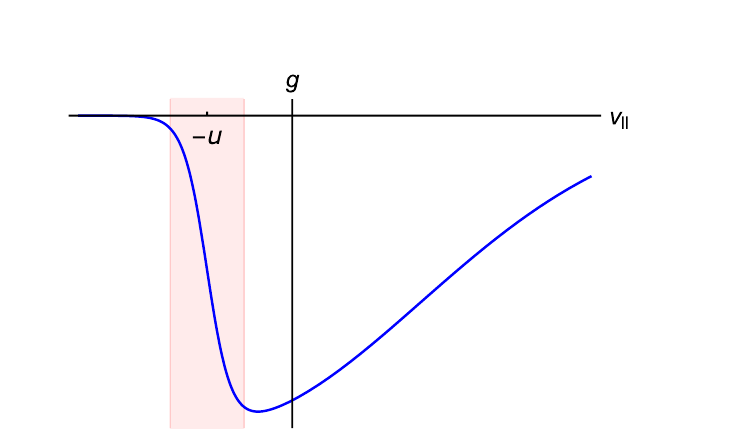}}
  \subfigure[]{\includegraphics[trim=0 0 0 0,clip,width=0.49\textwidth]{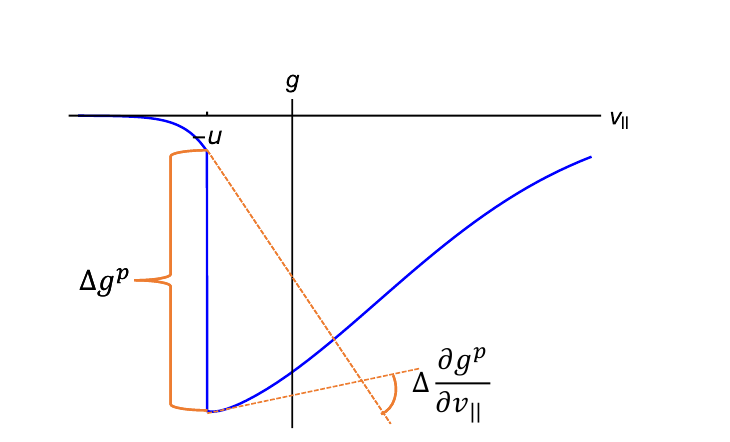}}
    \caption{(\textit{a}): This is a sketch of the distribution function $g$. The region of trapped-barely-passing particles (pink) is small whereas the passing region (white) covers most of velocity space \color{black}. (\textit{b}): The contribution coming from trapped-barely-passing particles is approximated as a discontinuity of the passing particle distribution function and its derivatives in velocity space.}
    \label{fig:gs}
\end{figure}

We will call the solution in the freely passing region, where $\abs{\vpu}\gg\sqrt{\epsilon}v_t$, the freely passing distribution function $g^p$, and the solution in the trapped-barely-passing region, where $\abs{\vpu}\sim\sqrt{\epsilon}v_t$, the trapped-barely-passing distribution function $g^t$. Note that, for convenience, we use the superscript $t$ for the trapped-barely-passing region even though $g^t$ also includes the distribution of barely-passing particles. The function $g^t$ only exists in a small region of phase space, where $\abs{v_{\parallel f}+u_f} \sim \sqrt{\epsilon} v_t$. Thus, the contribution of $g^t$ can be interpreted as a discontinuity in $g^p$. We will find that it is sufficient to set $g\approx g^p$ in the entire phase space and determine from the solution for $g^t$ the jump and derivative discontinuity conditions at $v_\parallel=-u$ for $g^p$. A sketch of $g$ and how $g^t$ is reduced to a discontinuity is shown in figure \ref{fig:gs}. \par 

Within the trapped-barely-passing region only -- the region shaded in pink in figure \ref{fig:gs}a -- we introduce the velocity variable $w\equiv v_{\parallel } + u\sim \sqrt{\epsilon} v_t$ which is defined such that, within the trapped-barely passing region, the region of overlap with the passing particle region maps to $w\rightarrow \pm \infty$, whereas from the point of view of the passing particle region, the region of overlap is still located at $v_\parallel+u\rightarrow 0$. The new variable $w$ effectively stretches out the trapped-barely passing region. \color{black} We require that the outer limiting solutions for $g^t$ match the two inner limiting solutions of $g^p$, such that
\begin{align}
    g^t(w\rightarrow\infty)=g^p(v_\parallel\rightarrow-u^+) &&\text{and} && g^t(w\rightarrow-\infty)=g^p(v_\parallel\rightarrow-u^-),
\end{align}
as well as
\begin{align}
    \pdv{g^t}{w}\bigg\rvert_{w\rightarrow\infty}=\pdv{g^p}{v_\parallel}\bigg\rvert_{v_\parallel\rightarrow-u^+}&&\text{and}&&\pdv{g^t}{w}\bigg\rvert_{w\rightarrow-\infty}=\pdv{g^p}{v_\parallel}\bigg\rvert_{v_\parallel\rightarrow-u^-}.
\end{align}
The jump condition at the trapped-passing boundary becomes
\begin{equation}\label{jump condition banana}
    \Delta g^p=g^t_0(w\rightarrow \infty)-g^t_0(w\rightarrow - \infty).
\end{equation}
The jump condition measures the difference between the co- and counter-moving barely passing particle distribution across the trapped-barely passing region.\\ \color{black} 
In order for this jump to remain finite, the derivative of $g^t_0$ must tend to zero at $\pm \infty$. The discontinuity condition in the derivatives thus requires the next order correction
\begin{equation}\label{discontinuity condition banana}
    \Delta \left(\pdv{g^p}{v_\parallel}\right)=\pdv{g^t_1}{w}\bigg\rvert_{w\rightarrow \infty}-\pdv{g^t_1}{w}\bigg\rvert_{w\rightarrow - \infty}.
\end{equation}
\par The jump and derivative discontinuity conditions follow from the solution of \eqref{C Sigma}, for which we need an expression for the ion-ion \color{black} collision operator. The lowest order solution is a Maxwellian, so we can linearise the collision operator around $f_M$ using \eqref{expansion f},
\begin{equation}
    C[f,f]\simeq C[f_M,g]+C[g,f_M]\equiv C^{(l)}[g] .
\end{equation}
Here, we have used that the collision operator acting on the Maxwellians vanishes. We neglect the smaller, nonlinear contribution $C[g,g]$. The linearised collision operator is
\begin{equation} \label{collision operator}
\begin{split}
   C^{(l)}[g] = & \lambda \bnabla_v\bcdot\left[\int\mathrm{d}^3v'\:f_M f'_M \bnabla_\omega\bnabla_\omega \omega \bcdot\left(\bnabla_v\left(\frac{g}{f_M}\right)-\bnabla_{v'}\left(\frac{g'}{f_M'}\right)\right)\right]\\
    = & \bnabla_v\bcdot\Bigg[f_M \mathsfbi{M} \bcdot\bnabla_v\left(\frac{g}{f_M}\right)-\lambda f_M\int_{V_\text{tbp}}\mathrm{d}^3v'\: f'_M \bnabla_\omega\bnabla_\omega \omega \bcdot \bnabla_{v'}\left(\frac{g^{t'}}{f'_M}\right)\\
    &-\lambda f_M\int_{V_\text{p}}\mathrm{d}^3v'\: f'_M \bnabla_\omega\bnabla_\omega \omega \bcdot \bnabla_{v'}\left(\frac{g^{p'}}{f'_M}\right)\Bigg],
\end{split}
\end{equation}
where $\lambda=2\pi Z^4e^4 \log\Lambda/m^2$ and $\log\Lambda$ is the Coulomb logarithm. The integrals are over the trapped-barely-passing region $V_\text{tbp}$ and the freely passing region $V_\text{p}$, respectively, and $\boldsymbol{\omega}=\boldsymbol{v}-\boldsymbol{v}'$. We have introduced the matrix
\begin{equation}
\begin{split}
    \mathsfbi{M}&=\lambda \int\mathrm{d}^3v'\:f'_M\bnabla_\omega\bnabla_\omega \omega\color{black}\\
    &=\frac{\nu_\perp}{4}\left(|\boldsymbol{v}-V_\parallel\boldsymbol{\hat{b}}|^2\mathsfbi{I}-(\boldsymbol{v}-V_\parallel\boldsymbol{\hat{b}})(\boldsymbol{v}-V_\parallel\boldsymbol{\hat{b}})\right)+\frac{\nu_\parallel}{2}(\boldsymbol{v}-V_\parallel\boldsymbol{\hat{b}})(\boldsymbol{v}-V_\parallel\boldsymbol{\hat{b}}),
    \end{split}
\end{equation}
\begin{align}
    \nu_\perp=3\sqrt{\frac{\pi}{2}}\nu\frac{\Xi(x)-\Psi(x)}{x^3}, &&
    \nu_\parallel=3\sqrt{\frac{\pi}{2}}\nu\frac{\Psi(x)}{x^3},&&\text{and}&&
    \nu= \frac{4\sqrt{\pi} Z^4e^4n\log\Lambda}{3T^{3/2}m^{1/2}},
\end{align}
where $x=\sqrt{m/(2T)}|\boldsymbol{v}-V_\parallel\boldsymbol{\hat{b}}|$, $\Xi(x)=\text{erf}(x)=(2/\sqrt{\pi})\int_0^x\exp(-y^2)\mathrm{d}y$, $\Psi(x)=(\Xi - x\Xi')/(2x^2)$. The term proportional to $\nu_\perp$ describes pitch angle scattering and the term proportional to $\nu_\parallel$ represents energy diffusion.\par
We proceed to find the correction $g$. We expand \eqref{C Sigma} in orders of $\sqrt{\epsilon}$ and find to \textit{O}$(\nu f_M/\sqrt{\epsilon})$ the jump condition $\Delta g^p$ in section \ref{sec: jump} and to \textit{O}$(\nu f_M)$ the derivative discontinuity condition $\Delta (\partial g^p/\partial v_\parallel)$ in section \ref{sec: deriv disc}. The distribution function $g^p$ as well as poloidal variations of density and potential enter at \textit{O}$(\sqrt{\epsilon}\nu f_M)$ and are presented in section \ref{sec: passing} and section \ref{sec: poloidal vari}


\subsection{Jump condition}\label{sec: jump}
The solution in the trapped-barely-passing region gives the jump and derivative discontinuity conditions for $g^p$. We start by finding an expression for the jump condition \eqref{jump condition banana} by collecting terms of order \textit{O}$(\nu f_M/\sqrt{\epsilon})$ in \eqref{C Sigma}. The results of this subsection were already derived in a similar way by \cite{shaing94resonance}. We reproduce the calculations to this order before presenting the higher order calculations where we find significant differences with previous work.
\par The equation to solve for $g_0^t$ is
\begin{equation} \label{Cg0t}
    \langle C^{(l)}[g]\rangle_\tau=0.
\end{equation}
Changing to the fixed-$\theta$ variables and keeping only terms of \textit{O}$(\nu f_M/\sqrt{\epsilon})$ of the collision operator in \eqref{collision operator} yields
\begin{equation}
\begin{split}
    C^{(l)}[g]  \simeq \bnabla_v w_f\bcdot \pdv{}{w_f}\left[f_M \mathsfbi{M}\bcdot\bnabla_v w_f\pdv{(g^t/f_M)}{w_f} \right].
  \end{split}
\end{equation}
Only the derivatives with respect to $w_f\equiv\vpu$ are kept because they are larger than the other velocity derivatives by $1/\sqrt{\epsilon}$. This is because in the trapped-barely-passing region $w_f\sim\sqrt{\epsilon}v_t$ and hence we assume $\partial g^t/\partial w_f\sim g^t/(\sqrt{\epsilon}v_t)$. Using fixed-$\theta$ variables is also convenient because the matching between the trapped-barely-passing and freely passing region will hold for all $\theta$. It follows from \eqref{vplusu trapped2} that
\begin{equation} \label{wf derivative}
    \bnabla_v w_f=\frac{\left[w+Su\left(B_f/B-1\right)\right]\boldsymbol{\hat{b}}-S\left(B_f/B-1\right)\boldsymbol{v}_\perp}{w_f}\simeq\frac{w}{w_f}\boldsymbol{\hat{b}}.
\end{equation}
Thus, the linear collision operator to lowest order is
    \begin{equation}
        \begin{split}
   C^{(l)}[g] \simeq &\frac{w}{w_f}\pdv{}{w_f}\left[M_\parallel \frac{w}{w_f}\pdv{g^t_{0}}{w_f}\right],
\end{split}
\end{equation}
where we have introduced the parallel component of $\mathsfbi{M}$
\begin{equation}\label{M parallel}
M_\parallel\equiv\boldsymbol{\hat{b}}\bcdot\mathsfbi{M}\bcdot \boldsymbol{\hat{b}}\simeq\frac{\nu_\perp}{2}\mu B+\frac{\nu_\parallel}{2}(u+V_{\parallel})^2.
\end{equation}
Here, we have used that $v_\parallel\simeq-u$ for trapped-barely-passing particles. The collision frequencies $\nu_\parallel$ and $\nu_\perp$ are evaluated at $x\simeq \sqrt{m[(u+V_\parallel)^2+2\mu B]/(2T)}$.
\par To determine $g_0^t$, we use \eqref{expansion f} and expand the lowest order solution around a Maxwellian in the variables $(\psi,v_\parallel,\mu)$
\begin{multline}
    f=f_{M_f}+h(\psi_f,\vp,\mu)\simeq f_M+(\psi_f-\psi)\Bigg[\pdv{}{\psi}\ln p+\frac{m(v_\parallel-V_\parallel)}{T}\pdv{V_\parallel}{\psi}\\
    +\left(\frac{m(v_\parallel-V_\parallel)^2}{2T}+\frac{m\mu B}{T}-\frac{5}{2}\right)\pdv{}{\psi}\ln T\Bigg]f_M-\frac{m(v_\parallel-V_\parallel)}{T}(\vp-v_\parallel)f_M+h(\psi,v_\parallel,\mu).
    \end{multline}
The radial derivative of the magnetic field is small and the term $m\mu/T(\partial B/\partial \psi)\sim \partial/\partial \psi \ln B\sim 1/(Ir)$ can be dropped. \color{black}
This result can be rewritten using the velocity variable $w=v_\parallel+u$, the relations \eqref{vbananatrapped2}, and the fact that $v_\parallel\simeq -u$ in the trapped-barely-passing region,
\begin{equation}
    \begin{split}
     f\simeq & f_M - \frac{I}{S\Omega}\left(w-w_f\right)\mathcal{D}f_M(v_\parallel=-u)+h,
    \end{split}
\end{equation}
where we have defined
\begin{equation}
    \mathcal{D}=\pdv{}{\psi}\ln{p} -\frac{m(u+ V_\parallel)}{T}\left(\pdv{V_\parallel}{\psi}-\frac{\Omega}{I}\right) +\left(\frac{m(u+V_\parallel)^2}{2T}+\frac{m\mu B}{T}-\frac{5}{2}\right)\pdv{}{\psi}\ln{T}.
\end{equation}

To avoid cluttering our notation, we will not distinguish between fixed-$\theta$ variables and $(\psi, v_\parallel, \mu)$  in most terms as they are almost the same. We will only keep the distinction between the two types of variables in places where they appear subtracted from each other, e.g. when we need $v_\parallel - \vp$ or $\psi - \psi_f$.\par
One can define the auxiliary function $\bar{h}$, which is a function of fixed-$\theta$ variables only, as
\begin{equation}
    \begin{split}
    \bar{h}=&h+\frac{I}{\Omega}\frac{w_f}{S}\mathcal{D} f_{M}(v_\parallel=-u),
     \end{split}
\end{equation}
and with that we find 
\begin{equation} \label{gt banana}
    \begin{split}
        g^t_0=&\bar{h}-\frac{I}{\Omega}\frac{w}{S}\mathcal{D}f_{M}(v_\parallel=-u).
    \end{split}
\end{equation}

\par The trapped-barely-passing region contains both barely-passing particles and trapped particles and we need to distinguish between the two. The trapped-barely-passing boundary for ions trapped on the low (high) field side for $S>0$ ($S<0$) and $\theta_f=0$ is
\begin{equation}\label{def wtpb}
    w_\text{tpb}^2=4 S\left[\left(\mu B+u^2\right)\frac{r}{R}-\frac{Ze\phi_c }{m}\right].
\end{equation}
The trapped-barely-passing boundary for ions trapped on the low (high) field side for $S>0$ ($S<0$) and $\theta_f=\pi$ is
\begin{equation}
    w_\text{tpb}^2=4 S\left[\frac{Ze\phi_c }{m}-\left(\mu B+u^2\right)\frac{r}{R}\right].
\end{equation}
A more detailed discussion about the distinction between the two cases, is presented in Appendix \ref{sec: FQ}.
For barely-passing particles, for which $w_f^2\geq w_\text{tpb}^2$ holds, one can change from transit averages to flux surface averages by using that
\begin{equation} \label{transit surface}
    \Bigg\langle\frac{w}{w_f}(...)\Bigg\rangle_\tau=\frac{1}{\tau}\int\frac{\mathrm{d}\theta}{w}qR\frac{w}{w_f}(...)=\frac{2\pi qR}{\tau w_f}\langle...\rangle_\psi
\end{equation}
where $\langle...\rangle_\psi=1/(2\pi)\int \mathrm{d}\theta(...)$ is the flux surface average. Then, using expression \eqref{gt banana} and $(\partial w/\partial w_f)\simeq w_f/w$, the transit averaged collision operator becomes
\begin{equation} \label{der g bar}
\begin{split}
\langle C^{(l)}[g]\rangle_\tau\simeq&\frac{2\pi qR}{\tau w_f}\pdv{}{w_f}\Biggl\lbrace M_\parallel\Bigg[\frac{\langle w\rangle_\psi}{w_f}\pdv{\bar{h}}{w_f}-\frac{I}{\Omega S}\mathcal{D}f_{M}(v_\parallel=-u)\Bigg]\Biggr\rbrace.
\end{split}
\end{equation}
For trapped particles, which obey $w_f^2\leq w_\text{tpb}^2$, the contribution $g^t_0-\bar{h}$ is odd in $w$ and hence it follows from \eqref{transit average trapped} and \eqref{gt banana} that
\begin{equation}\label{gt odd in v+u}
\begin{split}
    &\langle C^{(l)}[g-\bar{h}]\rangle_\tau=-\Bigg\langle\frac{w}{w_f}\pdv{}{w_f}\left[M_\parallel \frac{I}{\Omega S}\mathcal{D}f_{M}(v_\parallel=-u)\right]\Bigg\rangle_\tau \\
    &=\frac{1}{\tau}\int_{-\theta_b}^{\theta_b}\frac{\mathrm{d}\theta}{w_f}qR\pdv{}{w_f}\left[M_\parallel\frac{I}{\Omega S}\mathcal{D}f_{M}(v_\parallel=-u)\right]\\
    &-\frac{1}{\tau}\int_{-\theta_b}^{\theta_b}\frac{\mathrm{d}\theta}{w_f}qR\pdv{}{w_f}\left[M_\parallel \frac{I}{\Omega S}\mathcal{D}f_{M}(v_\parallel=-u)\right]=0.
\end{split}
\end{equation}
It then follows from \eqref{Cg0t} and \eqref{der g bar} that
\begin{equation}
  M_\parallel \frac{ \tau \langle w^2\rangle_\tau}{w_f}\pdv{\bar{h}}{w_f}=K,
\end{equation}
where $K$ is a constant. $M_\parallel$ is constant in $w_f$ and
\begin{equation}
     \frac{\tau \langle w^2\rangle_\tau}{w_f}=qR\int^{\theta_b}_{-\theta_b}\mathrm{d}\theta \frac{w}{w_f}=qR\int^{\theta_b}_{-\theta_b}\mathrm{d}\theta\sqrt{1-\kappa^2\sin^2(\theta/2)},
\end{equation}
where $\kappa^2$ is defined in \eqref{kappa positive}, such that for $w_f\rightarrow0$, $\kappa^2\rightarrow\infty$ as $\theta_b\rightarrow0$. Hence, $\tau \langle w^2\rangle_\tau/w_f\rightarrow 0$ for $w_f\rightarrow 0$ and consequently $K=0$ and $\partial \bar{h}/\partial w_f=0$. For trapped particles, we find from \eqref{gt banana} that
\begin{equation} \label{div gt banana2}
\begin{split}
    \pdv{g^t_{0}}{w_f}=&-\frac{I}{\Omega S}\frac{w_f}{w}\mathcal{D}f_{M}(v_\parallel=-u).
\end{split}
\end{equation}
\par
\begin{figure}
    \centering
    \includegraphics[trim=0 0 0 0,clip,width=0.6\textwidth]{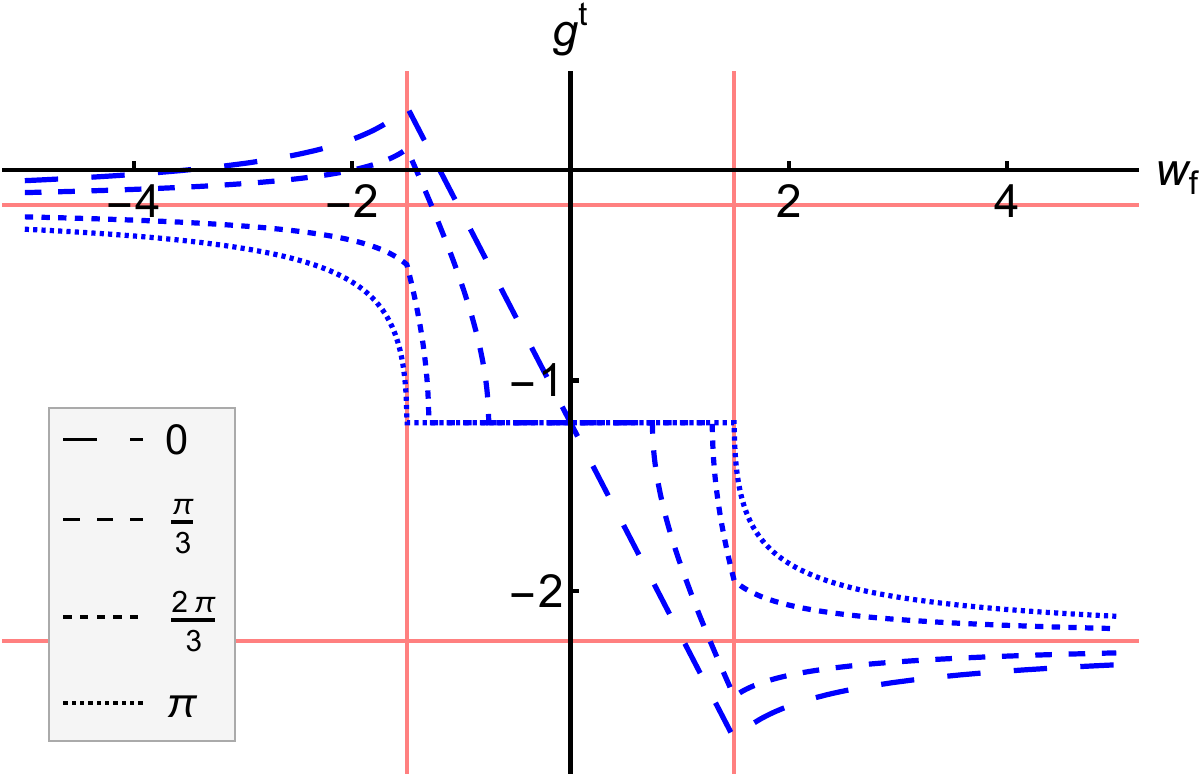}
    \caption{The distribution function $g^t$ in the trapped and barely-passing region is symmetric around $w=0$ and goes towards the same constants for any value of $\theta$ at $w\rightarrow\pm \infty$. Here, we chose $Iv_t\mathcal{D}f_M(v_\parallel=-u)/(\Omega S )=1$, $g(\psi_f,w_f=0,\mu)=-1.2$, $w_\text{tpb}=\pm1.5$, and $w_f$ is in units of thermal velocity. The jump is $\Delta g^p=-2.0685$.}
    \label{Gt beautiful}
\end{figure}

 The contribution $\langle C^{(l)}[g-\bar{h}]\rangle_\tau$ is not zero for barely-passing particles because particles do not bounce, so there is no change in the sign of $w$ and thus the transit average of a function that is odd in $w$ does not vanish. Using equation \eqref{der g bar} with the boundary condition $\partial g_0^t/\partial w_f\rightarrow 0$ for $w_f\rightarrow \infty$, we find that the derivative of the distribution function for barely-passing particles is
\begin{equation} \label{div gt banana}
\begin{split}
    \pdv{g^t_{0}}{w_f}=&\frac{I}{\Omega S}\left(\frac{w_f   }{\langle w \rangle_\psi}-\frac{w_f}{w}\right)\mathcal{D}f_{M}(v_\parallel=-u),
\end{split}
\end{equation}
where we have used $\partial w/\partial w_f\simeq w_f/w$. 
For the jump condition \eqref{jump condition banana} we need to integrate \eqref{div gt banana2} and \eqref{div gt banana} over $w_f$. We will show in section \ref{sec: passing} that in the freely passing particle region, the distribution function is independent of $\theta$ to lowest order and hence the jump condition must be independent of $\theta$ as well. Thus, the jump condition must satisfy
\begin{equation}\label{no theta}
    \Delta g^p=\int_{-\infty}^\infty\mathrm{d}w_f\:\pdv{g_0^t}{w_f}=\Bigg\langle\int_{-\infty}^\infty\mathrm{d}w_f\:\pdv{g_0^t}{w_f}\Bigg\rangle_\psi.
\end{equation}
We calculate this integral in Appendix \ref{sec: FQ} using the potential $\phi_\theta=\phi_c\cos\theta$ (see section \ref{sec: poloidal vari}). The final result is 
\begin{equation} \label{jump condition banana final}
\begin{split}
    \Delta g^p=&-2.758\frac{I}{ \Omega S}\sqrt{\abs{S\left[(\mu B+u^2)\frac{r}{R}-\frac{Ze}{m}\phi_c\right]}} \mathcal{D}f_M(v_\parallel=-u).
    \end{split}
\end{equation}
The distribution function $g^t$ in \eqref{A1} can be plotted using the integrals from Appendix \ref{sec: FQ}. The results for different values of $\theta$ are shown in figure \ref{Gt beautiful}. We find that the derivative is discontinuous at the trapped-passing boundary, and that the jump \eqref{jump condition banana final} is the same for any value of $\theta$.

\subsection{Derivative discontinuity condition} \label{sec: deriv disc}
We proceed to derive an expression for the discontinuity condition \eqref{discontinuity condition banana}. For the jump condition, we have to consider terms of \textit{O}$(\nu f_M/\sqrt{\epsilon})$. For the derivative discontinuity condition, we still consider the trapped-barely-passing particles but need to go to higher order in $\sqrt{\epsilon}$ and collect terms of \textit{O}$(\nu f_M)$. Going back to \eqref{Cg0t}, we perform the change of variables in the collision operator \eqref{collision operator} and only keep terms of \textit{O}$(\nu f_M)$ or larger to get
\begin{equation}\label{col transformation}
    \begin{split}
        C^{(l)}[g]\simeq & \frac{1}{\mathcal{J}}\pdv{}{w_f}\left[\mathcal{J} f_M \bnabla_v w_f\bcdot\mathsfbi{M}\bcdot\bnabla_v\left(\frac{g^t}{f_M}\right)\right]\\
        &+ \frac{1}{\mathcal{J}}\pdv{}{\mu}\left[\mathcal{J} f_M \bnabla_v \mu \bcdot \mathsfbi{M}\bcdot\bnabla_v w_f\pdv{(g_0^t/f_M)}{w_f}\right]\\
        &+\frac{1}{\mathcal{J}}\pdv{}{\psi_f}\left[\mathcal{J} f_M \bnabla_v \psi_f \bcdot \mathsfbi{M}\bcdot\bnabla_v w_f\pdv{(g_0^t/f_M)}{w_f}\right],
    \end{split}
\end{equation}
where
\begin{equation}\label{Jacobian}
    \begin{split}
\mathcal{J}=\det\left(\pdv{(\boldsymbol{r},\boldsymbol{v})}{(\psi_f,\theta,\zeta,w_f,\mu,\varphi)}\right)\simeq \frac{1}{\boldsymbol{B}\bcdot\bnabla \theta} \frac{1}{\bnabla_vw_f\bcdot(\bnabla_v \mu\times \bnabla_v\varphi)} \simeq\frac{qRw_f}{w}
\end{split}
\end{equation}
is the Jacobian (note that we used \eqref{wf derivative} to obtain the last equality), $\varphi$ is the gyroangle with $\nabla_v\varphi=\boldsymbol{\hat{b}}\times\boldsymbol{v}/v_\perp^2$ and
\begin{align}\label{gradients}
    \bnabla_v\mu=\frac{\boldsymbol{v}_\perp}{B}, && \bnabla_v\psi_f=\bnabla_v(\psi_f-\psi)\simeq\frac{I}{\Omega S}\left(\frac{w}{w_f}-1\right)\boldsymbol{\hat{b}}, 
\end{align}
for which we have used \eqref{vbananatrapped2}. The Maxwellians in the second and third term of \eqref{col transformation} are evaluated at $v_{\parallel }=-u$. Recall that the derivatives with respect to $w_f$ are bigger by $1/\sqrt{\epsilon}$ than the derivatives with respect to $\mu$ and $\psi_f$.\par 
We argued in \eqref{discontinuity condition banana} that the parallel velocity derivative of $g^t_1$ is required for the derivative discontinuity condition. This derivative is of order $\sqrt{\epsilon} f_M$ and hence $g^t_1$ only appears in the first term of \eqref{col transformation}, where the second derivative in parallel velocity of $g^t_1$ produces a term of \textit{O}$(\nu f_M)$. In all other terms that involve smaller derivatives with respect to $\mu$ and $\psi_f$, only $g^t_0$ enters to this order. We show in Appendix \ref{sec: transit average C} that taking the transit average of the collision operator yields
\begin{equation}\label{col transformation explicit}
\begin{split}
   &\langle C^{(l)}[g]\rangle_\tau \simeq \frac{1}{w_f \tau}\pdv{}{w_f}\left[f_Mw_f\tau\Bigg\langle\bnabla_vw_f\bcdot\mathsfbi{M}\bcdot\bnabla_v\left(\frac{g^t}{f_M}\right)\Bigg\rangle_\tau\right]\\
   &+\frac{1}{w_f \tau}\pdv{}{\mu}\left[2\mu f_Mw_f\tau M_{\perp }\Bigg\langle\frac{w}{w_f}\pdv{(g^t_0/f_M)}{w_f}\Bigg\rangle_\tau\right]\\
   &+\frac{1}{w_f\tau}\pdv{}{\psi_f}\left[f_Mw_f\tau\frac{I}{\Omega S}M_{\parallel } \Bigg\langle\left(\frac{w}{w_f}-1\right)\frac{w}{w_f}\pdv{(g^t_0/f_M)}{w_f}\Bigg\rangle_\tau\right]=0.
   \end{split}
\end{equation}
Here, we introduced the component of $\mathsfbi{M}$
\begin{equation}
M_\perp\equiv \frac{\boldsymbol{v}_\perp}{\abs{\boldsymbol{v}_\perp}^2}\bcdot\mathsfbi{M}\bcdot \boldsymbol{\hat{b}}\simeq(-u-V_{\parallel})\left(-\frac{\nu_\perp}{4}+\frac{\nu_\parallel}{2}\right),
\end{equation}
and set $v_\parallel=-u$ in the arguments of $\nu_\parallel$ and $\nu_\perp$, which is a good approximation in the trapped-barely-passing region. 

\par The first term in equation \eqref{col transformation explicit} contains the derivative of $g^t_1$ that is needed for the discontinuity condition. The distribution function for trapped-barely-passing particles, $g^t$, has to match with $g^p$ at the boundary between the trapped-barely-passing region and the freely passing region, and thus
\begin{equation}\label{dis gt to gp}
    w_f\Bigg\langle\bnabla_vw_f\bcdot\mathsfbi{M}\bcdot\bnabla_v\left(\frac{g^t}{f_M}\right)\Bigg\rangle_\tau\simeq  w\boldsymbol{\hat{b}}\bcdot\mathsfbi{M}\bcdot\bnabla_v\left(\frac{g^p}{f_M}\right)
\end{equation}
for $w\rightarrow\pm\infty$. Hence, the solution for the discontinuity condition \eqref{discontinuity condition banana} in the banana regime takes the form
\begin{equation} \label{dis condition banana final}
 \begin{split}
    \Delta&\left[ w\tau\boldsymbol{\hat{b}}\bcdot\mathsfbi{M}\bcdot\bnabla_v\left(\frac{g^p}{f_M}\right) \right]f_M \\
    =&-\pdv{}{\mu}\left[f_M \int_{-\infty}^{\infty}\mathrm{d}w_f \:w_f\tau 2\mu M_{\perp } \Bigg\langle\frac{w}{w_f}\pdv{(g^t_0/f_M)}{w_f}\Bigg\rangle_\tau\right]\\
    &-\pdv{}{\psi_f}\Bigg[f_M\frac{I}{\Omega S} \int_{-\infty}^\infty\mathrm{d}w_f\:w_f\tau M_{\parallel } \Bigg\langle\left(\frac{w}{w_f}-1\right)\frac{w}{w_f}\pdv{(g^t_0/f_M)}{w_f}\Bigg\rangle_\tau\Bigg],
     \end{split}
\end{equation}
where we have multiplied \eqref{col transformation explicit} by $w_f\tau$ and integrated over $w_f$. Note that on the left hand side of the equation $w\tau\simeq 2\pi qR$. Following the steps in Appendix \ref{sec: Delta derivative} and recalling \eqref{no theta}, we arrive at
\begin{equation}\label{Delta dgp}
    \Delta \left[\boldsymbol{\hat{b}}\bcdot\mathsfbi{M}\bcdot\bnabla_v\left(\frac{g^p}{f_M}\right)\right]f_M=-\pdv{}{\mu}\left(2\mu M_{\perp} \Delta g^p\right)+\pdv{}{\psi_f}\left(\frac{I}{\Omega S}M_{\parallel}\Delta g^p\right),
\end{equation}
where $\Delta g^p$ is given in \eqref{jump condition banana final}.\par
We have found the jump and derivative discontinuity conditions. Next, an equation for the freely passing region is derived which completes an approximate description of the entire velocity space. 

\subsection{The freely passing region}\label{sec: passing}
The freely passing particle distribution function enters to order \textit{O}$(\sqrt{\epsilon}\nu f_M)$ in \eqref{C Sigma}. The explicit expression of the collision operator in \eqref{collision operator} is substituted into the simplified drift kinetic equation \eqref{C Sigma}, which gives 
\begin{equation}\label{eqpassing0}
\begin{split}
 &\Bigg\langle\bnabla_v\bcdot\left[f_M \mathsfbi{M} \bcdot\bnabla_v\left(\frac{g^p}{f_M}\right)-\lambda f_M\int_{V_\text{p}}\mathrm{d}^3v'\: f'_M \bnabla_\omega\bnabla_\omega \omega \bcdot \bnabla_{v'}\left(\frac{g^{p'}}{f'_M}\right)\right]\Bigg\rangle_\tau\\
    &-\lambda\Bigg\langle \bnabla_v \bcdot\left[ f_M\int_{V_\text{tbp}}\mathrm{d}^3v'\:f'_M\bnabla_\omega\bnabla_\omega\omega\bcdot\bnabla_{v'}\left(\frac{g^{t'}}{f'_M}\right)\right]\Bigg\rangle_\tau=-\langle \Sigma \rangle_\tau.
    \end{split}
\end{equation}
The distribution function $g\sim\sqrt{\epsilon} f_M$ and the gradient acting on $g^{t'}$ gives a factor of $1/\sqrt{\epsilon}v_t$. In the third term on the right hand side $\bnabla_{v'}g^{t'}\sim f_M/v_t$ and $V_\text{tbp}\sim\sqrt{\epsilon}v_t^3$, so all three terms on the left hand side are of the order \textit{O}$(\sqrt{\epsilon}\nu f_M)$.\par 
We combine the first two terms in equation \eqref{eqpassing0} and define the linearised freely passing collision operator 
\begin{equation}\label{Cp linear}
    C_p^{(l)}[g]\equiv\bnabla_v\bcdot\Bigg[f_M \mathsfbi{M} \bcdot\bnabla_v\left(\frac{g^p}{f_M}\right)-\lambda f_M\int_{V_\text{p}}\mathrm{d}^3v' f'_M \bnabla_\omega\bnabla_\omega \omega \bcdot \bnabla_{v'}\left(\frac{g^{p'}}{f'_M}\right)\Bigg]
\end{equation}
to write \eqref{eqpassing0} as
\begin{equation}\label{eqpassing}
\begin{split}
 \langle C_p^{(l)}[g] \rangle_\tau  -\lambda \Bigg\langle \bnabla_v \bcdot\left[ f_M\int_{V_\text{tbp}}\mathrm{d}^3v'\:f'_M\bnabla_\omega\bnabla_\omega\omega\bcdot\bnabla_{v'}\left(\frac{g^{t'}}{f'_M}\right)\right]\Bigg\rangle_\tau=-\langle \Sigma \rangle_\tau.
    \end{split}
\end{equation}
This is the equation for the passing distribution function. Equation \eqref{eqpassing} has solvability conditions, which are the moment equations we calculate in section \ref{sec: Moment}. To obtain the moment equations, the jump and derivative discontinuity conditions in equations \eqref{jump condition banana final} and \eqref{dis condition banana final} are needed. \par

We are interested in the poloidal variations of density, mean parallel flow velocity, temperature, and electric potential, for which the $\theta$-dependent part of $g^p$, $g^p-\langle g^p \rangle_\psi$, is of interest. We argued that $h$ only depends on $\theta$ via the dependence of $\psi_f$ and $\vp$ on $\theta$. Since $g^p=h^p+f_{M_f}-f_M$ and $f_{M_f}-f_M\sim\epsilon f_M$, $g^p \simeq h^p$ to lowest order. The $\theta$-dependent part of $g^p$ is given by the next order,
\begin{equation} \label{theta dependent gp}
    \begin{split}
        &g^p-\langle g^p\rangle_\psi\simeq f_{M_f}-f_M-\langle f_{M_f}-f_M\rangle_\psi  \\
        \simeq & \left(\psi_f-\psi-\langle \psi_f-\psi\rangle_\psi\right)\pdv{f_M}{\psi}+\left(v_{\parallel f}-v_\parallel-\langle v_{\parallel f}-v_\parallel\rangle_\psi\right)\pdv{f_M}{v_\parallel}+\frac{m \mu}{T}\left(B-\langle B\rangle_\psi\right)f_M  \\
        =&-\frac{Ir}{\Omega R}\frac{\left(v_\parallel^2+\mu B\right) \cos{\theta}-Ze\phi_\theta R/(mr)}{v_\parallel +u}\Bigg[\pdv{}{\psi}\ln{p} +\frac{m(v_\parallel- V_\parallel)}{T}\left(\pdv{V_\parallel}{\psi}-\frac{\Omega}{I}\right)\\
        &+\left(\frac{m(v_\parallel-V_\parallel)^2}{2T}+\frac{m\mu B}{T}-\frac{5}{2}\right)\pdv{}{\psi}\ln{T}\Bigg]f_M-\frac{r}{R}\cos{\theta}\frac{m}{T}\left[v_\parallel(v_\parallel-V_\parallel)+\mu B \right]f_M,
    \end{split}
\end{equation}
where we have used the relations \eqref{vbananapassing2} and \eqref{psibananapassing2} as well as $B_f/B-\langle B_f/B\rangle_\psi=(r/R)\cos \theta$. The $\theta-$dependent part of the distribution function is of \textit{O}$(\epsilon f_M)$ and consequently the $\theta$-independent part of $g^p$ is bigger than $g^p-\langle g^p \rangle_\psi$ by order $\sqrt{\epsilon}$. In Appendix \ref{matching} we show that the $\theta$ dependent part of the solution for $g^t_0$ matches with \eqref{theta dependent gp}.\par

\subsection{Poloidal variations and electric potential}\label{sec: phi}\label{sec: poloidal vari}

In the tokamak core, trapped particles are located around $v_\parallel=0$, and for a Maxwellian with $V_\parallel=0$ the number of passing particles with $v_\parallel>0$ and $v_\parallel<0$ is the same to lowest order. The trapped-passing boundary in our ordering is shifted such that trapped particles are located around $v_\parallel=-u$. The lowest order distribution function is still a Maxwellian, but it has a mean parallel velocity $V_\parallel$. For $V_\parallel \neq -u$, this implies that the number of passing particles with $v_\parallel+u>0$ and $v_\parallel+u<0$ is different.
This discrepancy causes a poloidal variation in density, mean parallel velocity, temperature and poloidal potential.\par

\begin{figure}
  \subfigure[]{\includegraphics[trim=0 0 0 0,clip,width=0.47\textwidth]{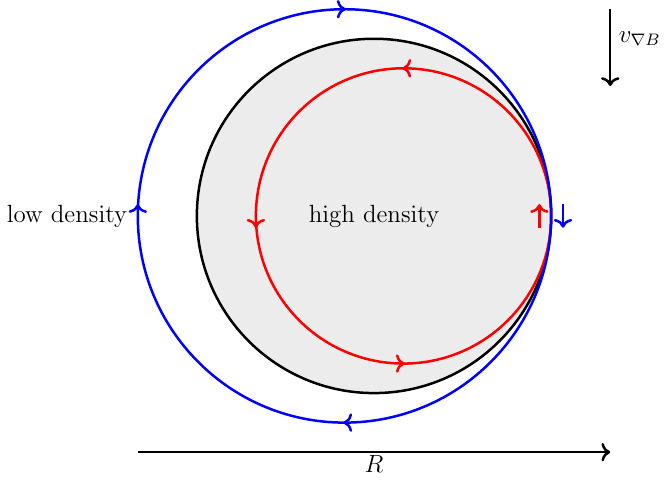}}
  \subfigure[]{\includegraphics[trim=0 0 0 0,clip,width=0.47\textwidth]{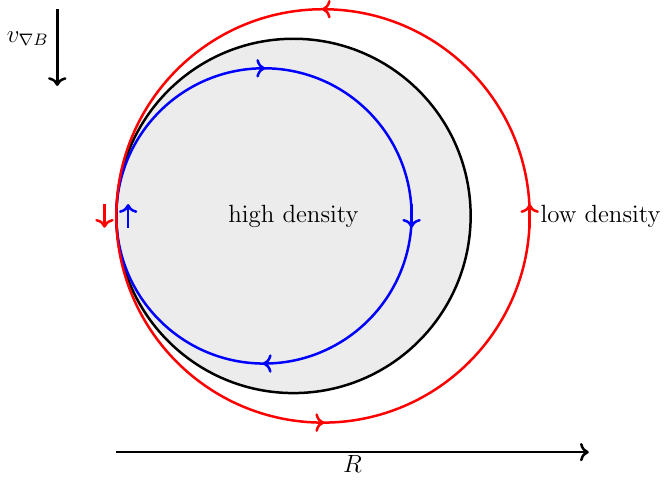}}
    \caption{(\textit{a}): At $\theta=0$, particles with a positive poloidal velocity (red) are pushed inwards, completing their orbits through a region of higher density, and particles with a negative poloidal velocity (blue) are pushed outwards, completing their orbits through a region of lower density. Hence, red particles are more numerous than blue particles. (\textit{b}): At $\theta=\pi$ on the same flux surface, the opposite is the case and there are fewer red particles than there are blue particles. If red particles are more numerous than blue particles and the density is higher at smaller radii, there will be a higher density at $\theta=0$ than at $\theta=\pi$ and there is poloidal variation of density within a flux surface. }
    \label{fig: PoloidalPicture}
\end{figure}

If, for example, the magnetic drifts are pointing downwards, as shown in figure \ref{fig: PoloidalPicture}, particles with a positive (negative) poloidal velocity are being pushed inwards (outwards) with respect to their flux surface at $\theta=0$ and outwards (inwards) at $\theta=\pi$. Let us assume a density gradient such that there is higher density inside a flux surface than there is outside. In this case, there are more particles with positive poloidal velocity at $\theta=0$ than there are particles with negative poloidal velocity (see figure \ref{fig: PoloidalPicture}a), because particles with positive poloidal velocity come from the high density region. At $\theta=\pi$, the opposite is true, because the orbits of particles with positive poloidal velocity come from the low density region (see figure \ref{fig: PoloidalPicture}b). Thus, for a shifted trapped-passing boundary in the strong gradient case, the number of particles with positive and negative poloidal velocity are different to lowest order in $\epsilon$ and $\rho/r$ and density varies poloidally within a flux surface. For comparison, the same effect occurs in standard low flow neoclassical theory, but the number of particles with positive and negative poloidal velocity is the same to lowest order in $\rho/r$ and these effects cancel out. The asymmetry in the passing particle distribution function in the strong gradient case gives a poloidal density variation of order \textit{O}$(\epsilon)$, whereas, in standard low flow neoclassical theory, the poloidal density variation is much smaller\color{black}. The same argument can be constructed for poloidal variation of temperature and mean parallel flow. \par
\begin{figure}
    \centering
    \includegraphics[trim=0 0 0 0,clip,width=0.5\textwidth]{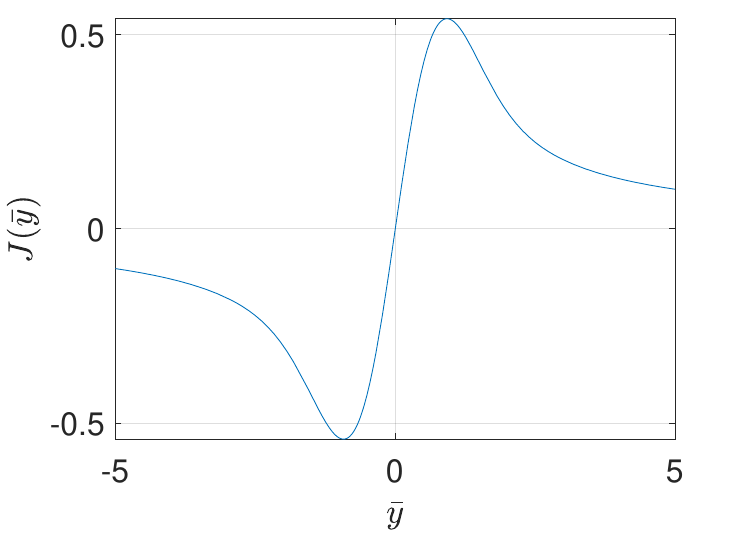}
    \caption{The function J defined in \eqref{J} as a function of $\bar{y}=\sqrt{m/(2T)}(u+V_\parallel)$.}
    \label{fig:J}
\end{figure}

The small poloidal variation of density, $n_\theta$, is
\begin{equation}\label{poloidal density}
    n_\theta(\psi,\theta)=\int\mathrm{d}^3v\: g-\Bigg\langle \int\mathrm{d}^3v\:g \Bigg\rangle_\psi.
\end{equation}
The integration is over the entire range of the parallel velocity and hence over both, the trapped-barely-passing and freely passing regions. The freely passing region is the part of velocity space for which $v_\parallel$ is not close to $-u$. Importantly, the freely passing distribution function \eqref{theta dependent gp} diverges at $v_\parallel=-u$. This divergence is picked up by the trapped distribution function $g^t$. As a result, the integration over phase space is split up into an integration over $g^t$ in the trapped-barely-passing region and a principle value integral over $g^p$ which captures the freely passing region while ignoring the divergence near $v_\parallel=-u$. Contribution from the divergence is accounted for by the integral of the distribution function $g^t$ in the trapped-barely-passing region. For trapped particles, it follows directly from \eqref{A1} that 
\begin{equation}
    \int\mathrm{d}\mu\int_{\text{trapped}}\mathrm{d}w\:2\pi B g^t-\Bigg\langle \int\mathrm{d}\mu\int_{\text{trapped}}\mathrm{d}w\:2\pi B g^t\Bigg\rangle_\psi=0.
\end{equation}
For barely passing and freely passing particles, the flux surface average of the density can be replaced by the integral over the flux surface averaged distribution function because the $\theta$-dependence of $B$ is small. Thus, \eqref{poloidal density} can be written as
\begin{equation}\label{ntheta2}
    n_\theta=\int\mathrm{d}\mu\int_{\text{barely-passing}}\mathrm{d}w\:2\pi B (g^t-\langle g^t\rangle_\psi)+\int\mathrm{d}\mu\:\left[\text{PV}\!\int\mathrm{d}v_\parallel\: 2\pi B(g^p-\langle g^p\rangle_\psi)\right],
\end{equation}
where the first term only contains the barely-passing particles. However, this contribution vanishes to lowest order because $g^t_0-\langle g^t_0\rangle_\psi$ is odd in $w$, which follows from \eqref{odd}. The integration of the second term in equation \eqref{ntheta2} is performed in Appendix \ref{sec: poloidal var}, where the $\theta$-dependent part of the distribution function is taken from \eqref{theta dependent gp}. The result is
\begin{equation}\label{poloidal density 2}
\begin{split}
   & n_\theta=- n\frac{Ir}{\Omega R}\Bigg\lbrace \sqrt{\frac{2T}{m}}J \Bigg[\left(\frac{mV_\parallel^2}{T}\cos\theta+\cos\theta -\frac{Ze\phi_\theta R}{Tr}\right)\left(\pdv{}{\psi}\ln p-\frac{3}{2}\pdv{}{\psi}\ln T\right)\\
   &+\cos\theta\pdv{}{\psi}\ln T \Bigg]+\left[1-2\sqrt{\frac{m}{2T}}(V_\parallel+u)J\right]\Bigg\lbrace(V_\parallel-u)\cos\theta\left(\pdv{}{\psi}\ln p-\frac{3}{2}\pdv{}{\psi}\ln T\right)\\
   &-(V_\parallel+u)\Bigg[\left(\frac{mV_\parallel^2}{2T}+\frac{1}{2}\right)\cos\theta-\frac{Ze\phi_\theta R}{2Tr}\Bigg]\pdv{}{\psi}\ln T+\left(\pdv{V_\parallel}{\psi}-\frac{\Omega}{I}\right)\Bigg[\Bigg(\frac{m u^2}{T}+1\\
   &-\frac{m(V_\parallel+u)^2}{T}\Bigg)\cos\theta-\frac{Ze\phi_\theta R}{Tr}\Bigg]\Bigg\rbrace+\left[1+2\frac{m(V_\parallel+u)^2}{2T}-4\left(\frac{m}{2T}\right)^{3/2}(V_\parallel+u)^3J\right]\\
   &\times\cos\theta\left(\pdv{V_\parallel}{\psi}-\frac{\Omega}{I}+\frac{V_\parallel-u}{2}\pdv{}{\psi}\ln T\right)\Bigg\rbrace-2n\frac{r}{R}\cos\theta,
    \end{split}
\end{equation}
where we introduced the function 
\begin{equation}\label{J}
    J=\frac{\sqrt{\pi} }{2}\exp\left(-\frac{m(u+V_\parallel)^2}{2T}\right)\erfi\left(\sqrt{\frac{m}{2T}}(u+V_\parallel)\right),
\end{equation}
which is plotted in figure \ref{fig:J} and $\erfi(x)=(2/\sqrt{\pi})\int_0^x\exp(t^2)\mathrm{d}t$.
The orbit width of passing particles is of order $\epsilon$ and hence the poloidal variation in density is of order $\epsilon$ as well.\par 
The poloidal variation in density creates a poloidal variation in electric potential $\phi_\theta$ that is determined via quasineutrality. Assuming a Boltzmann response of the electrons, the quasineutrality condition yields
\begin{equation}\label{quasineutrality banana}
    Z\int\mathrm{d}^3v \:g-\Bigg \langle Z\int\mathrm{d}^3v\:g \Bigg \rangle_\psi=\frac{en_e}{T_e}\phi_\theta.
\end{equation}
Looking at \eqref{poloidal density 2} we find that the potential has the form $\phi_\theta=\phi_c \cos\theta$, and the quasineutrality condition \eqref{quasineutrality banana} yields 
\begin{equation} \label{phi banana final}
    \begin{split}
        &\Bigg\lbrace\frac{en_e}{T_e}\!-\!\frac{Z^2ne I}{T\Omega}\!\left[\!\sqrt{\frac{2T}{m}}\!J\!\left(\!\pdv{}{\psi}\ln p\!-\!\frac{3}{2}\pdv{}{\psi}\ln T\right)\!+\!\left[1\!-\!2\sqrt{\frac{m}{2T}}(V_\parallel+u)J\right]\right(\pdv{V_\parallel}{\psi}\!-\!\frac{\Omega}{I}\!\\
        &-\!\frac{(V_\parallel+u)}{2}\pdv{}{\psi}\ln T\Bigg)\Bigg]\Bigg\rbrace\phi_c=-Zn\frac{Ir}{\Omega R}\Bigg\lbrace \sqrt{\frac{2T}{m}}J \Bigg[\left(\frac{mV_\parallel^2}{T}+1\right)\left(\pdv{}{\psi}\ln p-\frac{3}{2}\pdv{}{\psi}\ln T\right)\\
        &+\pdv{}{\psi}\ln T\Bigg]+\!\left[1\!-\!2\sqrt{\frac{m}{2T}}(V_\parallel+u)J\right]\!\Bigg[\!(V_\parallel\!-\!u)\!\left(\pdv{}{\psi}\!\ln \!p\!-\!\frac{3}{2}\pdv{}{\psi}\!\ln\! T\right)\!\\
        &+\!\left(\pdv{V_\parallel}{\psi}\!-\!\frac{\Omega}{I}\right)\!\left(\frac{mu^2}{T}\!+\!1\!-\!\frac{m(V_\parallel+u)^2 }{T}\right)\!-\frac{V_\parallel+u}{2}\left(\frac{mV_\parallel^2}{T}+1\right)\pdv{}{\psi}\ln T\Bigg]\\
        &+\left[1+2\frac{m}{2T}(V_\parallel+u)^2-4\left(\frac{m}{2T}\right)^{3/2}(V_\parallel+u)^3J\right]\left(\pdv{V_\parallel}{\psi}-\frac{\Omega}{I}+\frac{V_\parallel-u}{2}\pdv{}{\psi}\ln T\right)\Bigg\rbrace\\
        &-2Zn\frac{r}{R}.
    \end{split}
\end{equation}
For $\phi_c>0$, the maximum of the potential is on the low-field side of the plasma, so the potential can trap particles on the high-field side for $S>0$. For $\phi_c<0$ and $S>0$, the potential reaches its maximum on the high field side and it can trap particles on the low-field side if electrostatic trapping dominates over magnetic trapping and centrifugal force. \par
Charge exchange recombination spectroscopy measurements in both Alcator C-Mod \citep{churchill15, theiler14} and ASDEX-Upgrade \citep{cruz22} have observed poloidal variation in impurity density and temperatures in the pedestal of H-mode plasmas. These experiments also demonstrated that the main ion temperature and radial electric field cannot simultaneously be flux functions. This is consistent with our calculation and argumentation of poloidal variation in the electric potential and density.

We have found expressions for the distribution function in the passing region and the jump and derivative discontinuity condition given by the trapped-barely-passing region, and we have found the form of the poloidally varying component of the electric potential. These expressions are needed to calculate the solvability conditions for \eqref{eqpassing}.

\section{Moment equations} \label{sec: Moment}
In order to study the transport in the pedestal, we want to find particle, parallel momentum and energy fluxes and how they give rise to profiles of $n$, $T$, $u$, $V_\parallel$ and $\phi_c$. First, we integrate \eqref{eqpassing}, for which the jump and derivative discontinuity conditions are required, and find the solvability conditions, which are the equations for particle, parallel momentum and energy conservation.\par

The full derivation is explained in Appendix \ref{sec: transport equations}, where we show that the particle conservation equation
\begin{equation}\label{Particle conservation}
    \pdv{}{\psi_f}\left(-\frac{I }{\Omega m}F_{\parallel }\right)=\int \mathrm{d}^3v_f\:\langle \Sigma\rangle_\tau,
\end{equation}
is the result of integrating \eqref{eqpassing} over velocity space. Here,
\begin{equation}\label{F parallel definition}
    F_{\parallel }=-\int\mathrm{d}\mu \:\frac{2\pi m B}{S}M_{\parallel } \Delta g^p
\end{equation}
is the parallel force due to the friction between passing and trapped particles, and $\Delta g^p$ is the jump condition given in \eqref{jump condition banana final}. The integration over $\mathrm{d}^3v_f$ is an integration over velocity space in the fixed-$\theta$ variables, where $\mathrm{d}^3v_f=2\pi B\: \mathrm{d}\mu\text{ d}\vp$. The integration eliminated the contribution from the freely passing particle distribution $g^p$ to the particle transport and for this reason is a solvability condition: it must be satisfied regardless of the value of $g^p$. Trapped and barely passing particles dominate transport as we have estimated in section \ref{sec:phase_space}. 
We can compare \eqref{Particle conservation} to a typical continuity equation
\begin{equation}\label{Particle Transport}
    \pdv{\Gamma}{\psi_f}=\Bigg \langle \int\mathrm{d}^3v_f \:\Sigma\Bigg\rangle_\psi,
\end{equation}
where the term on the left hand side is the divergence in $\psi_f$ of a particle flux $\Gamma$ and the term on the right hand side is a source of particles. \color{black} It follows directly from \eqref{Particle conservation} and \eqref{Particle Transport} that
the neoclassical ion particle flux is
\begin{equation}\label{Gammac}
    \Gamma=-\frac{I}{m\Omega}F_{\parallel }.
\end{equation}
The parallel force $F_\parallel$ can drive a radial particle flux via an effect similar to the one that gives the Ware pinch \citep{ware70}. \par
The parallel momentum equation is the result of multiplying \eqref{eqpassing} by $m\vp$ and integrating over velocity space. The equation becomes
\begin{equation} \label{Momentum conservation}
    \pdv{}{\psi_f}\left(\frac{I }{\Omega}u F_{\parallel }\right)+F_{\parallel }=\gamma,
\end{equation}
where $\gamma=\int \mathrm{d}^3v \:m\vp \langle \Sigma\rangle_\tau$ is the parallel momentum input per unit volume. The calculation that leads to \eqref{Momentum conservation} is presented in Appendix \ref{sec: momentum transport}. 
We can use the particle flux \eqref{Gammac} in \eqref{Momentum conservation} and arrive at
\begin{equation}\label{gammac1}
    \pdv{}{\psi_f}\left(mu\Gamma\right)+\frac{m\Omega}{I}\Gamma=-\gamma,
\end{equation}
which is a relation purely between the particle flux, parallel momentum input and $u$. The first term on the left hand side of \eqref{gammac1} is the flux of parallel momentum carried by the trapped particles. The second term on the left is the force due to the friction between trapped and passing particles. The term on the right hand side of the equation is a source of parallel momentum.\par
As for the particle and parallel momentum equations, one can find the energy equation by multiplying \eqref{eqpassing} by $mv_f^2/2$ and integrating over velocity space to arrive at
\begin{equation} \label{Energy conservation}
    \pdv{}{\psi_f}\left(\frac{IT }{m\Omega}\Theta\right)-u F_{\parallel }=\int \mathrm{d}^3v_f \:\frac{mv_f^2}{2}  \langle\Sigma\rangle_\tau,
\end{equation}
where
\begin{equation}\label{Kf definition}
    \Theta=\int\mathrm{d}\mu\:\frac{2\pi mB}{S}\left(\frac{m\mu B}{T}+\frac{mu^2}{2T}\right)M_{\parallel } \Delta g^p.
\end{equation}
The energy flux $Q$ is defined similarly to $\Gamma$ as
\begin{equation}\label{Energy Transport}
     \pdv{Q}{\psi_f}-Ze\Gamma \pdv{\phi}{\psi_f}=\Bigg\langle\int\mathrm{d}^3v_f \: \frac{m v_f^2}{2}\Sigma\Bigg\rangle_\psi.
\end{equation}
A comparison to \eqref{Energy conservation} gives
\begin{equation}\label{qc1}
    Q=\frac{TI}{m\Omega}\Theta.
\end{equation}
The flux of energy on the left hand side of \eqref{Energy conservation} contains both convective energy flux, which is the energy carried by the particle flux, and a conduction energy flux. The second term on the left of \eqref{Energy Transport} is the work done by the radial electric field. The term on the right represents energy injection.\par 

The same equations for particle, parallel momentum and energy \eqref{Particle conservation}, \eqref{Momentum conservation} and \eqref{Energy conservation} can be found using moments of the original Fokker-Planck kinetic equation. At this point we can switch from fixed-$\theta$ variables to normal variables and drop the subscript $f$ because the difference is small in $\epsilon$.\par
We can substitute \eqref{M parallel} for $M_\parallel$  and the jump condition \eqref{jump condition banana final} into \eqref{F parallel definition} to find the particle flux from \eqref{Gammac}
\begin{equation}
      \begin{split}
        \Gamma=&-2.758\frac{I^2\pi  B }{\abs{S}^{3/2}\Omega^2}n\left(\frac{m}{2\pi T}\right)^{3/2} \sqrt{\frac{T}{m}}\int\mathrm{d}\mu\: \exp\left(-\frac{m(u+V_{\parallel })^2}{2T}-\frac{m\mu B}{T}\right)\\
        &\times \sqrt{\abs{\left(\frac{m\mu B}{T}+\frac{m u^2}{T}\right)\frac{r}{R}-\frac{Ze}{T}\phi_c}} \left(\nu_\perp\mu B+\nu_\parallel(u+V_{\parallel })^2\right)\mathcal{D}.
        \end{split}
\end{equation}

\begin{figure}
\centering
  \subfigure[]{\includegraphics[trim=0 0 0 0,clip,width=0.48\textwidth]{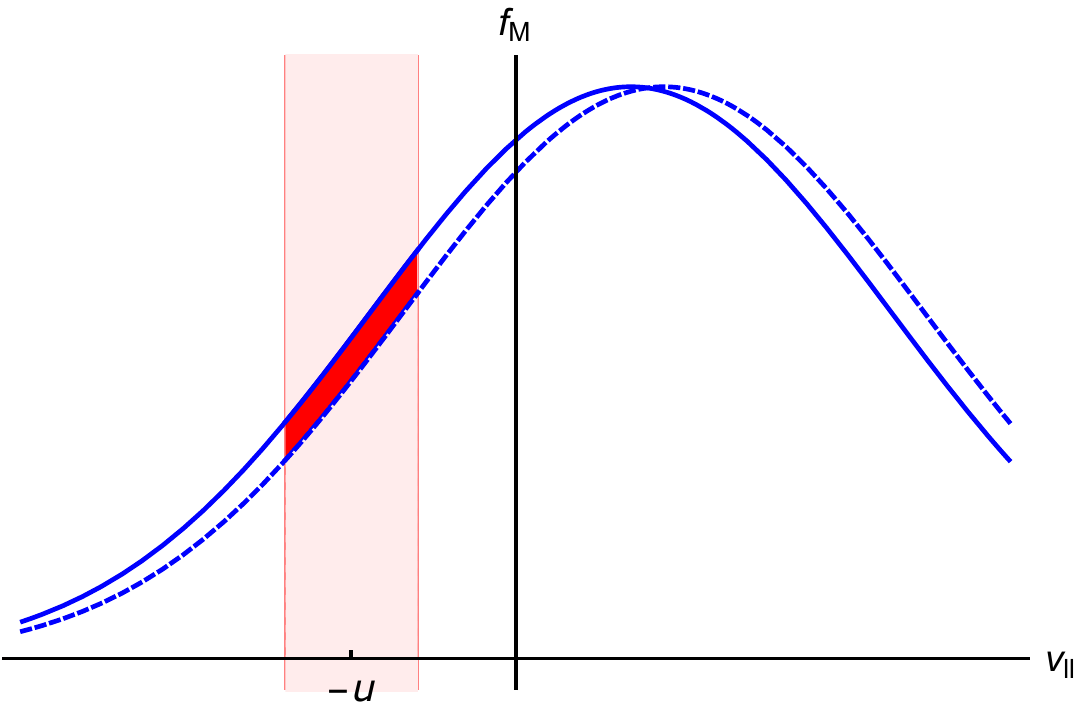}}
  \subfigure[]{\includegraphics[trim=0 0 0 0,clip,width=0.48\textwidth]{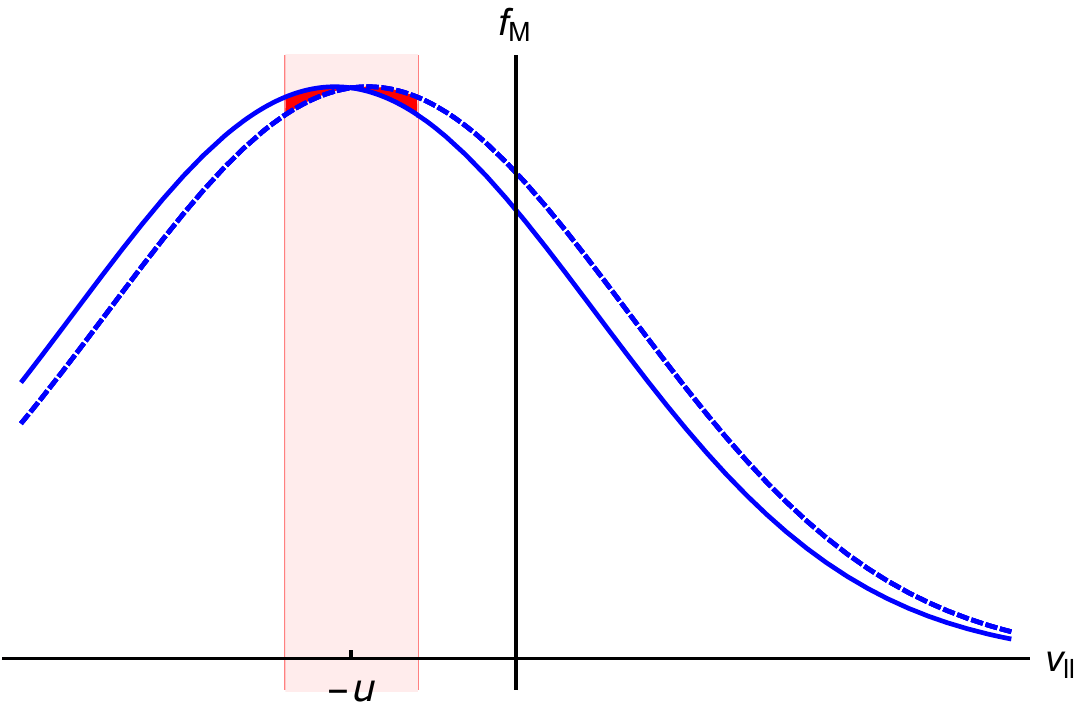}}
    \caption{(\textit{a}): A small shift in $V_\parallel$ for $V_\parallel$ not close to $-u$ going from one surface (solid line) to another flux surface (dashed line) causes a strong change of the number of trapped particles (red area between curves) in the trapped-barely-passing region (pink). (\textit{b}): A small shift in $V_\parallel$ for $V_\parallel$ close to $-u$ gives only a small change in the number of trapped-barely-passing particles (red areas between curves cancel) in the trapped-barely-passing region.}
    \label{fig: MaxShift}
\end{figure}

Integration over $x=\sqrt{m\mu B/T+m(u+V_{\parallel })^2/(2T)}$ gives the final form of $\Gamma$, 

\begin{equation}\label{Fparallel}
    \begin{split}
       \Gamma=-1.102\sqrt{\frac{r}{R}}\frac{\nu I^2  p}{\abs{S}^{3/2}m\Omega^2}\Bigg\lbrace & \left[\pdv{}{\psi}\ln p-\frac{m(u+V_{\parallel})}{T}\left(\pdv{V_{\parallel }}{\psi}-\frac{\Omega}{I}\right)\right]G_1(\bar{y},\bar{z})\\
        &-1.17\frac{1}{T}\pdv{T}{\psi} G_2(\bar{y},\bar{z})\Bigg\rbrace,
    \end{split}
\end{equation}
where $\bar{y}=\sqrt{m/(2T)}(u+V_{\parallel })$, $\bar{z}=m{u}^2/T-Ze\phi_c R/(T r)$,

\begin{align}\label{Gs}
    G_1(\Bar{y},\Bar{z})\equiv\frac{\int_{|\bar{y}|}^\infty\mathrm{d}x\:k(x,\bar{y},\bar{z})}{\int_0^\infty\mathrm{d}x\:0.5xe^{-x^2}\left[\Xi(x)-\Psi(x)\right]}=7.51\int_{|\bar{y}|}^\infty\mathrm{d}x\:k(x,\bar{y},\bar{z}), 
\end{align}
\begin{equation}\label{Gs2}
\begin{split}
    G_2(\Bar{y},\Bar{z})&\equiv\frac{\int_{|\bar{y}|}^\infty\mathrm{d}x \left(x^2-5/2\right)k(x,\bar{y},\bar{z})}{\int_0^\infty\mathrm{d}x\:0.5x\left(x^2-5/2\right) e^{-x^2}\left[\Xi(x)-\Psi(x)\right]}\\
    &=-6.40\int_{|\bar{y}|}^\infty\mathrm{d}x \left(x^2-\frac{5}{2}\right)k(x,\bar{y},\bar{z}),
    \end{split}
\end{equation}
 and
\begin{equation}\label{k}
    k(x,\bar{y},\bar{z})=\sqrt{\abs{x^2+\bar{z}-\bar{y}^2}}e^{-x^2}\left\lbrace\left(\frac{1}{2}-\frac{\bar{y}^2}{2x^2}\right)\left[\Xi(x)-\Psi(x)\right]+\frac{\bar{y}^2}{x^2}\Psi(x)\right\rbrace.
\end{equation}
The functions $G_1$ and $G_2$ are normalised to recover the standard neoclassical results when $\bar{y}=0=\bar{z}$, $G_1(0,0)=1=G_2(0,0)$. The neoclassical ion particle flux in \eqref{Fparallel} depends on the radial electric field through $u$ (see \eqref{def u}) and thus also through $\bar{y}$ and $\bar{z}$. We note that the term in \eqref{Fparallel} proportional to $[V_\parallel-(-u)]\Omega/I$ is particle flux due to the parallel friction between trapped particles located around $-u$ and the passing particles with a mean velocity $V_\parallel$. The term proportional to $(u+V_\parallel)\partial V_\parallel/(\partial \psi)$ is related to a shift in the Maxwellian and hence to the density gradient if the Maxwellian is not centered around the trapped region, i.e. if $V_\parallel+u$ is not small (see figure \ref{fig: MaxShift}). The remaining terms include the pressure and temperature gradients that usually drive radial particle flux but here are modified by the integrals $G_1$ and $G_2$. Note that also the poloidal potential affects transport as it enters in $\bar{z} $. \par
Similarly, $Q$ is
\begin{equation}\label{Q}
    \begin{split}
        Q=&\frac{mu^2}{2}\Gamma-1.463\sqrt{\frac{r}{R}}\frac{\nu I^2 p T}{\abs{S}^{3/2}m\Omega^2}\Bigg\lbrace\left[\pdv{}{\psi}\ln p-\frac{m(u+V_{\parallel })}{T}\left(\pdv{V_{\parallel }}{\psi}-\frac{\Omega}{I}\right)\right]H_1(\bar{y},\bar{z})\\
        &-0.25\frac{1}{T}\pdv{T}{\psi} H_2(\bar{y},\bar{z})\Bigg\rbrace,
    \end{split}
\end{equation}
where
\begin{align} \label{Hs}
    H_1(\Bar{y},\Bar{z})\equiv\frac{\int_{|\bar{y}|}^\infty\mathrm{d}x \left(x^2-\bar{y}^2\right)k(x,\bar{y},\bar{z})}{\int_0^\infty\mathrm{d}x\:0.5x^3e^{-x^2}[\Xi(x)-\Psi(x)]}=5.66\int_{|\bar{y}|}^\infty\mathrm{d}x \left(x^2-\bar{y}^2\right)k(x,\bar{y},\bar{z})
    \end{align}
    and
    \begin{equation}\label{Hs2}
    \begin{split}
    H_2(\Bar{y},\Bar{z})&\equiv\frac{\int_{|\bar{y}|}^\infty\mathrm{d}x \left(x^2-\bar{y}^2\right)\left(x^2-5/2\right)k(x,\bar{y},\bar{z})}{\int_0^\infty\mathrm{d}x\:0.5x^3\left(x^2-5/2\right)e^{-x^2}[\Xi(x)-\Psi(x)]}\\
    &=-22.63 \int_{|\bar{y}|}^\infty\mathrm{d}x \left(x^2-\bar{y}^2\right)\left(x^2-\frac{5}{2}\right)k(x,\bar{y},\bar{z}).
    \end{split}
\end{equation}
Again, we introduce a convenient normalisation such that $H_1(0,0)=1=H_2(0,0)$ in the standard neoclassical limit. The dependence of the neoclassical ion energy flux \eqref{Q} on the radial electric field is hidden in $u$, $\bar{y}$, and $\bar{z}$.\par

We have found explicit expressions for particle \eqref{Particle conservation}, parallel momentum \eqref{Momentum conservation} and energy conservation \eqref{Energy conservation}. Next, we want to compare our results to previous work. First, we take the high flow and low flow neoclassical limit, and then we give a comparison of our results to those by \cite{catto2013} and \cite{shaing12}. \par
In the high flow regime of the usual neoclassical theory \citep{hinton85}, $V_\parallel+u$ and all gradients as well as source terms are small. If we take this limit in \eqref{gammac1} and assume that the source of parallel momentum $\gamma$ is small, we find that
\begin{equation}
    \Gamma=0,
\end{equation}
which is consistent with the usual result in the high-flow regime \citep{hinton85, catto87}.
Using the particle flux equation \eqref{Fparallel}, $\bar{\Gamma}=0$ gives
\begin{equation}\label{high flow particle}
    \pdv{}{\psi}\ln p+\frac{m\Omega(V_\parallel +u)}{IT}=1.17\frac{G_2(\bar{y},\bar{z})}{G_1(\bar{y},\bar{z})}\pdv{}{\psi}\ln T.
\end{equation}
We can use this in \eqref{Q} to get the high flow energy flux
\begin{equation}\label{high flow Q}
    Q=-1.71\sqrt{\frac{r}{R}}\frac{I^2\nu T p}{m\Omega^2}\Delta \bar{Q}\pdv{}{\psi}\ln T,
\end{equation}
where
\begin{align}\label{Delta Q}
\Delta\bar{Q}\equiv\frac{H_1(\bar{y} ,\bar{z} )G_2(\bar{y} ,\bar{z} )-0.21H_2(\bar{y} ,\bar{z} )G_1(\bar{y} ,\bar{z} )}{G_1(\bar{y} ,\bar{z} )}\geq0.
\end{align}
The quantity $\Delta \bar{Q}$ is positive, which follows from
\begin{equation}
    \Delta \Bar{Q}=-4.82\frac{\left(\int_\abs{\bar{y} }^\infty\mathrm{d}x \:x^2 k(x,\bar{y} ,\bar{z} )\right)^2- \int^\infty_\abs{\bar{y} }\mathrm{d}x\:k(x,\bar{y} ,\bar{z} )\int_\abs{\bar{y} }^\infty\mathrm{d}x\:x^4k(x,\bar{y} ,\bar{z} )}{\int_\abs{\bar{y} }^\infty\mathrm{d}x \:k(x,\Bar{y},\Bar{z})
    }
\end{equation}

and the Cauchy-Schwarz inequality
\begin{equation}\label{Cauchy Schwarz}
    \begin{split}
    \left(\int_\abs{\bar{y} }^\infty\mathrm{d}x \:x^2 k(x,\bar{y} ,\bar{z} )\right)^2\leq \int^\infty_\abs{\bar{y} }\mathrm{d}x\:k(x,\bar{y} ,\bar{z} )\int_\abs{\bar{y} }^\infty\mathrm{d}x\:x^4k(x,\bar{y} ,\bar{z} ).
    \end{split}
\end{equation}
Here, $k(x,\bar{y} ,\bar{z} )$ is given in \eqref{k}. Note that $k>0$ because $\Xi-\Psi>0$, $\Psi>0$ and $x\geq\abs{\bar{y}}$.\par

The quasineutrality condition \eqref{phi banana final} gives the poloidally varying electric potential in the high flow limit,
\begin{equation} \label{high flow phi}
    \left(\frac{en_e}{T_e}+\frac{Z^2n_i e}{T}\right)\phi_c=-Zn_i\frac{r}{R}\frac{mu^2}{T}.
\end{equation}
The only contribution to the potential comes from the centrifugal force as all gradients and $m(V_\parallel +u)^2/T$ terms are small while $V_\parallel\simeq -u \sim v_t$.

\par The low flow neoclassical results can be retrieved by taking the limit of small radial electric field, $u/v_t \ll 1$, small mean parallel flow, $V_\parallel/v_t\ll1 $, and small gradients. It follows from \eqref{high flow phi} that the poloidal variation of the potential is small so that we can set $\bar{z}=0$ in the arguments of $G_1$, $G_2$, $H_1$, and $H_2$. Without a source of parallel momentum $\gamma=0$, equation \eqref{gammac1} gives $\Gamma=0$, so the mean parallel flow follows directly from \eqref{high flow particle}
\begin{equation}\label{neo V}
    V_{\parallel }=-\frac{IT}{m\Omega}\left(\pdv{}{\psi}\ln p+\frac{Ze}{T}\pdv{\Phi}{\psi}-1.17\pdv{}{\psi}\ln T\right).
\end{equation}
The neoclassical energy flux $Q$ then follows directly from \eqref{high flow Q} for $\Delta \bar{Q}=0.79$ and reads
\begin{equation}\label{neo q}
    Q=-1.35\sqrt{\frac{r}{R}}\frac{I^2\nu pT}{m \Omega ^2}\pdv{}{\psi }\ln T,
\end{equation}
in agreement with \cite{hinton85} and \cite{catto87}.
\par We can compare our results with those of \cite{catto2013} by taking the limit of small temperature gradient and small ${V}_\parallel $. We are able to retrieve the same energy flux if we set $\phi_\theta=0$, $\Gamma=0$ and correct an error in \cite{catto10} and pointed out by \cite{shaing12}. The calculation is presented in detail in Appendix \ref{sec: small temp}. \par 
The energy flux $Q$ in \eqref{Q} is proportional to $\abs{S}^{-3/2}$ and decays $\sim \exp(-\bar{y} ^2)$ which is consistent with the results of strong radial electric field and radial electric field shear obtained by \cite{Shaing92, shaing12}. We compare our results in the limit $(I/\Omega)(\partial V_\parallel/\partial \psi) \ll1$ and $\Gamma=0$ to those of \cite{shaing12} in Appendix \ref{sec: ShaingLimit}. We find the same particle and energy equations if we account for a discrepancy in the function $k(x,\bar{y} ,\bar{z} )$.\par

Comparisons to numerical results can be made in certain limits. The global code PERFECT requires weak temperature gradients and could be checked against our results in the limit of small temperature gradient \citep{landreman14}. Other codes such as the axisymmetric versions of XGC \citep{chang04}, Gkeyll \citep{hakim20}, and COGENT \citep{dorf12} could be used to reproduce aspects of the strong gradient fluxes and poloidal variation. In the next section, we impose radial force balance (see \eqref{radial force balance}) which needs to be reconsidered carefully when comparing the following results to numerical evaluations of fluxes.
\color{black}

\section{Transport equations and flux conditions} \label{sec: Transport}
We work with equations \eqref{gammac1}, \eqref{Fparallel} and \eqref{Q} to find relations between the particle flux $\Gamma$, the parallel momentum input $\gamma$, the energy flux $Q$, and the physical quantities $T$, $n$, $u$, $V_\parallel$, and $\phi_c$. Given $\Gamma$ and $Q$ as functions of $\psi$, and boundary conditions at the top or bottom of the transport barrier, we can integrate the equations to obtain the profiles of $T$, $n$, $u$, $V_\parallel$, and $\phi_c$.\par

So far, we have an equation for the particle flux \eqref{Fparallel}, the parallel momentum equation \eqref{gammac1}, the energy flux \eqref{Q} and quasineutrality \eqref{phi banana final}. We are missing an equation for the radial electric field to be able to relate $\Gamma$, $\gamma$ and $Q$ with $T$, $n$, $u$, $V_\parallel$ and $\phi_c$. The equation for the radial electric field is provided by the conservation of toroidal angular momentum, but the necessary derivation is beyond the scope of this paper. For the purpose of the following calculations, we assume that for the ions, the pressure gradient is the dominant contribution in the radial force balance \citep{McDermott09, viezzer2013, Kagan08}. Hence, we impose
\begin{equation}\label{radial force balance}
    Z e n \pdv{\Phi}{\psi}+\pdv{p}{\psi}=0,
\end{equation}
which can be written as
\begin{equation}\label{pressure balance 1}
    \pdv{}{\psi}\ln n=-\frac{\Omega m u}{IT}-\pdv{}{\psi}\ln T.
\end{equation}\par
We introduce the new, dimensionless quantities,
\begin{align}
    \bar{u}=\sqrt{\frac{m}{2T_0}}u,&&\bar{V}=\sqrt{\frac{m}{2T_0}}V_{\parallel },&&\bar{T}=\frac{T}{T_{0}},&& \bar{n}=\frac{n}{n_{0}},  && \bar{\phi}_c=\frac{Ze\phi_cR}{T_0r}
\end{align}
\begin{align}
   \pdv{}{\bar{\psi}}=\frac{I}{\Omega}\sqrt{\frac{2T_0}{m}}\pdv{}{\psi}, &&\bar{z}=\bar{T}^{-1}\left(2\bar{u}^2-\bar{\phi}_c\right), && \bar{y}=\frac{\bar{u}+\bar{V}}{\sqrt{\bar{T}}},
\end{align}
where $T_0$ is the ion temperature and $n_0$ the ion density at the boundary $\psi=0$. In the banana regime, the normalised fluxes are
\begin{align}
    \bar{\Gamma}=\frac{\Gamma}{n_0I\sqrt{\frac{2T_0r}{mR}}\frac{\nu_0}{\Omega}},&& \bar{Q}=\frac{Q}{n_0I\sqrt{\frac{2T_0r}{mR}}T_0\frac{\nu_0}{\Omega}},&&   \bar{\gamma}=\frac{\gamma}{\nu_0 n_0\sqrt{2mT_0r/R}},
\end{align}
where $\nu_0$ is the collision frequency at the boundary.
Changing to these dimensionless variables, we arrive at the following set of equations for the banana regime:
The particle flux equation from \eqref{Fparallel} and \eqref{pressure balance 1} is
\begin{equation} \label{V Banana}
\begin{split}
   \bar{\Gamma}=&-0.55\frac{\bar{n}^2}{\abs{S}^{3/2}\bar{T}^{3/2}}\Bigg\lbrace\left[-2\bar{u}-2(\bar{u}+\bar{V})\left(\pdv{\bar{V}}{\bar{\psi}}-1\right)\right]G_1(\bar{y},\bar{z}) -1.17\pdv{\bar{T}}{\bar{\psi}}G_2(\bar{y},\bar{z})\Bigg\rbrace.
\end{split}
\end{equation}
The parallel momentum equation from \eqref{gammac1} is
\begin{equation} \label{u Banana}
    \bar{u} \pdv{\bar{\Gamma}}{\bar{\psi}}+S\bar{\Gamma}=-\bar{\gamma}.
\end{equation}
The energy flux equation from \eqref{Fparallel}, \eqref{Q}, and \eqref{pressure balance 1} is
\begin{equation} \label{q Banana}
\begin{split}
    \bar{Q}=\bar{u}^2\bar{\Gamma}-0.73\frac{\bar{n}^2}{\abs{S}^{3/2}\bar{T}^{1/2}}\Bigg\lbrace\Bigg[-2\bar{u}-2(\bar{u}+\bar{V})\left(\pdv{\bar{V}}{\bar{\psi}}-1\right)\Bigg] H_1(\bar{y},\bar{z})-0.25\pdv{\bar{T}}{\bar{\psi}}H_2(\bar{y},\bar{z})\Bigg\rbrace.
        \end{split}
\end{equation}
The pressure balance equation from \eqref{pressure balance 1} gives
\begin{equation} \label{n Banana}
    \pdv{}{\bar{\psi}}\ln \bar{n}=-2\frac{\bar{u}}{\bar{T}}-\pdv{}{\bar{\psi}}\ln \bar{T};
\end{equation}
and the equation for the potential, which can be derived from \eqref{phi banana final}, is
\begin{equation}\label{z Banana}
\begin{split}
    &\Bigg\lbrace\! 1\!-\!\frac{Z T_{e}}{T_{0}}\!\frac{1}{\bar{T} }\!\left[\!\sqrt{\bar{T} }J\!\left(\!-2\frac{\bar{u} }{\bar{T} }\!-\!\frac{3}{2}\pdv{}{\bar{\psi} }\!\ln \!\bar{T} \!\right)\!+\!\left(\!1\!-\!2\frac{\bar{V} \!+\!\bar{u} }{\sqrt{\bar{T} }}\!J\!\right)\!\left(\!\pdv{\bar{V} }{\bar{\psi} }\!-\!1\!-\!\frac{\bar{V} +\bar{u} }{2}\!\pdv{}{\bar{\psi} }\!\ln\! T\!\right)\!\right]\!\Bigg\rbrace\bar{z} \\
    &=\frac{Z T_{e}}{T_{0}}\frac{1}{\bar{T} }\Bigg\lbrace \sqrt{\bar{T} }J\left[\left(2\frac{\bar{V} ^2-\bar{u} ^2}{\bar{T} }+1\right)\left(-2\frac{\bar{u} }{\bar{T} }-\frac{3}{2}\pdv{}{\bar{\psi} } \ln \bar{T} \right)+\pdv{}{\bar{\psi} }\ln \bar{T} \right]\\
    &+\left[1-2\frac{\bar{V} +\bar{u} }{\sqrt{\bar{T} }}J\right]\Bigg[(\bar{V} -\bar{u} )\left(-2\frac{\bar{u} }{\bar{T} }-\frac{3}{2}\pdv{}{\bar{\psi} } \ln \bar{T} \right)\\
    &+\left(\pdv{\bar{V} }{\bar{\psi} }-1\right)\left(1-2\frac{(\bar{V} +\bar{u} )^2}{\bar{T} }\right)-(\bar{V} +\bar{u} )\left(\frac{\bar{V} ^2-\bar{u} ^2}{\bar{T} }+\frac{1}{2}\right)\pdv{}{\bar{\psi} }\ln \bar{T} \Bigg]\\
    &+\left[1+2\frac{(\bar{V} +\bar{u} )^2}{\bar{T} }-4\frac{(\bar{V} +\bar{u} )^3}{\bar{T} ^{3/2}}J\right]\left(\pdv{\bar{V} }{\bar{\psi} }-1+\frac{\bar{V} -\bar{u} }{2}\pdv{}{\bar{\psi} }\ln \bar{T} \right)+2+2\frac{T_{0}}{ZT_e}\bar{u} ^2    \Bigg\rbrace.
    \end{split}
\end{equation}
The functions $J$, $G_1$, $G_2$, $H_1$, and $H_2$ are given in \eqref{J}, \eqref{Gs}, \eqref{Gs2}, \eqref{Hs}, and \eqref{Hs2}. This set of equations is the most important result of our calculation and allow a discussion of the neoclassical transport of ions in strong gradient regions.

We can integrate equations \eqref{V Banana}-\eqref{z Banana} relating $\bar{n}$, $\bar{T}$, $\bar{u}$, $\bar{V}$, and $\bar{z}$ numerically by imposing boundary conditions at the top of the transport barrier and specifying particle, parallel momentum and energy sources to find profiles in the pedestal. We discuss the implications for particle (section \ref{sec: Particle Flux}) and energy flux (section \ref{sec: Energy Flux}) before presenting some example profiles (section \ref{sec: Profiles}).

\subsection{Particle flux and parallel momentum injection} \label{sec: Particle Flux}
In order to understand the appearance of a neoclassical particle flux, we analyse the parallel momentum equation \eqref{u Banana}. In edge transport barriers, measurements of the radial electric field have shown that in the pedestal $\partial \phi/\partial \psi>0$ and thus $\bar{u} >0$ \citep{McDermott09}. We assumed that in the pedestal $\bar{u} \sim 1$. However, at the boundary to the large turbulent transport region, where our model connects to the usual neoclassical regime of small gradients in density and temperature, $\bar{u} \ll 1$. Thus, we are looking for solutions with a growing positive $\bar{u} $ as one moves into the transport barrier. 
Importantly, if there is no parallel momentum input, the particle flux must decay to ensure that $\bar{u} $ grows, because it follows from \eqref{u Banana} that
\begin{equation}\label{Gamma for gamma0}
    \bar{\Gamma} \propto \exp\left(-\int\mathrm{d}\bar{\psi}  \: \frac{S }{\bar{u} }\right).
\end{equation}
For $\bar{u} >0$ and $S >0$, the neoclassical particle flux $\bar{\Gamma}$ decreases and is even smaller inside a transport barrier than outside when $\bar{\gamma}=0$. \par

\begin{figure}
\centering
  \subfigure[]{\includegraphics[width=0.45\textwidth]{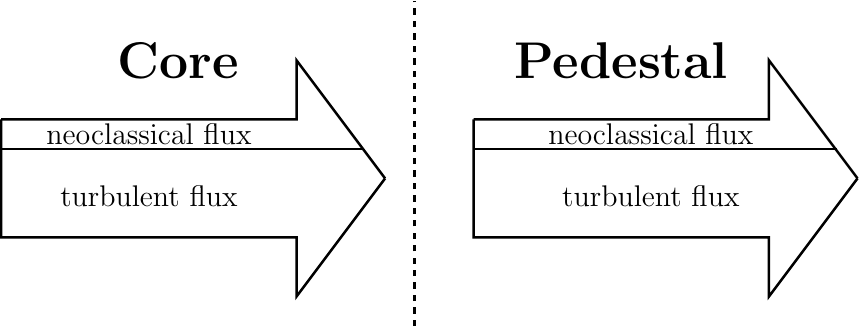}}
  \hspace{1cm}
  \subfigure[]{\includegraphics[width=0.45\textwidth]{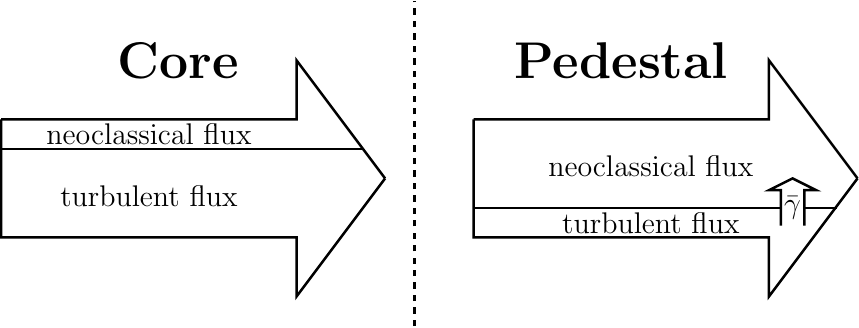}}
    \caption{(\textit{a}) The entire particle flux is carried by turbulence and the neoclassical particle flux stays negligible. (\textit{b}) Turbulence interact with neoclassical physics and supplies a parallel momentum source that allows a growing neoclassical particle flux.}
    \label{fig:CoreEdgeOpts}
\end{figure}

We argued in section \ref{sec:phase_space} that at the inner edge of a transport barrier there is a region of large turbulent transport and small collisional transport whereas in the transport barrier, we find a region of low turbulence. In order to keep up the same total flux, the neoclassical fluxes must increase and pick up the decreasing turbulent fluxes (see figure \ref{fig: CoreEdge}). However, this initial picture is too simple as it disagrees with our analysis of the decreasing particle flux. One option to solve the contradiction is that the particle flux is still carried by turbulence because the neoclassical fluxes never pick up the turbulent contribution. There must be enough turbulence in the transport barrier to carry the entire particle flux -- recall that at this point we are only discussing the particle flux and not the energy flux. So even if the entire particle flux is carried by turbulence, the energy flux could still be neoclassical (see figure \ref{fig:CoreEdgeOpts}a). The second option is that the particle flux is truly neoclassical in the transport barrier, but turbulence or impurities supply the necessary parallel momentum source $\gamma$ so that \eqref{Gamma for gamma0} is not valid. Somehow, and we can not specify at this point how exactly, turbulence or impurities interact with neoclassical transport and appear as a source of parallel momentum (see figure \ref{fig:CoreEdgeOpts}b). The difference between the two options is that in the first picture, the neoclassical particle flux is close to zero whereas in the second picture the particle flux is in large part neoclassical because turbulence or impurities produce $\bar{\gamma}$. This picture is consistent with previous results by \cite{landreman12} about the necessity of sources for non-zero steady state transport in the edge. Without the source, no ion neoclassical particle flux develops in the pedestal. \par
The neoclassical ion particle flux is larger than the electron particle flux by order \textit{O}$(\sqrt{m/m_e})$. Unlike in the weak gradient region, where intrinsic ambipolarity prevents different sizes of electron and ion particle fluxes, the neoclassical ion particle flux in the strong gradient region can be significantly larger than the neoclassical electron particle flux in the presence of sources if the total particle fluxes which include both the turbulent and neoclassical parts obey ambipolarity. Intrinsic ambipolarity \citep{ sugama98, parra09, calvo12} does not hold in the strong gradient limit where gradient length scales are of the order of $\rho_p$.
\color{black}

It is also worth pointing out that $\bar{\Gamma}$ and $\bar{\gamma}$ are necessarily of opposite sign if $\abs{\bar{\Gamma}}$ grows as one moves into the transport barrier. An outwards neoclassical ion particle flux requires a negative parallel momentum injection.


\subsection{Energy Flux}\label{sec: Energy Flux}
\begin{figure}
  \subfigure[]{\includegraphics[trim=0 0 0 0,clip,width=0.47\textwidth]{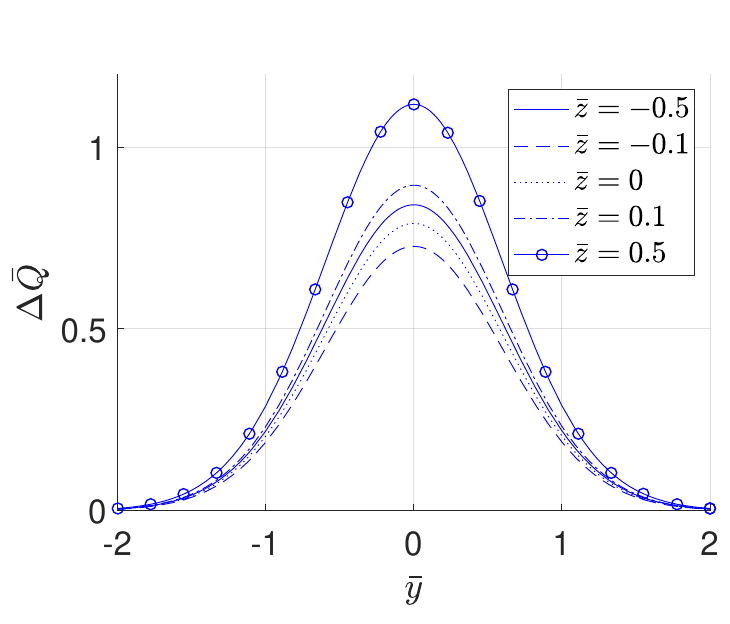}}
  \subfigure[]{\includegraphics[trim=0 0 0 0,clip,width=0.47\textwidth]{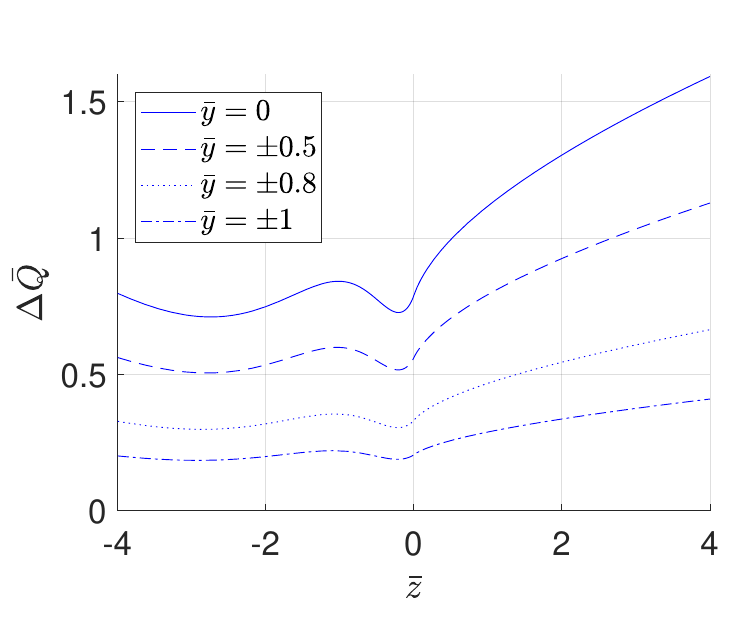}}
    \caption{(\textit{a}): The quantity $\Delta \bar{Q}$ in \eqref{Delta Q} as a function of $\bar{y} $ for different values of $\bar{z} $. (\textit{b}): The quantity $\Delta \bar{Q}$ as a function of $\bar{z} $ for different values of $\bar{y} $.}
    \label{fig:DeltaQ}
\end{figure}

Next, we want to discuss the energy flux equation \eqref{q Banana}. 
In transport barriers, $\bar{T} $ and $\bar{n} $ decrease. One can use this behaviour to estimate the energy flux in this case. Combining \eqref{V Banana} and \eqref{q Banana} to solve for $\partial \bar{T} /\partial \bar{\psi} $ as a function of $\bar{Q}$ and $\bar{\Gamma}$ yields
\begin{equation}\label{dTdpsi}
\begin{split}
    \pdv{\bar{T} }{\bar{\psi} }=&-\underbrace{1.17\frac{\abs{S} ^{3/2}\bar{T} ^{1/2}}{\bar{n} ^2}}_{>0}\underbrace{\frac{1}{\Delta\bar{Q}}}_{\geq 0} \Bigg\lbrace\bar{Q}-\left[ \bar{u} ^2+1.33\bar{T}  \frac{H_1(\bar{y} ,\bar{z} )}{G_1(\bar{y} ,\bar{z} )}\right]\bar{\Gamma}\Bigg\rbrace,
    \end{split}
\end{equation}
where $\Delta \bar{Q}$ was defined in \eqref{Delta Q}. Figure \ref{fig:DeltaQ} shows $\Delta \bar{Q}$ for different values of $\bar{y}$ and $\bar{z}$. It is large for small $\bar{y}$, symmetric in $\bar{y}$ with a maximum at $\bar{y}=0$, and asymmetric in $\bar{z}$ with larger values for $\bar{z}>0$. When $\bar{z}$ increases so does the number of trapped particles. Thus, $\Delta \bar{Q}$ is large when there are many trapped particles. \par

\begin{figure}
  \subfigure[]{\includegraphics[trim=0 0 0 0,clip,width=0.47\textwidth]{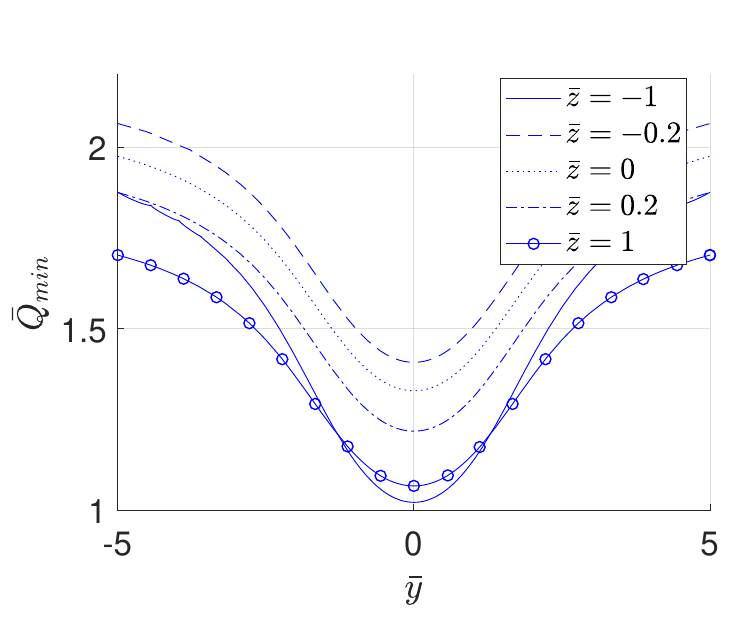}}
  \subfigure[]{\includegraphics[trim=0 0 0 0,clip,width=0.47\textwidth]{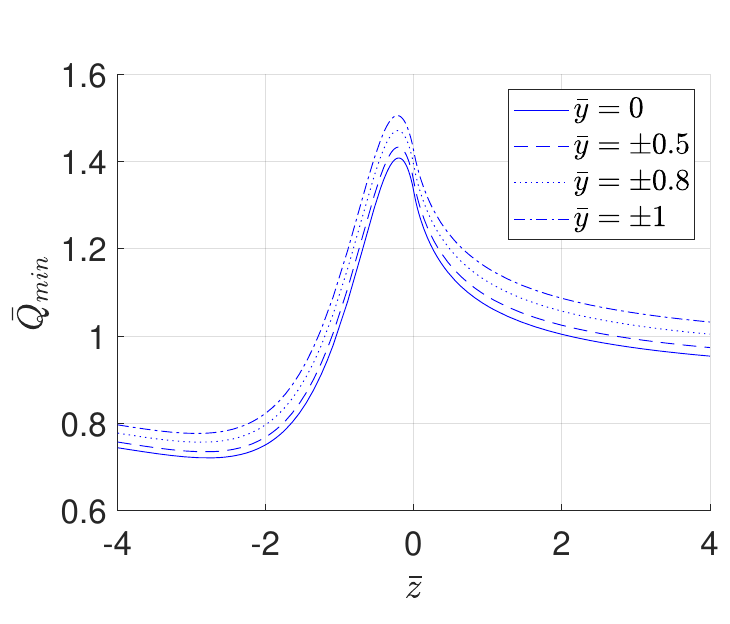}}
    \caption{(\textit{a}): The quantity $\bar{Q}_\text{min}$ defined in \eqref{Qmin} as a function of $\bar{y} $ for different values of $\bar{z} $, where $\bar{u}=0$, $\bar{T}=1$ and $\bar{\Gamma}=1$ (\textit{b}): The quantity $\bar{Q}_\text{min}$ as a function of $\bar{z} $ for different values of $\bar{y} $, where $\bar{u}=0$, $\bar{T}=1$ and $\bar{\Gamma}=1$.}
    \label{fig:Qmin}
\end{figure}

In order to get a negative temperature gradient, the expression in braces in \eqref{dTdpsi} must be positive. Thus, we find a lower bound for the energy flux
\begin{align}\label{Qmin}
    \bar{Q}\geq\bar{Q}_\text{min}=\left[\bar{u} ^2+1.33\bar{T}  \frac{H_1(\bar{y} ,\bar{z} )}{G_1(\bar{y} ,\bar{z} )}\right]\bar{\Gamma}.
\end{align}
The factor multiplying $\bar{\Gamma}$ is positive because $\bar{u}^2\geq 0$, $\bar{T}\geq 0$, and $k>0$ and $x\geq \abs{\bar{y}}$ in \eqref{Gs} and \eqref{Hs}. From this, we see that it is not possible to only have neoclassical particle flux and zero neoclassical energy flux. As long as there is neoclassical particle flux, energy will get advected by that particle flux.
Thus the energy flux will be in the same direction as the particle flux.  The quantity $\bar{Q}_\text{min}$ is shown in figure \ref{fig:Qmin}. It is large for large $\abs{\bar{y}}$ and small $\bar{z}$.
\par 

Surprisingly, a negative density gradient imposes an upper boundary for $\bar{Q}$. From \eqref{n Banana} it follows that for $\partial \bar{n}  /\partial \bar{\psi} <0$,
\begin{equation}
\begin{split}
    \frac{2\bar{n} }{\bar{T} }\bar{u }>&1.17 \frac{\abs{S} ^{3/2}}{\bar{n} \bar{T}^{1/2} }\frac{\bar{Q}-\bar{Q}_\text{min}}{\Delta \bar{Q}}
    \end{split}
\end{equation}
and thus we find that in order for $\bar{T} $ and $\bar{n} $ to decay simultaneously, the neoclassical energy flux has to be
\begin{align}\label{qBound}
    \bar{Q}_\text{min}<\bar{Q}  <\bar{Q}_\text{min}+1.71\frac{\bar{n}^2 \bar{u} }{\bar{T} ^{1/2}\abs{S} ^{3/2}}\Delta \bar{Q}.
    \end{align}
For zero neoclassical particle flux, the maximum energy flux for decaying density and temperature profiles is
\begin{equation}\label{Qmax}
    \bar{Q}_\text{max}=1.71\frac{\bar{n}^2 \bar{u} }{\bar{T} ^{1/2}\abs{S} ^{3/2}}\Delta \bar{Q}.
\end{equation}
If the density falls off faster than the temperature in such a way that $\bar{n} ^2/\sqrt{\bar{T} }\rightarrow0$, which can be expressed as 
\begin{equation}\label{L condition}
    L_{\bar{n}}<4L_{\bar{T}},
\end{equation}
then the upper bound of the energy flux in \eqref{qBound} also decreases unless it is compensated by a stronger growth in  $\bar{u}\Delta\bar{Q}/\abs{S}^{3/2}$. In most H-mode pedestals, \eqref{L condition} is observed \citep{Viezzer18,viezzer17}. It follows, that in order to achieve a growing neoclassical energy flux, it is necessary that $\bar{u}\Delta\bar{Q}/\abs{S}^{3/2}$ increases. Thus, the radial electric field seems to play an important role for the neoclassical energy flux at the top of transport barriers. Note, however, that the result in \eqref{qBound} relies strongly on the assumption made in \eqref{radial force balance} between the pressure gradient and the electric field, which is only applicable in the pedestal and not self-consistently derived. \color{black} A more thorough discussion of this relation will be necessary and we leave it for future work. For now, using \eqref{pressure balance 1}, the estimate \eqref{qBound} holds. We already argued in section \ref{sec: Particle Flux} that $\bar{u}$ is positive and growing at the transition from core to pedestal. The quantity $\Delta \bar{Q}$ is large for large $\abs{\bar{z}}$ and small $\bar{y}$ (see figure \ref{fig:DeltaQ}). This is consistent because large $\abs{\bar{z}}$ leads to an increased number of trapped particles. Transport is dominated by trapped particles, so more trapped particles allow for a larger energy flux. Small $\bar{y}$ likewise maximises the number of trapped particles because the trapped region is located close to the maximum of the lowest order Maxwellian.\par
In I-mode pedestals, the temperature falls off much faster than the density \citep{walk14}. In this case, \eqref{L condition} would not necessarily hold and the neoclassical heat flux could grow with a weaker radial electric field than in H-mode.

\subsection{ Example Profiles}\label{sec: Profiles}
To show some example solutions of \eqref{V Banana}-\eqref{z Banana}, we can take profiles of ion and electron temperature and density loosely based on those measured by \cite{viezzer17}. With these profiles, we calculate fluxes, velocities and electric potential.\par 
The integration of the mean parallel flow turns out to be very sensitive to the boundary conditions and source terms. Thus, we leave the discussion of the mean parallel flow solutions for future work, and instead only consider cases of known mean parallel flow. The two profiles we discuss for $\Bar{V}$ are the "high flow" case and the "low flow" case. Here, "high flow" and "low flow" only refers to the relationship between the mean parallel flow and the gradients of the density, temperature and potential and not to the usual stricter limits that we have discussed at the end of section \ref{sec: Moment}. \par
For the "high flow" profile, we set
\begin{equation}\label{V-u}
    \bar{V}=-\bar{u}.
\end{equation}
In this case, there is no friction between trapped and passing particles and the particle flux due to a shift in the Maxwellian is small because $f_M$ is centered around the trapped particle region (see discussion below \eqref{k}). For the "low flow" profile, we replace condition \eqref{V-u} with the usual neoclassical solution \eqref{neo V}.\par

The profile of $\bar{u} $ follows directly from assumption \eqref{n Banana} and consequently $\bar{V}$ is given by \eqref{V-u} for the first case or \eqref{neo V} for the second case. The quantities $\bar{T}$, $\bar{n}$, $\bar{V}$, and $\bar{u}$ based on realistic profiles or assumptions are presented in figure \ref{fig: Profiles}. The input profiles are further discussed in Appendix \ref{sec: pedestal profiles}.\par

\begin{figure}
    \centering
     \includegraphics[trim=0 0 0 0,clip,width=\textwidth]{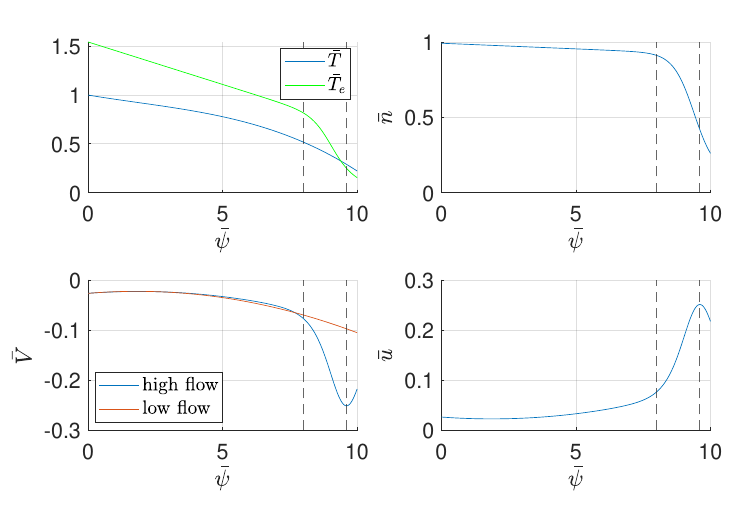}
    \caption{Input profiles of ion temperature, electron temperature $\bar{T}_e=T_e/T_{0}$, and density based on the profiles reported by \cite{viezzer17}, as well as the corresponding $\bar{u} $ and $\bar{V}$. The red profile for $\bar{V}$ is the usual neoclassical result for the mean parallel velocity as given by \eqref{neo V} and the blue curve is the "high flow" profile as given by \eqref{V-u}. Vertical dashed lines indicate the position of the top of the pedestal $\bar{\psi}=0.8$ and the point of maximum pressure gradient and minimum radial electric field $\bar{\psi}=0.965$.}
    \label{fig: Profiles}
\end{figure}
\begin{figure}
    \centering
     \includegraphics[trim=0 0 0 0,clip,width=\textwidth]{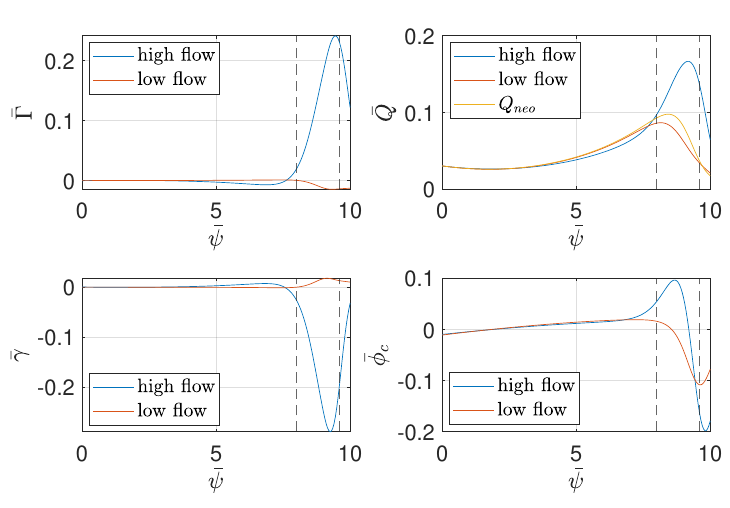}
    \caption{Calculated fluxes and poloidally varying potential from the profiles in figure \ref{fig: Profiles}. The blue profiles are the solutions with condition \eqref{V-u} whereas the red profiles show the solution with the usual neoclassical parallel velocity \eqref{neo V}. The yellow energy flux is the usual neoclassical result \eqref{neo q}. Vertical dashed lines highlight the top of the pedestal $\bar{\psi}=0.8$ and the point of maximum pressure gradient and minimum radial electric field $\bar{\psi}=0.965$. \color{black}}
    \label{fig: Fluxes}
\end{figure}

\color{black} The graphs in figure \ref{fig: Profiles} and figure \ref{fig: Fluxes} show the transition between core and pedestal nicely in the sense that at $\bar{\psi} =0$, which corresponds to $\rho_\text{pol}=0.8$ in \cite{viezzer17}, the profiles of density and temperature are still relatively flat. We see the expected growth of $\bar{u} $ in the strong gradient region starting at $\bar{\psi}=0.8$ (first dashed line in figure \ref{fig: Fluxes}) which relaxes when the pressure gradient reduces again beyond the dashed line at $\bar{\psi}=0.965$\color{black}. For $\bar{V} $, we see the difference between "high flow" and standard "low flow" neoclassical theory. The solution for $\bar{V}$ from \eqref{V-u} exceeds the standard "low flow" neoclassical result in the pedestal by about a factor of two but becomes as small as the standard "low flow" neoclassical result at the boundary to the core.\par

Equation \eqref{z Banana} gives $\bar{z} $, with which $\bar{\Gamma}$ can be calculated from \eqref{V Banana}. Then, the energy flux can be calculated using \eqref{q Banana}. Lastly, the parallel momentum input that is necessary to sustain the particle flux follows from \eqref{u Banana}. The four graphs for $\bar{\Gamma}$, $\bar{Q}$, $\bar{\gamma}$, and $\bar{\phi}_c$ are presented in figure \ref{fig: Fluxes}. \par
\color{black} The poloidally varying part of the potential is much stronger for $\bar{V}=-\bar{u}$, and changes sign in the pedestal region. The neoclassical particle flux, which is close to zero in the \color{black} core requires parallel momentum input to grow. In the case with condition \eqref{V-u}, the \color{black}  particle flux and the parallel momentum input are much bigger than for the case with the usual neoclassical mean parallel velocity \eqref{neo V}. Note that, even for the "low flow" neoclassical mean parallel velocity, the parallel momentum input and the particle flux are non-zero. Interestingly, the neoclassical particle flux and parallel momentum source in the pedestal for \eqref{neo V} are of opposite sign to the case with condition \eqref{V-u}.
\color{black} The energy flux of the "high flow" case matches the standard "low flow" neoclassical result close to the inner boundary but further into the pedestal it grows faster with radius. In the case where we set the parallel velocity to be \eqref{neo V}, the energy flux is smaller than the usual neoclassical result $Q_\text{neo}$ of \eqref{neo q}. The prefactor $\bar{n} ^2/\sqrt{\bar{T}} $ in \eqref{q Banana} decays in the strong gradient region for the example profiles of density and temperature, so \eqref{L condition} is satisfied, and the energy flux decays after $\bar{u}$ has reached its maximum. This is consistent with our discussion in section \ref{sec: Energy Flux} and the \color{black} observation that the energy transport in pedestals reaches significant neoclassical levels only in the middle of a pedestal and not at the top and bottom \citep{Viezzer2020}. If instead we had chosen profiles with a stronger temperature gradient such that $L_{\bar{n}}>4L_{\bar{T}}$, we could have been able to retrieve a growing energy flux throughout the pedestal. \par
\begin{figure}
    \centering
     \includegraphics[trim=0 0 0 0,clip,width=\textwidth]{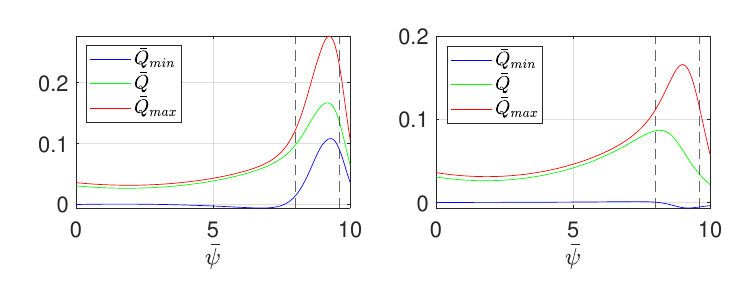}
    \caption{ Energy flux with upper and lower bounds \eqref{qBound} in (\textit{a}): the "high flow" case, and (\textit{b}): the "low flow" case.}
    \label{fig: Qbounds}
\end{figure}

\color{black} In figure \ref{fig: Qbounds} we show the energy fluxes and their respective lower bound \eqref{Qmin} and upper bound \eqref{Qmax}. In both cases, the energy flux is close to the upper bound in the flat gradient region. The lower bound stays close to zero where the particle flux is small.

\section{Conclusions} \label{sec: Results}
The core is a region of strong turbulent transport. With the transition into a transport barrier such as the pedestal, turbulence gets quenched and we argue that in order to keep up the total flux, the neoclassical fluxes must increase. This assumption is supported by experiments such as the ones by \cite{Viezzer18}, where it was demonstrated that the heat diffusivity reaches neoclassical levels in the pedestal. This opens the possibility of interaction between turbulent and neoclassical transport which we account for by keeping a source term that represents external particle, momentum and energy injection as well as interaction with turbulence. A random walk estimate was performed to predict the size of this source and to show that trapped particles give the main contribution to particle and energy transport. \par

We have extended neoclassical theory to transport barriers by choosing gradients to be of the same size as the poloidal gyroradius and expanded in large aspect ratio and low collisionality. A new set of variables, the fixed-$\theta$ variables, were derived from conserved quantities and confirmed that particles are trapped for $v_\parallel+u\sim \sqrt{\epsilon}v_t$. \par 

A change of variables to fixed-$\theta$ variables allowed for a convenient reduction of the drift kinetic equation, to which a Maxwellian is the solution to lowest order. We have discussed the trapped-barely-passing and freely passing regions in the banana regime. The drift kinetic equation can be solved for the trapped, barely-passing, and freely passing regions by expanding in $\sqrt{\epsilon}$. The phase space region of trapped and barely-passing particles is very narrow for large aspect ratio tokamaks and can be treated as a discontinuity in the freely passing region. The only information needed from the trapped-barely-passing region is the jump \eqref{jump condition banana final} and derivative discontinuity condition \eqref{dis condition banana final}. Additionally, one can find expressions for the poloidal variations of density \eqref{poloidal density 2} and potential \eqref{phi banana final} which have been observed previously \citep{theiler14, churchill15, cruz22}. Particles can get trapped on the high field side because the poloidally varying part of the potential can oppose the magnetic mirror and centrifugal forces. When integrating over velocity space, it is necessary to keep track of particles trapped on either side.\par

One can take moments of the freely passing particle equation \eqref{eqpassing} using the jump and derivative discontinuity condition to find the particle, parallel momentum and energy conservation equations \eqref{Particle conservation}, \eqref{Momentum conservation} and \eqref{Energy conservation}. From these equations, one can identify the neoclassical particle flux \eqref{Gammac} and neoclassical energy flux \eqref{qc1}. We find that the poloidally varying potential affects neoclassical fluxes and that the transport is dominated by trapped particles, which have a parallel velocity close to $-u$. The fluxes match with the usual neoclassical results in the appropriate limits. They equally match with the results for strong density and electric potential gradients derived by \cite{catto10} after we account for the missing orbit squeezing factor in the energy flux calculation, which was previously pointed out by \citet{shaing12}. In the limit of small mean velocity gradient and zero poloidal potential, we identify a previously noted discrepancy with \cite{shaing12}, but are otherwise able to reproduce their results. \par

The parallel momentum equation proves that a parallel momentum source is required to get a non-zero neoclassical particle flux. When there is no external parallel momentum source or sink in the edge (such as impurities or neutral beam injection), this implies that either turbulence does not decay and carries the particle flux throughout the transport barrier or that there is a mechanism by which turbulence supplies parallel momentum to neoclassical transport and the particle flux is indeed partially neoclassical. \par 

For the energy flux, we provided upper and lower bounds in relation to the particle flux to ensure decaying profiles of temperature and density (see \eqref{qBound}). The maximum energy flux can be achieved for $\bar{V}+\bar{u}=0$ and large $\bar{z}$. We also found that in pedestals a radially growing radial electric field is needed to obtain a radially growing neoclassical energy flux that substitutes the decreasing turbulent energy flux.\par 

We compared the high flow case $\bar{V}=-\bar{u}$ to the standard low flow neoclassical mean parallel velocity \eqref{neo V} to find fluxes for the realistic profiles of temperature and density presented in figure \ref{fig: Profiles}, which are similar to those measured by \cite{viezzer17}. We showed that for $\bar{V}=-\bar{u}$ the non-zero neoclassical particle flux, the energy flux, the mean parallel flow, and the poloidal variation exceed the usual neoclassical values in the strong gradient region. The neoclassical energy flux and especially the neoclassical particle flux are significantly smaller in the low flow case, but non-zero. 


\section*{Funding}
This work was supported by the U.S. Department of Energy (F.I.P., contract number DE-AC02-09CH11466) and (P.C., contract number DE-FG02-91ER-54109). The United States Government retains a non-exclusive, paid-up, irrevocable, world-wide license to publish or reproduce the published form of this manuscript, or allow others to do so, for United States Government purposes. S.T. was also supported by the German Academic Scholarship Foundation. 

\section*{Declaration of Interests}
The authors report no conflict of interest.

\section*{Data availability statement}
The code used to generate the figures in this paper is available in the DataSpace of Princeton University at http://arks.princeton.edu/ark:/88435/dsp0137720g96v.

\appendix
\section{Orbits}\label{sec: orbits}
\subsection{Freely passing particles}\label{sec: fixed theta passing}
For freely passing particles, we assume that $v_\parallel-v_{\parallel f}\sim \epsilon v_t$ and $\psi-\psi_f\sim \epsilon \rho_p R B_p$. The calculation that follows will prove these estimates correct. Subtracting the right hand side of \eqref{energy} from the left hand side yields
\begin{equation} \label{energy 2}
    \vp(v_\parallel-\vp)+\mu(B-B_f)+\frac{Ze}{m}\left[(\psi-\psi_f)\pdv{\phi}{\psi}\Bigg \rvert_{\psi_f}+(\phi_\theta-\phi_{\theta f})\right]=0,
\end{equation}
and rearranging \eqref{angular} gives
\begin{equation} \label{angular 2}
    \psi-\psi_f=\frac{I}{\Omega_f}(v_\parallel-\vp)-\frac{I}{\Omega_f}\vp\left(\frac{B}{B_f}-1\right),
\end{equation}
where $I$ is constant in $\psi$ at least to order $\textit{O}(\epsilon^2)$ and hence can be considered a function of $\psi_f$ throughout this work\color{black}. Equations \eqref{energy 2} and \eqref{angular 2} can be combined to solve for $v_\parallel-\vp$ and $\psi-\psi_f$. Using the definition for $u$ in \eqref{def u}, the deviations of parallel velocity and canonical angular momentum within the trajectory of one passing particle are
\begin{equation} \label{vbananapassing}
    v_\parallel-v_{\parallel f}\simeq-\frac{(\mu B_f-v_{\parallel f} u_{f})\left(B/B_f-1\right)+Ze\left(\phi_\theta-\phi_{\theta f}\right)/m}{v_{\parallel f}+u_f}\sim \epsilon v_t
\end{equation}
and
\begin{equation}\label{psibananapassing}
   \psi-\psi_f\simeq-\frac{I}{\Omega_f}\frac{(\mu B_f+v^2_{\parallel f})\left(B/B_f-1\right) +Ze(\phi_\theta-\phi_{\theta f})/m}{v_{\parallel f}+u_f}\sim \epsilon \rho_p R B_p.
\end{equation}
The deviations of parallel velocity and flux function from their values at $\theta_f$ are of \textit{O}$(\epsilon)$ and hence consistent with our initial assumption. We can invert expressions \eqref{vbananapassing} and \eqref{psibananapassing} to obtain $\vp$ and $\psi_f$ from the particle coordinates at any given $\theta$ by interchanging the fixed-$\theta$ and particle variables,
\begin{equation} \label{vbananapassing2}
    \vp-v_{\parallel}\simeq-\frac{(\mu B-v_{\parallel} u)\left(B_f/B-1\right)+Ze\left(\phi_{\theta f}-\phi_\theta\right)/m}{v_{\parallel }+u}\sim \epsilon v_t,
\end{equation}
\begin{equation}\label{psibananapassing2}
   \psi_f-\psi\simeq-\frac{I}{\Omega}\frac{(\mu B+v^2_{\parallel})\left(B_f/B-1\right) +Ze(\phi_{\theta f}-\phi_\theta)/m}{v_{\parallel }+u}\sim \epsilon \rho_p R B_p.
\end{equation}

\subsection{Trapped-barely-passing particles}\label{sec: fixed theta trapped}
The deviations of $v_\parallel$ and $\psi$ from $\vp$ and $\psi_f$ are larger in the trapped-barely-passing region and thus the Taylor expansion of $\phi$ must include the second derivative in order to collect all terms to \textit{O}$(\epsilon)$. We assume that $v_\parallel-\vp\sim \sqrt{\epsilon}v_t$ and $\psi-\psi_f\sim\sqrt{\epsilon}\rho_p R B_p$. Hence, \eqref{energy 2} becomes
\begin{equation}
    \frac{1}{2}(v_\parallel^2-\vp^2)+\mu(B-B_f)+\frac{Ze}{m}\left[(\psi-\psi_f)\pdv{\phi}{\psi}\Bigg \rvert_{\psi_f}\!+\frac{1}{2}(\psi-\psi_f)^2\pdv[2]{\phi}{\psi}\Bigg \rvert_{\psi_f}\!+(\phi_\theta-\phi_{\theta f})\right]\!=\!0
\end{equation}
which, inserting \eqref{angular 2}, reads
\begin{equation}
    \begin{split}
    &\frac{1}{2}(v_\parallel^2-\vp^2)+(\mu B_f-\vp u_f)\left(\frac{B}{B_f}-1\right)+u_f(v_\parallel-\vp)\\
    &+\frac{1}{2}(S_f-1)\left(v_\parallel^2-2v_\parallel\vp+\vp^2\right)+\frac{Ze}{m}(\phi_\theta-\phi_{\theta f})=0,
    \end{split}
\end{equation}
where we have used the squeezing factor $S_f$ as introduced in \eqref{def S}. 
Further simplifications lead to
\begin{equation}
\begin{split}
    &\frac{1}{2}S_f\left[\left(v_\parallel-\vp+\frac{\vp+u_f}{S_f}\right)^2-\left(\vp-\frac{\vp+u_f}{S_f}\right)^2\right]+\left(\mu B_f-\vp u_f\right)\left(\frac{B}{B_f}-1\right)\\
    &=\frac{1}{2}(2-S_f)\vp^2+u_f\vp-\frac{Ze}{m}\left(\phi_\theta-\phi_{\theta f}\right)
    \end{split}
\end{equation}
and finally
\begin{equation}
    v_\parallel-\!\vp\!=\!-\!\frac{\vpu}{S_f}\!\pm\! \sqrt{\frac{1}{S_f}\!\left[\frac{(\vpu)^2}{S_f}\!-\!2(\mu B_f\!-\!\vp u_f)\!\left(\frac{B}{B_f}\!-\!1\right)\!-\!2\frac{Ze}{m}(\phi_\theta\!-\!\phi_{\theta f})\right]}.
\end{equation}
It is useful to calculate $v_\parallel+u$,
\begin{equation}
\begin{split}
    v_\parallel+u&=(v_\parallel-\vp)+(u-u_f)+(\vpu)\\
    &\simeq(v_\parallel-\vp)+(\vpu)+(\psi-\psi_f)\pdv{u}{\psi}\Bigg\rvert_{\psi_f}\\
    &\simeq S_f(v_\parallel-\vp)+(\vpu).
\end{split}
\end{equation}
With this result, we can write
\begin{align}\label{vbananatrapped}
    v_\parallel-v_{\parallel f}=-\frac{v_{\parallel f} + u_f}{S_f}+\frac{v_\parallel+u}{S_f}\sim\sqrt{\epsilon}v_t && \text{and} &&  \psi-\psi_f=\frac{I}{\Omega_f}(v_\parallel-v_{\parallel f})\sim \sqrt{\epsilon} \rho_p R B_p,
\end{align}
where
\begin{equation}\label{vplusu trapped}
    v_\parallel+u=\pm \sqrt{(v_{\parallel f}+u_f)^2-2S_f\left[(\mu B_f-u_fv_{\parallel f })\left(\frac{B}{B_f}-1\right)+\frac{Ze}{m}(\phi_\theta-\phi_{\theta f})\right] }\sim\sqrt{\epsilon}v_t.
\end{equation}
This expression describes the trapped-barely-passing boundary and was first derived in this form by \citet{shaing94resonance}. The deviations of the parallel velocity and radial position are of \textit{O}$(\sqrt{\epsilon})$ and thus bigger than for passing particles, which is consistent with our initial assumption.\par
The solution in the trapped-barely-passing region matches with the solution in the freely passing region in the limit 
\begin{equation}
(\vpu)^2\gg 2S_f\left[(\mu B_f-u_fv_{\parallel f })\left(B/B_f-1\right)+Ze(\phi_\theta-\phi_{\theta f})/m\right],
\end{equation}
since
\begin{equation}
    v_\parallel+u \simeq (\vpu)\Bigg \lbrace 1-\frac{S_f\left[(\mu B_f-u_fv_{\parallel f })\left(B/B_f-1\right)+Ze(\phi_\theta-\phi_{\theta f})/m\right]}{(v_{\parallel f}+u_f)^2}\Bigg \rbrace.
\end{equation}
\par We can invert relations \eqref{vbananatrapped} to obtain
\begin{align}\label{vbananatrapped2}
    \vp-v_{\parallel }=-\frac{v_{\parallel } + u}{S}+\frac{\vp+u_f}{S}\sim\sqrt{\epsilon}v_t && \text{and} &&  \psi_f-\psi=\frac{I}{\Omega}(\vp-v_{\parallel})\sim \sqrt{\epsilon} \rho_p R B_p,
\end{align}
where
\begin{equation}\label{vplusu trapped2}
    \vp+u_f=\pm \sqrt{(v_{\parallel }+u)^2-2S\left[(\mu B-u v_{\parallel })\left(\frac{B_f}{B}-1\right)+\frac{Ze}{m}(\phi_{\theta f}-\phi_\theta)\right] }.
\end{equation}

\section{Matching of $\theta$-dependent parts of $g_0$} \label{matching}
One can use \eqref{div gt banana} to prove that the $\theta$-dependent parts of the distribution functions in the freely passing and trapped-barely-passing regime match. Following \eqref{div gt banana}, the function $g^t$ can be written as
\begin{equation}\label{A1}
\begin{split}
    g^t=&\frac{I}{\Omega S}\left[G(\psi_f,w_f,\mu)-w\right]\mathcal{D} f_{M}(v_\parallel=-u)+\textit{O}(\epsilon f_M),
\end{split}
\end{equation}
where $G-w\sim\sqrt{\epsilon}v_t$. We neglected the distinction between $\psi$ and $\psi_f$ in the Maxwellian and in $\mathcal{D}$ and thus terms of order $\epsilon f_M$ in deriving \eqref{div gt banana}. For barely-passing particles, \eqref{div gt banana} gives $G(\psi_f,w_f,\mu)$ to be 
\begin{equation}
\begin{split}
    G(\psi_f, w_f, \mu)&=G(\psi_f,w_f=0,\mu)+\int_{w_\text{tpb}}^{w_f}\mathrm{d}w'_f\frac{w_f'}{\langle w'\rangle_\psi},
    \end{split}
\end{equation}
where the trapped-barely-passing boundary $w_\text{tpb}\sim\sqrt{\epsilon}v_t$ is defined in \eqref{def wtpb}. For trapped particles $G(\psi_f,w_f,\mu)=G(\psi_f, w_f=0, \mu)$.\par 
We proceed to calculate the $\theta$-dependent piece of $g^t$ when $g^t$ is written as a function of $\psi$ and $w$ instead of $\psi_f$ and $w_f$. We calculate the $\theta$-dependent piece in the overlap region between the trapped-barely-passing and the freely passing regions. We show that $g^t$ is independent of $\theta$ to lowest order in $\epsilon$, and we calculate the next order $\theta$-dependent piece, which is of order $(\epsilon v_t/w) f_M$. Note that we can calculate this small correction despite the fact that we neglect terms small in $\epsilon$ throughout the article because its size is large by a factor of $1/w$ and $\epsilon f_M \ll (\epsilon v_t/w) f_M \ll \sqrt{\epsilon} f_M$ in this region. We start by expanding \eqref{A1} around $\psi$ and $w$,

\begin{equation} \label{A1 expanded}
    g^t\simeq\frac{I}{\Omega S}\left[(\psi_f-\psi)\pdv{G}{\psi}+(w_f-w)\pdv{G}{w}+G(\psi,w,\mu)-w\right]\mathcal{D}f_M(v_\parallel=-u)+\textit{O}(\epsilon f_M).
\end{equation}
Equation \eqref{div gt banana} shows that, for $\abs{w} \rightarrow \infty$, $w\simeq\langle w\rangle_\psi$ and $G(\psi_f, w_f, \mu) - w$ becomes a bounded function of order $\sqrt{\epsilon} v_t$ that only depends on $\psi_f$ and $\mu$. Hence, $G(\psi,w,\mu)\simeq \langle G(\psi,w,\mu) \rangle_\psi$ and the $\theta$-dependent piece in the overlap region between the trapped-barely-passing and the freely passing regions becomes
\begin{equation}\label{AppendixW}
\begin{split}
    g^t-\langle g^t\rangle_\psi\simeq& \frac{I}{\Omega S}\Bigg[\left(\psi_f-\psi-\langle\psi_f-\psi\rangle_\psi\right)\pdv{G}{\psi}+(\vp-v_\parallel-\langle \vp -v_\parallel \rangle_\psi\\
    &+u_f-u-\langle u_f-u\rangle_\psi)\pdv{G}{w}\Bigg] \mathcal{D}f_{M}(v_\parallel=-u)+\textit{O}(\epsilon f_M).
\end{split}
\end{equation}
We have argued above \eqref{AppendixW} that $G=w+\textit{O}(\sqrt{\epsilon} v_t)$ and thus $\partial G/\partial\psi\sim \sqrt{\epsilon}v_t/(RB_p\rho_p)$ and $\partial G/\partial w \simeq1$. We arrive at
\begin{equation}\label{AppendixW2}
    g^t-\langle g^t\rangle_\psi\simeq \frac{I}{\Omega S}\left(\vp-v_\parallel-\langle \vp -v_\parallel \rangle_\psi+u_f-u-\langle u_f-u\rangle_\psi\right)\mathcal{D}f_{M}(v_\parallel=-u)+\textit{O}(\epsilon v_t).
\end{equation}
For $\abs{w_f}\rightarrow \infty$ we can use \eqref{vplusu trapped2} to write, 
\begin{equation}
\begin{split}
    w_f-w&\simeq\pm \sqrt{w^2-2S\left[(\mu B+u^2)\left(\frac{B_f}{B}-1\right)+\frac{Ze}{m}(\phi_{\theta f}-\phi_\theta)\right]}-w\\
    &\simeq -\frac{S\left[(\mu B+u^2)\left(B_f/B-1\right)+Ze(\phi_{\theta f}-\phi_\theta)/m\right]}{w}
    \end{split}
\end{equation}
which simplifies equation \eqref{AppendixW2} to
\begin{equation}
\begin{split}
    g^t-\langle g^t\rangle_\psi \xrightarrow{|w_f|\rightarrow \infty}& -\frac{Ir}{\Omega R}\cos{\theta}\frac{u^2+\mu B-ZeR\phi_c/(mr)}{w}\mathcal{D}f_M(v_\parallel=-u)\sim \epsilon\frac{v_t}{w}f_M.
\end{split}
\end{equation}
Thus, $g^t-\langle g^t \rangle_\psi$ is indeed of order $\epsilon (v_t/w) f_M\gg \epsilon f_M$ and it matches with $f_{M_f}-f_M-\langle f_{M_f}-f_M \rangle_\psi$ in \eqref{theta dependent gp} for small $w$, as desired.\par 
In general, for barely-passing particles one can write 
\begin{equation}\label{poloidal gt bp}
\begin{split}
    g^t_0-\langle g^t_0\rangle_\psi=&\frac{I}{\Omega S}\left(G-\langle G \rangle _\psi\right)\mathcal{D} f_M.
\end{split}
\end{equation}
This function is odd in $w$ since
\begin{equation}
    G \simeq G(\psi,w_f=0,\mu)+\int_{w_\text{tpb}}^{w_f}\mathrm{d}w'_f\frac{w_f'}{\langle w'\rangle_\psi},
\end{equation}
giving
\begin{equation}\label{odd}
G-\langle G \rangle _\psi \simeq \int_{w_\text{tpb}}^{w_f}\mathrm{d}w'_f\frac{w'_f}{\langle w'\rangle_\psi}-\Bigg\langle \int_{w_\text{tpb}}^{w_f}\mathrm{d}w'_f\frac{w_f'}{\langle w'\rangle_\psi}\Bigg \rangle_\psi,
\end{equation}
which is odd in $w_f$ and hence in $w$.

\section{Transit average of the collision operator}\label{sec: transit average C}
The higher order collision operator in fixed-$\theta$ variables is given in \eqref{col transformation}. To calculate the derivative discontinuity condition, one has to solve \eqref{Cg0t} and thus take the transit average of the collision operator. \par 
We proceed to show that the transit average of \eqref{col transformation} leads to equation \eqref{col transformation explicit}. The drift kinetic equation in fixed-$\theta$ variables can be written as
\begin{equation}
    \dot{\theta}\pdv{f}{\theta}+\dot{\psi}_f\pdv{f}{\psi_f}+\dot{v}_{\parallel f}\pdv{f}{\vp}+\dot{\mu}\pdv{f}{\mu}+\dot{\varphi}\pdv{f}{\varphi}=C[f,f],
\end{equation}
where the dotted quantities obey phase space conservation
\begin{equation}\label{phase space conserv}
    \pdv{}{\psi_f} \left(\mathcal{J}\dot{ \psi}_f\right)+\pdv{}{\theta}\left(\mathcal{J}\dot{\theta}\right)+\pdv{}{\vp}\left(\mathcal{J}\dot{v}_{\parallel f}\right)+\pdv{}{\mu}\left(\mathcal{J}\dot{\mu}\right)+\pdv{}{\varphi}\left(\mathcal{J}\dot{\varphi}\right)=0.
\end{equation}
From the definition of $\psi_f$ and $\vp$, it follows that $\dot{\psi}_f=0$ and $\dot{v}_{\parallel f}=0$. Furthermore, conservation of magnetic moment gives $\dot{\mu}=0$. The gyrophase can be defined to higher order such that both $\dot{\varphi}$ and $\mathcal{J}$ are independent of gyrophase to all orders \citep{parra2008}. Hence, \eqref{phase space conserv} reduces to
\begin{equation} \label{J theta}
    \pdv{}{\theta}\left(\mathcal{J}\dot{\theta}\right)=0.
\end{equation}
We find that $\mathcal{J}\dot{\theta}$ is independent of $\theta$. \par 
Equation \eqref{J theta} can be used to take the transit average of \eqref{col transformation}, noting that
\begin{equation} \label{J trans swap}
\begin{split}
    \Bigg\langle\frac{1}{\mathcal{J}}\pdv{}{w_f}\left[\mathcal{J}(...)\right]\Bigg\rangle_\tau&=\frac{1}{\tau}\int\frac{\mathrm{d}\theta}{\dot{\theta}\mathcal{J}}\pdv{}{w_f}\left[\mathcal{J}(...)\right]=\frac{1}{\tau \dot{\theta}\mathcal{J}}\pdv{}{w_f}\left[\dot{\theta}\mathcal{J}\int\frac{\mathrm{d}\theta}{\dot{\theta}}(...)\right]\\
    &=\frac{1}{\tau \dot{\theta}\mathcal{J}}\pdv{}{w_f}\left[\tau \dot{\theta}\mathcal{J}\langle(...)\rangle_\tau\right]\simeq\frac{1}{w_f\tau}\pdv{}{w_f}\left[w_f\tau\langle(...)\rangle_\tau\right],
    \end{split}
\end{equation}
where we used \eqref{Jacobian} as well as $\dot{\theta}\simeq w/qR$ in the last step.\par
Taking the transit average inside the derivatives via \eqref{J trans swap} and using \eqref{gradients} in \eqref{col transformation} yields \eqref{col transformation explicit}.

\section{Integration over the distribution function}\label{sec: FQ}
The integration of the distribution function \eqref{div gt banana} for the jump and derivative discontinuity condition requires the calculation of terms such as
\begin{equation}\label{Georgia}
\begin{split}
 \Bigg\langle\int_{-\infty}^\infty\mathrm{d}w_f \pdv{g^t_0}{w_f}\Bigg\rangle_\psi \propto&\int_{ \text{barely-passing}}\mathrm{d}w_f\left(\frac{w_f}{\langle w\rangle_\psi}-\Big\langle\frac{w_f}{w}\Big\rangle_\psi\right)-\Bigg\langle\int_{ \text{trapped}}\mathrm{d}w_f\frac{w_f}{w}\Bigg\rangle_\psi.
\end{split}
\end{equation}
In \eqref{phi banana final} we show that $\phi_\theta=\phi_c\cos\theta$ in the banana regime, and using \eqref{B} and \eqref{vplusu trapped}, we get
\begin{equation}
    \frac{w}{w_f}\simeq \sqrt{1-2S_f(\cos\theta_f-\cos\theta)\left[(\mu B_f+u_f)^2\frac{r}{R}-\frac{Ze}{m}\phi_c\right]}.
\end{equation}
For $\theta_f=0$, this can be written as
\begin{align}\label{sin^2}
    \frac{w}{w_f}\simeq\sqrt{1-\kappa^2\sin^2\left(\frac{\theta}{2}\right)} 
    \end{align}
 where 
 \begin{equation}\label{kappa positive}
 \kappa^2=\frac{r}{R}\frac{4S_f\left[\mu B_f+u_f^2-ZeR\phi_c/(mr)\right]}{w_f^2}.
 \end{equation}
As a result,
\begin{equation}\label{w/wf}
    \frac{\langle w\rangle_\psi}{w_f}=\frac{2}{\pi}E(\kappa)
\end{equation}
with $E(\kappa)=\int_0^{\pi/2}\mathrm{d}\alpha \sqrt{1-\kappa^2\sin^2\alpha}$ the elliptic integral of the second kind. With these definitions, trapped particles are characterized by $1<\kappa<\infty$, and barely-passing particles are defined by $0<\kappa<1$, which is in agreement with \eqref{trapping condition2}. Thus the integration in \eqref{Georgia} over the barely passing region is from $0$ to $1$ and over the trapped region is from $1$ to $\infty$. However, this calculation only holds for $\kappa^2>0$, which is not always true. In fact, $\kappa^2>0$ does not capture all trapped particles but only particles that are trapped on the low field side for $S_f>0$. If $\phi_c$ is strong enough, it can overcome the centrifugal force and accumulate particles on the inboard side. Particles trapped on the high field side will only exist for $\phi_c>mu_f^2r/(ZeR)$. For $S_f<0$, these particles are captured in our definition for $\kappa^2$ in \eqref{kappa positive}. If we set $\vp$ and $\psi_f$ to be the particle velocity and position at $\theta_f=0$, trapped particles on the high field side that satisfy
\begin{equation}\label{high field mu}
    \mu<\frac{Ze\phi_c R}{m B_f r}-\frac{u_f^2}{B_f}
\end{equation}
for $S_f>0$, as well as trapped particles trapped on the low field side for $S_f<0$ are being missed out. For these particles, $\kappa^2$ in \eqref{kappa positive} would go negative. Thus, one must also consider the choice $\theta_f=\pi$, for which \eqref{sin^2} turns into
\begin{equation}\label{cos^2}
    \frac{w}{w_f}\simeq\sqrt{1-\kappa^2\cos^2\left(\frac{\theta}{2}\right)},
\end{equation}
and $\kappa^2$ is defined as
\begin{equation}\label{kappa negative}
    \kappa^2=\frac{r}{R}\frac{4S_f\left[ZeR\phi_c/(mr)-(\mu B_f+u_f^2)\right]}{w_f^2}.
\end{equation}
Using the substitution $\alpha=\pi/2-\theta/2$ in \eqref{cos^2}, one arrives at the same expression for $\langle w\rangle_\psi/w_f$ as in \eqref{w/wf} but with $\kappa^2$ as defined in \eqref{kappa negative}.

\subsection{Jump Condition}\label{sec: C1}
The integration for the particles that are trapped on the low (high) field side for $S_f>0$ ($S_f<0$) yields
\begin{equation}\label{trapped Delta}
   \Bigg\langle \int_{\text{trapped}}\mathrm{d}w_f\frac{w_f}{w}\Bigg\rangle_\psi=\langle 2| w|\rangle_\psi\Big\rvert_{\kappa=1}=\frac{8}{\pi}\sqrt{S_f\left[(\mu B_f+u^2)\frac{r}{R}-\frac{Ze}{m}\phi_c\right]},
\end{equation}
where the factor of 2 comes from including both possible signs of $w$. For the integration over the barely-passing region, we make a change of variables from $w_f$ to $\kappa$ using that
\begin{equation}
    \dv{\kappa}{w_f}=\pm\frac{\sqrt{4S_f\left[(\mu B_f+u_f^2)r/R-Ze\phi_c/m\right]}}{w_f^2}=\pm \frac{\kappa^2}{\sqrt{4S_f\left[(\mu B_f+u_f^2)r/R-Ze\phi_c/m\right]}}
\end{equation}
so that the integral can be written as
\begin{equation}\label{barely-passing Delta}
\begin{split}
   &  \int_{ \text{barely-passing}}\mathrm{d}w_f\left(\frac{w_f}{\langle w\rangle_\psi}-\Big\langle\frac{w_f}{w}\Big\rangle_\psi\right)\\
    &=4\sqrt{S_f\left[(\mu B_f+u_f^2)\frac{r}{R}-\frac{Ze}{m}\phi_c\right]}\int_0^1\frac{\mathrm{d}\kappa} {\kappa^2}\left(\frac{\pi}{2E(\kappa)}-\frac{2}{\pi}
    K(\kappa)\right)
\end{split}
\end{equation}
with the elliptic function of the first kind.
\begin{equation}
    K(\kappa)=\int_0^{\pi/2}\frac{\mathrm{d}\alpha}{\sqrt{1-\kappa^2\sin^2\alpha}}.
\end{equation}
Again, one factor of two comes from keeping track of both signs of $w$.
For particles obeying relation \eqref{high field mu}, the same calculations can be carried out and combining the two results \eqref{trapped Delta} and \eqref{barely-passing Delta} for particles trapped on either side and $S_f$ of either sign yields
\begin{equation}\label{div gt integral}
\begin{split}
     &\Bigg\langle\int_{-\infty}^\infty\mathrm{d}w_f\pdv{g^t_0}{w_f}\Bigg\rangle_\psi = -2.758\frac{I}{S \Omega}\sqrt{\abs{S\left[(\mu B+u^2)\frac{r}{R}-\frac{Ze}{m}\phi_c\right]}}\mathcal{D} f_M(v_\parallel=-u).
     \end{split}
\end{equation}
We note that the magnetic field $B_f$ is different at $\theta_f=0$ and $\theta_f=\pi$, but the difference is small in $\epsilon$ as shown in section \ref{sec:banana_variables}. At this point, we have dropped the subscript $f$ because the difference is small in epsilon.
\subsection{Derivative discontinuity condition}\label{sec: Delta derivative}
In order to calculate the derivative discontinuity condition, one has to calculate integrals of the form
\begin{equation}
    \int_{-\infty}^{\infty}\mathrm{d}w_f\:w_f\tau\Bigg\langle\frac{w}{w_f}\pdv{g^t_0}{w_f}\Bigg\rangle_\tau
\end{equation}
and 
\begin{equation}
    \int_{-\infty}^{\infty}\mathrm{d}w_f\:w_f\tau\Bigg\langle\left(\frac{w}{w_f}-1\right)\frac{w}{w_f}\pdv{g^t_0}{w_f}\Bigg\rangle_\tau
\end{equation}
For barely-passing particles, \eqref{transit surface} is applicable, so
\begin{equation}\label{Bp0}
    \Bigg\langle\frac{w^2}{w_f^2}\pdv{g^t_0}{w_f}\Bigg\rangle_\tau=\frac{2\pi qR}{\tau w_f}\Bigg\langle \frac{I}{\Omega S}\left(\frac{w}{\langle w\rangle_\psi}-1\right)\mathcal{D}  f_{M}(v_\parallel=-u)\Bigg\rangle_\psi=0,
\end{equation}
and 
\begin{equation}\label{pas dis derivative}
    \int_{\text{barely-passing}}\mathrm{d}w_f\:w_f\tau\Bigg\langle\frac{w}{w_f}\pdv{g^t_0}{w_f}\Bigg\rangle_\tau= \int_{\text{barely-passing}} \mathrm{d}w_f\:2\pi qR\Bigg\langle \pdv{g^t_0}{w_f}\Bigg\rangle_\psi.
\end{equation}
This integral was calculated in \eqref{barely-passing Delta}.
\par For trapped particles,
\begin{equation}\label{P0}
\begin{split}
     \Bigg\langle\frac{w^2}{w_f^2}\pdv{g^t_0}{w_f}\Bigg\rangle_\tau=0
\end{split}
\end{equation}
because $\partial g^t_0/\partial w_f$ is odd in $w$ and it follows from \eqref{transit average trapped} that transit averages over functions that are odd in $w$ are zero for trapped particles. The remaining term is
\begin{equation}\label{trap dis derivative}
    \int_{\text{trapped}}\mathrm{d}w_f\:w_f\tau\Bigg\langle \frac{w}{w_f}\pdv{g^t_0}{w_f}\Bigg\rangle_\tau=\Bigg\langle\int_{\text{trapped}}\mathrm{d}w_f\: 2\pi qR \pdv{g_0^t}{w_f}\Bigg\rangle_\psi.
\end{equation}
This integral was calculated in \eqref{trapped Delta}. Summing the contributions from barely-passing particles \eqref{pas dis derivative} and trapped particles \eqref{trap dis derivative}, we arrive at the expression for the derivative discontinuity condition in \eqref{Delta dgp}.

\section{Poloidal variation of the density}\label{sec: poloidal var}
The poloidal variation of the density follow from the $\theta$-dependent part of $g^p$. In order to find the poloidally varying part of the density in \eqref{poloidal density}, we need to calculate the integral
\begin{equation}
\begin{split}
    & \int \mathrm{d}\mu\left[\text{PV}\int\mathrm{d} v_\parallel \:2\pi B(g^p-\langle g^p\rangle_\psi)\right]\\
    &=- 2\pi B\int \mathrm{d}\mu \Bigg \lbrace\text{PV}\int \mathrm{d} v_\parallel\frac{Ir}{\Omega R}\frac{\left(v_\parallel^2+\mu B\right) \cos{\theta}-ZeR\phi_\theta/(mr)}{v_\parallel +u}\Bigg[\pdv{}{\psi}\ln{p} \\
    &+\frac{m(v_\parallel- V_\parallel)}{T}\left(\pdv{V_\parallel}{\psi}-\frac{\Omega}{I}\right) +\left(\frac{m(v_\parallel-V_\parallel)^2}{2T}+\frac{m\mu B}{T}-\frac{5}{2}\right)\pdv{}{\psi}\ln{T}\Bigg]f_M\\
    &-\frac{r}{R}\cos{\theta}\frac{m}{T}\left[v_\parallel(v_\parallel-V_\parallel)+\mu B \right]f_M\Bigg\rbrace.
\end{split}
\end{equation}
To calculate this integral, we first define
\begin{equation}
    \mathcal{I}=\text{PV}\int\mathrm{d}v_\parallel\frac{f_M}{v_\parallel+u}=\text{PV}\int\mathrm{d}\xi\frac{f_M}{\xi+y},
\end{equation}
where $\xi\equiv v_\parallel-V_\parallel$ and $y\equiv u+V_\parallel$. The first derivative of $\mathcal{I}$ with respect to $y$ is
\begin{equation}
\begin{split}
    \pdv{\mathcal{I}}{y}&=-\text{PV}\int\mathrm{d}\xi \frac{f_M}{(\xi+y)^2}=\text{PV}\int\mathrm{d}\xi\pdv{}{\xi}\left(\frac{1}{\xi+y}\right)f_M\\
    &=\text{PV}\int\mathrm{d}\xi\frac{\xi}{\xi+y}\frac{m}{T}f_M=\frac{m}{T}\int \mathrm{d}\xi\: f_M-\text{PV}\int\mathrm{d}\xi\frac{y}{\xi+y}\frac{m}{T}f_M\\
    &=\frac{n}{2\pi}\left(\frac{m}{T}\right)^2\exp\left(-\frac{m\mu B}{T}\right)-\frac{m}{T}y\mathcal{I},
 \end{split}
\end{equation}
which gives
\begin{equation}
    \pdv{}{y}\left[\mathcal{I}\exp\left(\frac{my^2}{2T}\right)\right]=\frac{n}{2\pi}\left(\frac{m}{T}\right)^2\exp\left(-\frac{m\mu B}{T}\right)\exp\left(\frac{my^2}{2T}\right).
\end{equation}
For $y=0$,
\begin{equation}
    \mathcal{I}(y=0)=\text{PV}\int\mathrm{d}\xi\frac{f_M}{\xi}=0,
\end{equation}
which can be used as a boundary condition. Thus, the solution for $\mathcal{I}$ is
\begin{equation}\label{w0}
\begin{split}
    \mathcal{I}=&\frac{n}{2\pi}\left(\frac{m}{T}\right)^2\exp\left(-\frac{m\mu B}{T}\right)\exp\left(-\frac{my^2}{2T}\right)\int_0^{y}\mathrm{d}t\:\exp\left(\frac{mt^2}{2T}\right)\\
    \equiv&\frac{2n}{\pi}\left(\frac{m}{2T}\right)^{3/2}\exp\left(-\frac{m\mu B}{T}\right)J,
    \end{split}
\end{equation}
where $J$ is given in \eqref{J}.\par
Furthermore, we find that
\begin{equation}\label{w1}
\begin{split}
    \text{PV}\int\mathrm{d}\xi\frac{\xi}{\xi+y}f_M&=\int\mathrm{d}\xi \: f_M -(V_\parallel+u)\mathcal{I}=\frac{n}{2\pi}\frac{m}{T}\exp\left(-\frac{m \mu B}{T}\right)-(V_\parallel+u)\mathcal{I}
    \end{split}
\end{equation} 
\begin{equation}
    \begin{split}
    \text{PV}\int\mathrm{d}\xi\frac{\xi^2}{\xi+y}f_M
    &=-(V_\parallel+u)\frac{n}{2\pi}\frac{m}{T}\exp\left(-\frac{m \mu B}{T}\right)+(V_\parallel+u)^2\mathcal{I}
    \end{split}
\end{equation} 
\begin{equation}
    \begin{split}
     \text{PV}\int\mathrm{d}\xi\frac{\xi^3}{\xi+y}f_M
     &=\frac{n}{2\pi}\exp\left(-\frac{m \mu B}{T}\right)\left[1+(V_\parallel+u)^2\frac{m}{T}\right]-(V_\parallel+u)^3\mathcal{I}
         \end{split}
\end{equation} 
\begin{equation} \label{w4}
\begin{split}
    \text{PV}\int\mathrm{d}\xi\frac{\xi^4}{\xi+y}f_M
    &=-(V_\parallel+u)\frac{n}{2\pi}\exp\left(-\frac{m \mu B}{T}\right)\left[1+(V_\parallel+u)^2\frac{m}{T}\right]+(V_\parallel+u)^4\mathcal{I}.
    \end{split}
\end{equation} 
Expressions \eqref{w0}-\eqref{w4} can be used to calculate $n_\theta$ in \eqref{poloidal density 2}.

\section{Derivation of transport equations}\label{sec: transport equations}
In this section we show the derivation of the moment equations \eqref{Particle conservation}, \eqref{Momentum conservation} and \eqref{Energy conservation} in more detail. A conventional moment approach \citep{parra2008} is not useful when $u\sim v_t$ and $\abs{S-1}\sim 1$, as radial scale lengths must be of order of the poloidal ion gyroradius.
\subsection{Particle Conservation}\label{sec: particle transport}
For particle conservation, one can start by integrating \eqref{eqpassing} over velocity space
\begin{multline} \label{particle eq integral}
    \int\mathrm{d}^3v_f\:\langle C_p^{(l)}[g]\rangle_\tau -\int\mathrm{d}^3v_f \:\lambda \Bigg\langle\bnabla_v\bcdot\left[f_M\int_{V_\text{tbp}}\mathrm{d}^3v'\:f_M'\bnabla_\omega\bnabla_\omega\omega\bcdot\bnabla_{v'}\left(\frac{g_0^{t'}}{f'_M}\right)\right]\Bigg\rangle_\tau\\
    =-\int\mathrm{d}^3v_f\:\langle \Sigma\rangle_\tau,
\end{multline}
where the passing collision operator of \eqref{Cp linear} in the fixed-$\theta$ variables is
\begin{equation}
\begin{split}
   &\langle C_p^{(l)}[g]\rangle_\tau \simeq \frac{1}{w_f\tau}\pdv{}{w_f}\left[f_Mw_f\tau\Bigg\langle \frac{w}{w_f}\boldsymbol{\hat{b}}\bcdot\mathsfbi{M}\bcdot\bnabla_v\left(\frac{g^p}{f_M}\right)\Bigg\rangle_\tau\right]\\
   &+\frac{1}{w_f \tau}\pdv{}{\mu}\left[f_M w_f\tau  \Bigg\langle\frac{\boldsymbol{v}_\perp}{B}\bcdot \mathsfbi{M}\bcdot\bnabla_v\left(\frac{g^p}{f_M}\right)\Bigg\rangle_\tau\right]\\
   &+\frac{1}{w_f\tau}\pdv{}{\psi_f}\left[f_M w_f\tau\frac{I}{\Omega S} \Bigg\langle\left(\frac{w}{w_f}-1\right)\boldsymbol{\hat{b}}\bcdot\mathsfbi{M}\bcdot\bnabla_v\left(\frac{g^p}{f_M}\right)\Bigg\rangle_\tau\right]\\
   &-\lambda \Bigg\lbrace\Bigg\langle \frac{1}{w_f\tau}\pdv{}{w_f}\left[f_Mw_f\tau\frac{w}{w_f}\boldsymbol{\hat{b}}\bcdot\int\mathrm{d}^3v'\:f_M'\bnabla_\omega\bnabla_\omega\omega\bcdot\bnabla_{v'}\left(\frac{g^{p'}}{f_M'}\right)\right]\Bigg\rangle_\tau\\
   &+\Bigg\langle \frac{1}{w_f\tau}\pdv{}{\mu}\left[f_Mw_f\tau\frac{\boldsymbol{v}_\perp}{B}\bcdot\int\mathrm{d}^3v'\:f_M'\bnabla_\omega\bnabla_\omega\omega\bcdot\bnabla_{v'}\left(\frac{g^{p'}}{f_M'}\right)\right]\Bigg\rangle_\tau\\
   &+\Bigg\langle\! \frac{1}{w_f\tau}\pdv{}{\psi_f}\left[f_Mw_f\tau\frac{I}{\Omega S}\!\left(\frac{w}{w_f}\!-\!1\right)\!\boldsymbol{\hat{b}}\!\bcdot\!\int\mathrm{d}^3v'\:f_M'\bnabla_\omega\bnabla_\omega\omega\bcdot\!\bnabla_{v'}\left(\frac{g^{p'}}{f_M'}\right)\right]\Bigg\rangle_\tau\Bigg\rbrace.
   \end{split}
\end{equation}
In the passing region, $(w/w_f-1)$ is small in $\epsilon$ and therefore the terms including the derivatives in $\psi_f$ are negligible. One can change from transit averages to flux surface averages using \eqref{transit surface}. The simplified collision operator becomes
\begin{equation} \label{Collision operator simplified}
\begin{split}
    &\langle C_p^{(l)}[g]\rangle_\tau=\pdv{}{w_f}\left[f_M\Bigg \langle \boldsymbol{\hat{b}}\bcdot\mathsfbi{M}\bcdot\bnabla_v\left(\frac{g^p}{f_M}\right)\Bigg\rangle_\psi\right]+\pdv{}{\mu}\left[f_M \Bigg\langle\frac{\boldsymbol{v}_\perp}{B}\bcdot \mathsfbi{M}\bcdot\bnabla_v\left(\frac{g^p}{f_M}\right)\Bigg\rangle_\psi\right]\\
    &-\lambda\Bigg\lbrace\pdv{}{w_f}\left[f_M\Bigg\langle \boldsymbol{\hat{b}}\bcdot\int\mathrm{d}^3v'\:f_M'\bnabla_\omega\bnabla_\omega\omega\bcdot\bnabla_{v'}\left(\frac{g^{p'}}{f_M'}\right)\Bigg\rangle_\psi\right]\\
    &+\pdv{}{\mu}\left[f_M\Bigg\langle \frac{\boldsymbol{v}_\perp}{B}\bcdot\int\mathrm{d}^3v'\:f_M'\bnabla_\omega\bnabla_\omega\omega\bcdot\bnabla_{v'}\left(\frac{g^{p'}}{f_M'}\right)\Bigg\rangle_\psi\right]\Bigg\rbrace.
    \end{split}
\end{equation}
Integrating \eqref{Collision operator simplified} over velocity space gives the first term in \eqref{particle eq integral}. The integration over $\mu$ cancels the respective derivative terms in \eqref{Collision operator simplified} and the integration in $w_f$ cancels the respective derivative acting on the Maxwellian in the third term in \eqref{Collision operator simplified}. The only term left is
\begin{equation}
    \int \mathrm{d}^3v_f\:\langle C_p^{(l)}[g]\rangle_\tau=\int\mathrm{d}\mu \int\mathrm{d}w_f \:2\pi B \pdv{}{w_f}\left[f_M\Bigg\langle \boldsymbol{\hat{b}}\bcdot\mathsfbi{M}\bcdot\bnabla_v\left(\frac{g^p}{f_M}\right)\Bigg\rangle_\psi\right],
\end{equation}
where we have used that $\mathrm{d}^3v_f\simeq \mathrm{d}\mu\text{ d}w_f\:2\pi B $. The derivative is acting on the passing particle distribution function, which has a discontinuity at $w_f=0$. 
 We arrive at
\begin{equation}\label{Particle Delta}
    \int \mathrm{d}^3v_f\:\langle C_p^{(l)}[g]\rangle_\tau=-\int \mathrm{d}\mu \: 2\pi B \Delta\left[f_M\Bigg \langle \boldsymbol{\hat{b}}\bcdot \mathsfbi{M}\bcdot\bnabla_v\left(\frac{g^p}{f_M}\right)\Bigg \rangle_\psi\right],
\end{equation}
where the integrand on the right hand side is given by equation \eqref{Delta dgp}. For the second term in \eqref{particle eq integral}, one can follow the same steps and write the velocity divergence in terms of the fixed-$\theta$ variables. As the derivatives are not acting on the trapped distribution function but on the Maxwellian, there is no discontinuity and the integration cancels all terms in it.\par
Next, the derivative discontinuity condition in \eqref{Delta dgp} is substituted into \eqref{Particle Delta}. The integration cancels the derivative in $\mu$ and we find
\begin{equation}
\begin{split}
\int\mathrm{d}\mu\:2\pi B \pdv{}{\psi_f} \Bigg[\frac{I}{\Omega S} M_{\parallel }\Delta g^p\Bigg]=\int\mathrm{d}^3v_f\:\langle \Sigma\rangle_\tau.
  \end{split}
\end{equation}
for the particle equation. With the definition of the parallel friction force in \eqref{F parallel definition} we arrive at \eqref{Particle conservation}.

\subsection{Parallel momentum conservation}\label{sec: momentum transport}
One can follow the same procedure for the derivation of the parallel momentum and energy equations.
For parallel momentum conservation, we multiply \eqref{eqpassing} by $m v_{\parallel f}$ and integrate over velocity space
\begin{multline} \label{momentum eq integral}
        \int\mathrm{d}^3v_f\: m\vp\langle C_p^{(l)}[g]\rangle_\tau\\
        -\int\mathrm{d}^3v_f \:m\vp \lambda\Bigg\langle\bnabla_v\bcdot\left[f_M\int_{V_\text{tbp}}\mathrm{d}^3v'f_M'\bnabla_\omega\bnabla_\omega\omega\bcdot\bnabla_{v'}\left(\frac{g_0^{t'}}{f'_M}\right)\right]\Bigg\rangle_\tau
        =-\int\mathrm{d}^3v_f\: m\vp\langle \Sigma\rangle_\tau.
\end{multline}    
For the first term,  one can use the expression in \eqref{Collision operator simplified}. Again, the integrals over $\mu$ cancel the derivatives in $\mu$, and the only remaining terms are

\begin{equation}
\begin{split}
     &\int\mathrm{d}^3v_f\: m\vp\langle C_p^{(l)}[g]\rangle_\tau=\int\mathrm{d}\mu\int\mathrm{d}w_f\:2\pi B m \vp \pdv{}{w_f}\left[f_M\Bigg \langle \boldsymbol{\hat{b}}\bcdot\mathsfbi{M}\bcdot\bnabla_v\left(\frac{g^p}{f_M}\right)\Bigg\rangle_\psi\right]\\
    &-\lambda  \int\mathrm{d}\mu\int\mathrm{d}w_f\:2\pi B m \vp \pdv{}{w_f}\left[f_M\Bigg\langle \boldsymbol{\hat{b}}\bcdot\int\mathrm{d}^3v'f_M'\bnabla_\omega\bnabla_\omega\omega\bcdot\bnabla_{v'}\left(\frac{g^{p'}}{f_M'}\right)\Bigg\rangle_\psi\right].
    \end{split}
\end{equation}
Integrating by parts leaves us with
\begin{equation}\label{nocheins}
\begin{split}
    &\int\mathrm{d}^3v_f\:m\vp\langle C_p^{(l)}[g]\rangle_\tau=\int\mathrm{d}\mu\:2\pi Bmu\Delta \left[f_M\Bigg \langle \boldsymbol{\hat{b}}\bcdot \mathsfbi{M}\bcdot\bnabla_v\left(\frac{g^p}{f_M}\right)\Bigg \rangle_\psi\right]\\
    &-\int\mathrm{d}\mu\int\mathrm{d}w_f\:2\pi Bm  \left[f_M\Bigg \langle \boldsymbol{\hat{b}}\bcdot \mathsfbi{M}\bcdot\bnabla_v\left(\frac{g^p}{f_M}\right)\Bigg \rangle_\psi\right]\\
   & -\lambda \int\mathrm{d}\mu \int\mathrm{d}w_f\:2\pi B m \left[f_M\Bigg\langle \boldsymbol{\hat{b}}\bcdot\int\mathrm{d}^3v'\:f_M'\bnabla_\omega\bnabla_\omega\omega\bcdot\bnabla_{v'}\left(\frac{g^{p'}}{f_M'}\right)\Bigg\rangle_\psi\right],
    \end{split}
\end{equation}
where we have used that $\vp\simeq -u_f\simeq -u$ in the trapped-barely-passing region. The integrand of the first integral in \eqref{nocheins} is given by equation \eqref{Delta dgp}.
The last two terms in \eqref{nocheins} can be seen to cancel by recalling the definition of $\mathsfbi{M}$ in \eqref{collision operator}. The only term that we are left with is
\begin{equation}
    \int\mathrm{d}^3v_f\:m\vp\langle C_p^{(l)}[g]\rangle_\tau=\int\mathrm{d}\mu\:2\pi B mu\Delta \left[f_M\Bigg \langle \boldsymbol{\hat{b}}\bcdot \mathsfbi{M}\bcdot\bnabla_v\left(\frac{g^p}{f_M}\right)\Bigg \rangle_\psi\right].
\end{equation}
Substituting the derivative discontinuity condition \eqref{Delta dgp}, we find
\begin{equation}
\begin{split}
  \int\mathrm{d}^3v_f\:m\vp\langle C_p^{(l)}[g]\rangle_\tau=  \int\mathrm{d}\mu\:2\pi B m u\pdv{}{\psi_f} \Bigg[\frac{I}{\Omega S} M_{\parallel } \Delta g^p \Bigg].
    \end{split}
\end{equation}
Taking the derivative with respect to $\psi_f$ outside of the integral and using \eqref{F parallel definition}, one arrives at
\begin{equation}\label{momentum term 1}
    -\int\mathrm{d}^3v_f \:m\vp\langle \Sigma\rangle_\tau=-\pdv{}{\psi_f}\left(\frac{I}{\Omega}uF_{\parallel }\right)+(S-1)F_{\parallel }.
\end{equation}
The second term in \eqref{momentum eq integral} can be integrated by parts to find, upon using \eqref{wf derivative} with $w\simeq w_f$,
\begin{equation}\label{momentum term 2}
\begin{split}
    &-\int\mathrm{d}^3v_f \:m\vp \lambda\Bigg\langle\bnabla_v\bcdot\left[f_M\int\mathrm{d}^3v'\:f_M'\bnabla_\omega\bnabla_\omega\omega\bcdot\bnabla_{v'}\left(\frac{g_0^{t'}}{f'_M}\right)\right]\Bigg\rangle_\tau\\
    &=\int \mathrm{d}\mu\int \mathrm{d}w_f\:2\pi B m \lambda \boldsymbol{\hat{b}}\bcdot \Bigg \langle \left[f_M\int\mathrm{d}^3v'\:f_M'\bnabla_\omega\bnabla_\omega\omega\bcdot\bnabla_{v'}\left(\frac{g_0^{t'}}{f'_M}\right)\right]\Bigg \rangle_\psi\\
    &\simeq 2\pi B m\int\mathrm{d}\mu\int\mathrm{d}w_f\:  f_M M_{\parallel }\Bigg\langle\pdv{(g^t_0/f_M)}{w_f}\Bigg\rangle_\psi=-S F_{\parallel }.
    \end{split}
\end{equation}
Combining \eqref{momentum term 1} and \eqref{momentum term 2} gives the parallel momentum equation in the form of \eqref{Momentum conservation}.

\subsection{Energy conservation}\label{sec: energy transport}
The energy equation requires a multiplication of \eqref{eqpassing} by $mv_f^2/2$ and integration over velocity space
\begin{multline} \label{energy eq integral}
        \int\mathrm{d}^3v_f\: \frac{mv_f^2}{2}\langle C_p^{(l)}[g]\rangle_\tau \\
        -\int\mathrm{d}^3v_f \:\frac{mv_f^2}{2} \lambda\Bigg\langle\bnabla_v\bcdot\left[f_M\int_{V_\text{tbp}}\mathrm{d}^3v'\:f_M'\bnabla_\omega\bnabla_\omega\omega\bcdot\bnabla_{v'}\left(\frac{g_0^{t'}}{f'_M}\right)\right]\Bigg\rangle_\tau =-\int\mathrm{d}^3v_f \frac{mv_f^2}{2}\langle \Sigma\rangle_\tau.
\end{multline}    

Once again we can use \eqref{Collision operator simplified} for the first term in \eqref{energy eq integral} and integrate by parts to arrive at
\begin{equation}\label{gyro C}
    \begin{split}
         &\int\mathrm{d}^3v_f\: \frac{mv_f^2}{2}\langle C_p^{(l)}[g]\rangle_\tau=-\int\mathrm{d}\mu \:\left(\frac{mu^2}{2}+m\mu B\right)2\pi B \Delta\left[f_M\Bigg \langle \boldsymbol{\hat{b}}\bcdot\mathsfbi{M}\bcdot\bnabla_v\left(\frac{g^p}{f_M}\right)\Bigg\rangle_\psi\right]\\
         &-\int\mathrm{d}\mu\int\mathrm{d}w_f\:2\pi m B \vp f_M\Bigg\langle\boldsymbol{\hat{b}}\bcdot\mathsfbi{M}\bcdot\bnabla_v\left(\frac{g^p}{f_M}\right)\Bigg\rangle_\psi\\
         &-\int\mathrm{d}\varphi\int\mathrm{d}\mu\int\mathrm{d}w_f\: m B^2f_M \Bigg\langle\frac{\boldsymbol{v}_\perp}{B}\bcdot \mathsfbi{M}\bcdot\bnabla_v\left(\frac{g^p}{f_M}\right)\Bigg\rangle_\psi\\
         &+\lambda\int\mathrm{d}\varphi\int\mathrm{d}\mu\int\mathrm{d}w_f \: m B (\vp \boldsymbol{\hat{b}}+\boldsymbol{v}_\perp)\bcdot \left[f_M\Bigg\langle \int\mathrm{d}^3v'\:f'_M\bnabla_\omega\bnabla_\omega\omega\bcdot\bnabla_{v'}\left(\frac{g^{p'}}{f'_M}\right)\Bigg\rangle_\psi\right].
    \end{split}
\end{equation}
We kept the integration over gyrophase in the last two terms of \eqref{gyro C} because they seemingly depend on gyrophase via $\boldsymbol{v}_\perp$. However, this dependence cancels, because we can use that
\begin{equation} \label{omega trick}
    (\vp \boldsymbol{\hat{b}}+\boldsymbol{v}_\perp)\bcdot\bnabla_\omega\bnabla_\omega\omega=\boldsymbol{v}\bcdot\bnabla_\omega\bnabla_\omega \omega=\boldsymbol{v}'\bcdot\bnabla_\omega\bnabla_\omega\omega
\end{equation}
and integrate over $\mathrm{d}^3v_f$ to get $\mathsfbi{M}$. We are left with
\begin{equation}
    \int\mathrm{d}^3v_f\: \frac{mv_f^2}{2}\langle C_p^{(l)}[g]\rangle_\tau=-\int\mathrm{d}\mu \:\left(\frac{mu^2}{2}+m\mu B\right)2\pi B \Delta\left[f_M\Bigg \langle \boldsymbol{\hat{b}}\bcdot\mathsfbi{M}\bcdot\bnabla_v\left(\frac{g^p}{f_M}\right)\Bigg\rangle_\psi\right].
\end{equation}
The derivative discontinuity condition \eqref{Delta dgp} can be used to yield
\begin{equation}
\begin{split}
    &\int\mathrm{d}^3v_f\: \frac{mv_f^2}{2}\langle C_p^{(l)}[g]\rangle_\tau=\int\!\mathrm{d}\mu\int\!\mathrm{d}w_f\:2\pi m B^2\mu \pdv{}{\mu}\left(2\mu M_{\perp }\Delta g^p\right)\\
    &-\int\!\mathrm{d}\mu\: 2\pi B m \frac{u^2+2\mu B}{2}\pdv{}{\psi_f}\left(\frac{I}{\Omega S}M_{\parallel }\Delta g^p\right).
    \end{split}
\end{equation}
We can integrate by parts in the first term and take the derivative with respect to $\psi_f$ out of the integral and find
\begin{equation}\label{energy term 1}
    \int\mathrm{d}^3v_f\: \frac{mv_f^2}{2}\langle C_p^{(l)}[g]\rangle_\tau=-\int\mathrm{d}\mu\:4\pi mB^2 \mu M_{\perp }\Delta g^p-(S-1)u F_{\parallel }-\pdv{}{\psi_f}\left(\frac{I}{\Omega}\Theta\right),
\end{equation}
where we introduced the heat viscous force $\Theta$ defined in \eqref{Kf definition}.

The second term in \eqref{energy eq integral} can be integrated by parts to give
\begin{equation}
\begin{split}
 &-\int\mathrm{d}^3v_f \:\frac{mv_f^2}{2} \lambda\Bigg\langle\bnabla_v\bcdot\left[f_M\int\mathrm{d}^3v'\:f_M'\bnabla_\omega\bnabla_\omega\omega\bcdot\bnabla_{v'}\left(\frac{g_0^{t'}}{f'_M}\right)\right]\Bigg\rangle_\tau\\
 &=\int\mathrm{d}\varphi\int\mathrm{d}\mu\int\mathrm{d}w_f\: B m \lambda\left(\vp\boldsymbol{\hat{b}}+\boldsymbol{v}_\perp\right)\bcdot\left[f_M\Bigg \langle \int\mathrm{d}^3v'\:f_M'\bnabla_\omega\bnabla_\omega\omega\bcdot\bnabla_{v'}\left(\frac{g_0^{t'}}{f'_M}\right)\Bigg\rangle_\psi\right].
 \end{split}
\end{equation}
The integrations over $\boldsymbol{v}$ and $\boldsymbol{v}'$ can be swapped using relation \eqref{omega trick}, which then gives $\mathsfbi{M}$. As a result
\begin{equation}
\begin{split} \label{energy term 2}
     &-\int\mathrm{d}^3v_f \:\frac{mv_f^2}{2} \lambda\Bigg\langle\bnabla_v\bcdot\left[f_M\int\mathrm{d}^3v'\:f_M'\bnabla_\omega\bnabla_\omega\omega\bcdot\bnabla_{v'}\left(\frac{g_0^{t'}}{f'_M}\right)\right]\Bigg\rangle_\tau\\
     &=\int\mathrm{d}\mu\int\mathrm{d}w_f\:2\pi B m f_M\Bigg\langle \boldsymbol{v}_f \bcdot \mathsfbi{M}\bcdot\bnabla_{v}\left(\frac{g^t_0}{f_M}\right)\Bigg\rangle_\psi\\
     &=uS F_{\parallel }+\int\mathrm{d}\mu\:4\pi mB^2\mu M_{\perp }\Delta g^p.
\end{split}
\end{equation}

The two terms containing $M_{\perp }$ cancel when substituting \eqref{energy term 1} and \eqref{energy term 2} into \eqref{energy eq integral}, which leaves us with energy equation \eqref{Energy conservation}.

\section{Comparison to previous work} 
\subsection{Small temperature gradient limit}\label{sec: small temp}
Equations \eqref{Fparallel} and \eqref{Q} can reproduce the results for the ion energy flux in the banana limit derived by \cite{catto2013} when taking the limit of small temperature gradient, small particle flux, small mean velocity and small mean velocity gradient. To lowest order, \eqref{Fparallel} gives
\begin{equation}
    \pdv{}{\psi}\ln n+\frac{m\Omega u} {IT}=0.
\end{equation}
To next order,
\begin{equation}\label{GammaCatto1}
    \Gamma=-1.102\sqrt{\frac{R}{r}}\frac{\nu I q p}{\abs{S}^{3/2}m\Omega^2}\left[\left(\pdv{}{\psi}\ln T-\frac{mu}{T}\pdv{V_\parallel}{\psi}\right)G_1(\bar{y},\bar{z})-1.17\frac{1}{T}\pdv{T}{\psi}G_2(\bar{y},\bar{z})\right].
\end{equation}
Similarly, the energy flux reduces to
\begin{multline}\label{QCatto1}
    Q=\frac{mu^2}{2}\Gamma-1.463\sqrt{\frac{R}{r}}\frac{qI\nu p}{\abs{S}^{3/2}\Omega^2}\frac{T}{m}\Bigg[\left(\pdv{}{\psi}\ln T-\frac{mu}{T}\pdv{V_{\parallel }}{\psi}\right)H_1(\bar{y},\bar{z})\\
    -0.25\frac{1}{T}\pdv{T}{\psi} H_2(\bar{y},\bar{z})\Bigg],
\end{multline}
    
We can solve \eqref{GammaCatto1} for $T^{-1}(\partial T/\partial \psi)-(mu/T)(\partial V_\parallel/\partial \psi)$ and substitute this into \eqref{QCatto1}.
\begin{equation}\label{Catto full q}
    Q=\left(\frac{mu^2}{2}+1.33T\frac{H_1}{G_1}\right)\Gamma-1.71\sqrt{\frac{R}{r}}\frac{qI\nu p }{\abs{S}^{3/2}\Omega^2m}\pdv{T}{\psi}\Delta \bar{Q},
\end{equation}
where $\Delta \bar{Q}$ was defined in \eqref{Delta Q}.
Furthermore, \cite{catto2013} assumed $\Gamma=0$ and no poloidally varying potential. Neglecting the poloidal potential variation is consistent with our model as it follows from \eqref{phi banana final} that for small temperature gradient, $V_\parallel \ll v_t$ and $\Gamma=0$, the electric potential $\phi_c=0$ and hence $\bar{z}=2\bar{u} ^2 $, where $\bar{u}=u/v_t$. We impose these restriction on \eqref{Catto full q} in order to get an energy flux consistent with the energy flux of Catto, and find
\begin{equation}\label{qMine}
    Q=-1.71\sqrt{\frac{R}{r}}\frac{qI\nu p}{\abs{S}^{3/2}\Omega^2 m}\Delta\bar{Q}\pdv{T}{\psi}.
\end{equation}
\par The energy flux in \cite{catto2013} is
\begin{equation}
    Q=-1.35\sqrt{\frac{R}{r}}\frac{qI\nu p }{\abs{S}^{1/2}\Omega^2m}L(\bar{u}^2)\pdv{T}{\psi}
\end{equation}
where
\begin{multline}
    L=1.53e^{-\bar{u}^2}\int_0^\infty\mathrm{d}\bar{\mu}\:(\bar{\mu}+2\bar{u}^2)^{3/2}(\bar{\mu}+2\bar{u}^2-\sigma)(\bar{\mu}+\bar{u}^2)^{-3/2}\lbrace \bar{\mu}\left[\Xi(x)-\Psi(x)\right]\\
    +2\bar{u}^2\Psi(x)\rbrace \exp(-\bar{\mu}),
\end{multline}
\begin{equation}
    \sigma=\frac{\int_0^\infty\mathrm{d}\bar{\mu} \:e^{-\bar{\mu}}(\bar{\mu}+2\bar{u}^2)^{3/2}\left[\bar{\mu}\nu_\perp(x)+2\bar{u}^2\nu_\parallel (x)\right]}{\int_0^\infty\mathrm{d}\bar{\mu}\:e^{-\bar{\mu}}(\bar{\mu}+2\bar{u}^2)^{1/2}\left[\bar{\mu}\nu_\perp(x)+2\bar{u}^2\nu_\parallel(x)\right]},
\end{equation}
and $\bar{\mu}=m\mu/(BT)\simeq x^2-\bar{u}^2$. One can write
\begin{equation}
    \sigma=\frac{\int_{\abs{\bar{u}}}^\infty\mathrm{d}x\:(x^2+\bar{u}^2)k(x,\bar{u})}{\int^\infty_{\abs{\bar{u}}}\mathrm{d}x \:k(x,\bar{u})}
\end{equation}
and
\begin{equation}
    L(\bar{u}^2)=6.12\int_{\abs{\bar{u}}}^\infty\mathrm{d}x \left(x^4+2x^2\bar{u}^2+\bar{u}^4-\sigma (x^2+\bar{u}^2)\right)k(x,\bar{u})=1.27\Delta\bar{Q}.
\end{equation}
Finally, the energy flux of \cite{catto2013} is
\begin{equation}\label{qPaper}
    Q=-1.71\sqrt{\frac{R}{r}}\frac{qI\nu p}{\abs{S}^{1/2}\Omega^2m}\Delta\bar{Q}\pdv{T}{\psi}.
\end{equation}
The energy fluxes in \eqref{qMine} and \eqref{qPaper} differ by a factor of $1/S$. However, when the energy flux was calculated in equation (38) in \cite{catto2013} and previously in \cite{catto10}, this factor had been missed as already pointed out by \citet{shaing12}. The energy flux can be obtained from the lowest order moment $I v_f^2 w_f /B$ of the drift kinetic equation \eqref{kinetic equation simplified}, 
\begin{equation}
\Bigg\langle \int d^3v_f \frac{w_f I}{B} v_f^2 \frac{w}{qR}\pdv{f}{\theta}\Bigg\rvert_{\vp,\psi_f}\Bigg\rangle_\psi=\Bigg\langle   \int d^3v_f\frac{w_fI}{B}v_f^2 C[f,f]\Bigg\rangle_\psi.
\end{equation}
One can integrate the left hand side by parts in $\theta$ to find
\begin{equation}
    \Bigg\langle \int d^3v_f \frac{w_f I}{B} v_f^2 \frac{w}{qR}\pdv{f}{\theta}\Bigg\rvert_{\vp,\psi_f}\Bigg\rangle_\psi=-\Bigg\langle \int d^3v_f \frac{w_f I}{B} \frac{v_f^2}{qR}\pdv{w}{\theta}\Bigg\rvert_{\vp,\psi_f}f\Bigg\rangle_\psi.
\end{equation}
Using \eqref{vplusu trapped} for
\begin{equation}
    \pdv{w}{\theta}\Bigg\rvert_{\vp,\psi_f}=-\frac{S_f}{w}\left[\left(\mu B_f+u_f^2\right)\frac{r}{R}\sin\theta+\frac{Ze}{m}\pdv{\phi_\theta}{\theta}\right] \simeq\frac{S}{w}\frac{\Omega }{I}qR\boldsymbol{v}_d\bcdot\bnabla\psi,
\end{equation}
we find the squeezing factor that was lost in \cite{catto2013}. The collision operator conserves energy, so $w_f$ on the right hand side can be reduced to $\vp$ and we arrive at
\begin{equation} \label{almost there now}
   - \Bigg\langle \frac{ZeS}{mc}\int d^3v_f \frac{w_f}{w} v_f^2 f \boldsymbol{v}_d\bcdot\bnabla\psi\Bigg\rangle_\psi=\Bigg \langle \int d^3v_f\frac{\vp I}{B}v_f^2 C[f,f]\Bigg\rangle_\psi.
\end{equation}
The energy flux in \cite{catto2013} is defined as
\begin{equation} \label{heat flux def}
   Q=\frac{IqR}{r} \Bigg\langle \int d^3v\:\frac{mv^2}{2}f\boldsymbol{v}_d\bcdot\bnabla\psi\Bigg\rangle_\psi.
\end{equation}
We can combine \eqref{almost there now} and \eqref{heat flux def}, and use that in the trapped region $d^3v\simeq d^3v_f\: w_f/w$ and that the collision operator conserves momentum to arrive at
\begin{equation} \label{better q}
   Q=-\frac{1}{S}\frac{mcI^2qRT}{Zer}\Bigg\langle \int \frac{d^3v}{B}\left(\frac{mv^2}{2T}-\frac{5}{2}\right)v_{\parallel } C[f,f]\Bigg\rangle_\psi,
\end{equation}
where we changed back to particle variables again and dropped the subscript $f$.
With \eqref{better q} instead of equation (48) in \cite{catto2013}, the additional squeezing factor that we get is retrieved and the result of \eqref{qPaper} is corrected to agree with \eqref{qMine} .\par

\subsection{Small mean parallel velocity gradient} \label{sec: ShaingLimit}
We take the limit of small mean parallel velocity gradient and vanishing particle flux $\Gamma=0$. In this limit we can compare our equations for particle flux \eqref{Fparallel} and energy flux \eqref{Q} with those presented in \cite{shaing12}. \par
We start by noting that the poloidal variation of the potential was neglected in \cite{shaing12}. However, taking the limit of small mean parallel velocity gradient in \eqref{phi banana final} does not give $\phi_c=0$ so the contribution from $\phi_\theta$ should have been kept. \par 
The first necessary step is to relate the functions $G_1$, $G_2$, $H_1$, $H_2$ with the functions $\mu_{1i}$, $\mu_{2i}$ and $\mu_{3i}$ used in \cite{shaing12}. Restricting our results to the case where $S>0$, we find that 
\begin{align} \label{Gtomu1}
        G_1=0.90\sqrt{\frac{R}{r}}\frac{S ^{3/2}}{\nu }\mu_{1i}, && H_1=0.68\sqrt{\frac{R}{r}}\frac{S ^{3/2}}{\nu } \left[\mu_{2i}-\left(\bar{y} ^2-\frac{5}{2}\right)\mu_{1i}\right], 
        \end{align}\begin{align}\label{Gtomu2}
         G_2=-0.77\sqrt{\frac{R}{r}}\frac{S ^{3/2}}{\nu }\mu_{2i} ,&& H_2=-2.74\sqrt{\frac{R}{r}}\frac{S ^{3/2}}{\nu }\left[\mu_{3i}-\left(\bar{y} ^2-\frac{5}{2}\right)\mu_{2i}\right],
\end{align}
if we make the replacement
\begin{equation} \label{submu}
    x\left(1-3\frac{\bar{y} ^2}{x^2}\right)^2\left(1+\frac{\bar{y}^2 }{x^2}\right)^{-3/2}\longrightarrow \sqrt{\abs{x^2+\bar{z} -\bar{y} ^2}}
\end{equation}
in the definition of $\mu_{ji}$ for $j=1,2,3$ in equation (52) in  \cite{shaing12}. Note, that we use $x$ and $\bar{y} $ as defined in our calculation in section \ref{sec: Moment} and not as in \cite{shaing12}. The discrepancy is caused by a combination of two effects. The poloidal variation of the electric field has been neglected reducing $\bar{z} $ to $\bar{z} =m\bar{u} ^2/T $. Second, the trapped particle distribution function in \eqref{div gt banana2} is different from the one in \cite{shaing12}. Our expressions \eqref{div gt banana} and \eqref{div gt banana2} almost match with the result in equation (40) in \citep{shaing94resonance}, which is 
\begin{equation}\label{Shaing g}
    \pdv{g^t_0}{w_f}=-\frac{I}{\Omega S}\frac{v^2-3(u+V_{\parallel })^2}{v^2+(u+V_{\parallel })^2}\left(\frac{w_f}{w}-H\frac{w_f}{\langle w\rangle_\psi}\right)\mathcal{D}f_{M}(v_\parallel=-u),
\end{equation}
where $H=0$ for trapped particles and $H=1$ for barely-passing particles. Equations \eqref{div gt banana} and \eqref{div gt banana2} differ from \eqref{Shaing g} by a factor of $(v^2-3(u+V_{\parallel })^2)/(v^2+(u+V_{\parallel })^2)$. This discrepancy was already pointed out in the Appendix of \cite{catto2013}. This discrepancy can be traced back to the moment approach used in \cite{ shaing12} for which one assumes that
\begin{equation}\label{Shaing Bs}
    v_\parallel\boldsymbol{\hat{b}}\bcdot\bnabla(v_\parallel B)-v_\parallel B^2\boldsymbol{\hat{b}}\bcdot\bnabla\left(\frac{v_\parallel}{B}\right)=2v_\parallel^2\boldsymbol{\hat{b}}\bcdot\bnabla B
\end{equation}
is small. However, this assumption only holds for $v_\parallel\sim\sqrt{\epsilon} v_t$, which is true only for weak radial electric fields in conventional neoclassical theory. In the case where the potential gradient is large such that $u\sim v_\parallel\sim v_t$, the trapped-barely-passing region is shifted to $w\sim\sqrt{\epsilon}v_t$ but $v_\parallel \sim v_t$ and \eqref{Shaing Bs} cannot be neglected.

\par
Once we correct for this discrepancy and make the substitution \eqref{submu}, we can compare terms in the parallel viscous force $\langle \boldsymbol{B}\bcdot\bnabla\bcdot\boldsymbol{\pi}^{\text{SH}}\rangle$ and heat viscous force $\langle \boldsymbol{B}\bcdot\bnabla\bcdot\boldsymbol{\Theta}^{\text{SH}}\rangle$ in equation (45) and (46) of \cite{shaing12} with our forces $F_\parallel$ and $\Theta$. We find
\begin{equation} \label{parallelviscousF}
    \langle \boldsymbol{B}\bcdot\bnabla\bcdot\boldsymbol{\pi}^{\text{SH}}\rangle=B F_{\parallel }
\end{equation}
and
\begin{equation} \label{viscous heat}
    \langle \boldsymbol{B}\bcdot\bnabla\bcdot\boldsymbol{\Theta}^{\text{SH}}\rangle=-B\Theta+B\left(\frac{mV_\parallel(2u+V_\parallel)    }{2T }-\frac{5}{2}\right)F_{\parallel }.
\end{equation}
\citet{shaing12} state that $F_\parallel=0$, so that these expressions reduce to
\begin{equation} \label{GammaShaing}
    0=\langle \boldsymbol{B}\bcdot\bnabla\bcdot\boldsymbol{\pi}^{\text{SH}}\rangle
\end{equation}
and the heat viscous force is related to the energy flux by
\begin{equation}
    Q=-\frac{q TR }{ m \Omega  Br }  \langle \boldsymbol{B}\bcdot\bnabla\bcdot\boldsymbol{\Theta}^{\text{SH}}\rangle.
\end{equation}
Note that setting $F_\parallel=0$ is necessary to match the energy flux $Q$.\par 
Explicitly, equation \eqref{Gammac} for $\Gamma=0$ and $\Omega/(I v_t)(\partial V/\partial \psi) \ll1$ reduces to
\begin{equation}\label{ShaingLimitParticle}
    \left[\pdv{}{\psi}\ln p +\frac{m\Omega (u+V_\parallel)}{IT}\right]G_1= 1.17 \pdv{T }{\psi }G_2,
\end{equation}
which can be rewritten using \eqref{Gtomu1} and \eqref{Gtomu2},
\begin{equation}
    \frac{V_{\parallel }+u }{B }=-\frac{IcT }{Z^2eB ^2}\left(\pdv{}{\psi }\ln p +\frac{\mu_{2i}}{\mu_{1i}}\pdv{}{\psi }\ln T \right),
\end{equation}
and is the same as equation (65) in \cite{shaing12}.
Similarly, the energy flux \eqref{Kf definition} simplifies to
\begin{equation}\label{ShaingLimitEnergy}
    Q=-1.71\sqrt{\frac{R}{r}}\frac{p  q \nu I }{m \Omega^2 S ^{3/2}}\Delta \bar{Q}\pdv{T}{\psi },
\end{equation}
where we have substituted \eqref{ShaingLimitParticle} into \eqref{Kf definition}. We can now compare these expressions with corresponding equations (65) and (67) in \cite{shaing12}. The energy flux \eqref{ShaingLimitEnergy} can be written as
\begin{equation}
    Q=-pmq\frac{c^2 IR}{Z^2e^2B^2 r}\mu_{3i}\left(1-\frac{\mu_{2i}^2}{\mu_{1i}\mu_{3i}}\right)\pdv{T}{\psi },
\end{equation}
which is the same as equation (67) in \cite{shaing12}. Hence, the particle flux equation and energy flux equation give the same result as the one in \cite{shaing12} in the limit $\Omega/(I v_t)(\partial V/\partial \psi) \ll1$ and $\Gamma=0$ if the factors in $\mu_{ji}$ are corrected as indicated in \eqref{submu}. \par

\section{Pedestal profiles}\label{sec: pedestal profiles}
The realistic pedestal profiles of density, ion and electron temperature that we use to calculate example fluxes in section \ref{sec: Profiles} are shown in figure \ref{fig: Profiles}. The profiles are based on those measured by \cite{viezzer17}. The functions we use are
\begin{equation}\label{density in}
    \bar{n}=a_1+a_2\tanh[a_3(\bar{\psi}-a_4)]+a_5\bar{\psi},
\end{equation}
\begin{equation}
    \bar{T}=b_1+b_2\bar{\psi}+b_3\bar{\psi}^2+b_3\bar{\psi}^3,
\end{equation}
and
\begin{equation}\label{Te in}
     \bar{T}_e=c_1+c_2\tanh[c_3(\bar{\psi}-c_4)]+c_5\bar{\psi},
\end{equation}
where the numerical parameters are given in table \ref{table param}.

\begin{table}
\begin{center}
\begin{tabular}{l|lll|lll|l}
$a_1$ & 0.5951  &  & $b_1$ & 1.0000       &  & $c_1$ & 1.2648  \\ 
$a_2$ & 0.3965  &  & $b_2$ & -0.0459 &  & $c_2$ & -0.2798 \\
$a_3$ & -1.2929 &  & $b_3$ & 0.0038 &  & $c_3$ & 1.3578  \\ 
$a_4$ & 9.3942  &  & $b_4$ & -0.0007 &  & $c_4$ & 9.0470  \\ 
$a_5$ & -0.0075 &  &       &         &  & $c_5$ & -0.0871 \\ 
\end{tabular}

\caption{Numerical values for the parameters of the functions in \eqref{density in}-\eqref{Te in}.}
\label{table param}
\end{center}
\end{table}
\bibliographystyle{jpp}

\bibliography{BibPaper1}
\end{document}